\documentclass[final]{siamltex}
\usepackage{amssymb}
\usepackage{latexsym}
\input{epsf}
\newtheorem{theo}{Theorem}
\newcommand{\vs}{\vspace{2mm}\newline\noindent}
\newcommand{\p}{.13\textwidth}
\newcommand{\s}{.75\textwidth}

\newcommand {\beq} {\begin{equation}}
\newcommand {\enq} {\end{equation}}
\newcommand {\ra} {\rangle}
\newcommand {\la} {\langle}

\newtheorem{lemm}{Lemma}

\newtheorem{deff}{Definition}

\newcommand{\C}{{\cal C}}
\newcommand{\LL}{{\bf L}}

\newcommand{\calE}{{\cal E}}
\newcommand{\calR}{{\cal R}}

\newcommand{\calH}{{\cal H}}
\newcommand{\calF}{{\cal F}}
\newcommand{\calT}{{\cal T}}
\newcommand{\calG}{{\cal G}}
\newcommand{\calL}{{\cal L}}

\newcommand{\calM}{{\cal M}}
\newcommand{\calN}{{\cal N}}

\def\C{{\bf C}}

\newcommand{\Tr}{\mathop{\rm Tr}\nolimits}
\newcommand{\PR}{\mathop{\rm Pr}\nolimits}

\title{Fault-Tolerant Quantum Computation With Constant Error Rate
\thanks{A preliminary version of this paper, under the name ``Fault-Tolerant Quantum Computation With Constant Error'', was published in {\em Proceedings of the 29th Annual ACM Symposium on Theory of Computing (STOC)}
 1997 }}
\author{Dorit Aharonov\thanks{School of Mathematics, Institute for Advanced Studies, Princeton, New Jersey}
 \and Michael Ben-Or\thanks{Department of Computer Science, The Hebrew University, Jerusalem, Israel} }
\date{18 June 1999}
\begin{document}
\maketitle

{~}

{~}

\parbox{\p}{Key Words:\newline\hfill}\parbox{\s}{Quantum computation, 
Noise and Decoherence, Density matrices, Concatenated
 quantum
error correcting codes, Polynomial codes, Universal quantum gates}
\vs
%\parbox{\p}{AMS Subject Classification:\newline\hfill}\parbox{\s}{}
%\vs

{~}

\begin{abstract}
Shor has showed how to perform fault tolerant quantum
computation when the probability for an error in a qubit or a gate,
$\eta$, decays with the size of the computation polylogarithmically,
an assumption which is physically unreasonable.  This paper improves
this result and shows that quantum computation can be made robust
against errors and inaccuracies, when the error rate, $\eta$, is
smaller than a constant threshold, $\eta_c$.  The cost is
polylogarithmic in space and time.  The result holds for a very
general noise model, which includes probabilistic errors, decoherence,
amplitude damping, depolarization, and systematic inaccuracies in the
gates.  Moreover, we allow exponentially decaying correlations between
the errors both in space and in time.  Fault tolerant computation can
be performed with any universal set of gates.  The result also holds
for quantum particles with $p>2$ states, namely qupits, and is also
generalized to one dimensional quantum computers with only nearest
neighbor interactions.  No measurements, or classical operations, are
required during the quantum computation.

We use Calderbank-Shor Steane (CSS) quantum error correcting codes,
generalized to qupits, and introduce a new class of CSS codes over
$F_p$, called polynomial codes.  It is shown how to apply a universal
set of gates fault tolerantly on states encoded by general CSS codes,
based on modifications of Shor's procedures, and on states encoded by
polynomial codes, where the procedures for polynomial codes have a
simple and systematic structure based on the algebraic properties of
the code. Geometrical and group theoretical arguments are used to
prove the universality of the two sets of gates.  Our key theorem
asserts that applying computation on encoded states recursively
achieves fault tolerance against constant error rate.  The
generalization to general noise models is done using the framework of
quantum circuits with density matrices.  Finally, we calculate the
threshold to be $\eta_c\simeq 10^{-6}$, in the best case.

The paper contains new and significantly simpler proofs
for most of the known results which we use. For example, we give a simple
proof that it suffices to correct bit and phase flips, we significantly
simplify Calderbank and Shor's original proof of the correctness of CSS
codes. We also give a simple proof of the fact that two-qubit gates
are universal. The paper thus provides a self contained and complete
proof for universal fault tolerant quantum computation.
\end{abstract}

%\vspace*{3in}
%\pagebreak
\section{Outline} 
Quantum computation
 has recently gained a lot of attention, 
 due to oracle results\cite{simon, bv,bert3} and  
quantum algorithms\cite{deutsch3, simon, grover1},
in particular Shor's factorization algorithm \cite{shor1}. 
These results
indicate a possibility that quantum computers are exponentially more powerful than 
classical computers.  
It is yet unclear whether and how quantum computers will
 be physically realizable\cite{cirac1, cory4,gershenfeld, jones2,lloyd3,loss,pellizzari,privman,steane6}
but as any physical system, they {\em in principle}
will be subjected to noise, such as decoherence
\cite{zurek1,stern1},
and inaccuracies.
Thus, the question of correcting noise cannot be separated from the complexity 
questions. 
Without error corrections, the effect of noise will accumulate and 
ruin the entire computation\cite{unroh1,decoherence,palma,miquel1,barenco6},
and  hence the computation must be protected.
For classical circuits, von Neumann has shown already in $1956$ 
 that the computation can be made robust 
to noise\cite{neumann}. 
However, the similar question for quantum systems is much more complicated. 
Even the simpler question of protecting  quantum information is harder 
than the  classical analogue because one must also protect the
 quantum correlations between the quantum bits (qubits).
It was argued by scientists that due to the continuity of quantum states, and to the 
fact that quantum states cannot be cloned\cite{wootters}, it will be impossible
to protect quantum information from noise\cite{unroh1, landauer1}. 
Despite these pessimistic beliefs, Shor discovered
a scheme to reduce the effect of decoherence\cite{shor2}. 
Immediately after that,  Calderbank and Shor\cite{calshor} and Steane\cite{steane1}
 showed
 that good quantum error correcting codes 
exist, a result which was followed by many explicit examples 
of quantum codes
(e.g.\cite{laflamme2,steane1}). A
 theory for quantum error correcting codes
was developed\cite{knill3}, and a group theoretical framework for almost all codes
was found
\cite{calderbank3,gf4,gottesman1}. 
The existence of quantum error correcting codes, however, 
 does not imply the existence of noise resistant 
  quantum computation, since  due to computation  the faults propagate. 
One must be able to compute without allowing the errors to
 propagate too much, while error corrections should be made in the presence of noise. 
 Recently Shor\cite{shor3}
 showed how to use quantum codes in order to perform 
fault tolerant quantum computation in the presence of probabilistic
errors,  when the {\it error rate}, or the  
 fault probability each time step, per qubit or gate,
 is polylogarithmically small.
This assumption is physically unrealistic. 
In this paper we improve this result  
and show how to perform fault tolerant quantum 
 computation in the presence of constant error rate,
 as long as the error rate
 $\eta$, is smaller than 
some constant threshold, $\eta_0$.
The result holds for a very general noise model, which includes,
 besides 
%independent and quasi independent
 probabilistic errors, also
 decoherence, amplitude and phase damping,
 depolarization, and 
  systematic inaccuracies in the gates.
Fault tolerance can be achieved using any universal set of gates.  
The result also holds when working with quantum particles of more than 
two states, instead of qubits. 
Our scheme can be generalized  to work also in the case in which 
the quantum computer is a one dimensional array of 
qubits or qupits, with only nearest neighbor interactions. 
No measurements, or classical operations, 
 are required during the quantum computation.
The cost is polylogarithmic in the depth  and size of the quantum circuit. 
The main assumption on the noise is that of locality, i.e. 
the noise process in different
 gates and qubits is independent in time and space. 
This assumption can be slightly relaxed, by allowing exponentially decaying correlations in both space and time. 
Such assumptions are made also in the classical scenario, and are likely to 
hold in physical realizations of quantum computers. 
Thus, this paper settles the question of quantum computation 
in the presence of local noise.

Let us first describe the computational model with which we work. 
The standard  model\cite{deutsch1,deutsch2,yao} of quantum circuits with unitary gates, 
allows only unitary operations on qubits. 
However,  noisy quantum systems are not isolated from 
other quantum systems, usually referred to as the environment. 
 Their interactions with the environment are unitary, but when restricting 
the system to the  quantum computer alone the operation on the 
system is no longer unitary, and the state of the quantum circuit 
is no longer a vector in the Hilbert space. 
It is possible to stay in the unitary model, keeping track of the 
state of the environment. However, this environment is not part 
of the computer, and it is assumed that we have no control 
or information on its state. 
We find it more elegant to follow the framework of the physicists, and to work   
 within the model of quantum circuits with mixed states defined 
by Aharonov, Kitaev and Nisan\cite{aharonov3}.  
In this model, the state of the set of qubits is always defined: It is 
a  probability distribution over pure states, i.e. a 
{\it  mixed  state}, or a density matrix,  
 and not merely a pure state as in
the standard model. 
The quantum gates in this model are not necessarily unitary: 
any  physically allowed operator on qubits is a quantum gate. 
In general, it is easiest to think of a general 
physical operator  as a unitary operation on the system of qubits and any number of
 extra qubits, followed by an operator which  discards the extra qubits. 
In particular, one can describe in this model
 an operation which adds a blank qubit 
to the system, or discards a qubit. 
The model of quantum circuits with mixed states is equivalent in computational power 
to the standard model of quantum circuits\cite{aharonov3}, but is more appropriate to work with when 
dealing with errors.

The noise process can be described naturally in this model as follows. 
Since noise  is a dynamical process which depends on time,
the circuit will be leveled, i.e. gates will be applied at discrete  time steps.
Between the time steps, we add the noise process. 
The simplest model for noise is the  probabilistic 
process: Each time step, 
 each qubit or gate  undergoes  a fault (i.e. an arbitrary quantum operation)
 with independent probability $\eta$,
 and $\eta$ is referred to as the {\it error  rate}.
The probabilistic noise process can be generalized to 
 a more realistic model of noise, which is the following: 
Each qubit or (qubits participating in the same gate), each time step, 
 undergoes a physical operator which is at most $\eta$
far from the identity, in some  metric on operators. 
This model includes, apart from  probabilistic errors, also
 decoherence, 
amplitude and phase damping and systematic inaccuracies in the gates. 
Two important  assumptions were  made in this definition:
Independence between different faults in space, i.e. locality,
and independence in time, which is called 
the Markovian assumption.
%\cite{lindbland}.
It turns out that for our purposes, we can  
release these two restrictions slightly, to allow 
exponentially decaying correlations in both time and space,  
and all the results of this paper will still hold. 
However, we will turn to the generalized noise model
 in a very late stage of this
 paper. 
Meanwhile,  it is simpler to keep in mind the independent probabilistic noise model.

Before we explain how to make quantum circuits fault tolerant, let us 
discuss how to  protect quantum information against noise using
  quantum error correcting codes. 
As in classical linear block codes, a quantum error correcting code
 encodes the state of each qubit 
on a {\it block} of, say, $m$ qubits. 
The encoded state is sometimes called the {\it logical} state. 
The code is said to correct $d$ errors if the logical state is
 recoverable given that  
not more than $d$ errors occurred in each block.
The difference from classical codes is that  quantum superpositions should 
be recoverable as well, so one should be able to protect 
quantum coherence, and  not only basis states. 
Calderbank and Shor\cite{calshor} and Steane \cite{steane1} 
were the first to construct such quantum codes, which are now called CSS codes.  
The  basic ingredients are two observations. 
First, the most general fault, or quantum operation,
  on a qubit  can be described as a linear combination of 
four simple operations: The identity, i.e. no error at all, 
 a bit flip ($|0\ra \leftrightarrow |1\ra$), 
 a phase flip ($|0\ra\longmapsto  |0\ra, |1\ra\longmapsto  -|1\ra$)  
 or  both a bit flip and a phase flip. 
We give here a simple proof of the fact, first proven by Bennett {\it at. al.}\cite{bennett14}, 
 that it suffices to correct only these four basic errors. 
In order to correct  bit flips, one can use analogues of
classical error correcting codes. To correct phase flips, 
 another observation comes in handy: A phase flip 
   is actually a bit flip in the Fourier transformed basis, and hence, 
one can correct bit flips  in the original basis, and then correct 
bit flips in the Fourier transformed basis, which translates to correcting 
phase flips in the original basis. 
In this paper we give an alternative proof for CSS codes, which 
generalized these ideas to quantum particles with $p> 2$ states, 
and is significantly simpler than the original proof of Calderbank and Shor\cite{calshor}. 

We define a new class of quantum codes, which are called 
{\it polynomial codes.}
The idea is based on a theorem by Schumacher and Nielsen\cite{schumacher2}
which asserts that a quantum state can be corrected only if 
no information about the state has leaked to the environment 
through the noise process. 
More precisely, if the reduced density matrix on any $t$ qubits 
does not depend on the logical qubit, 
then there exists a unitary operation which 
recovers the original state even if the environment
 interacted with $t$ qubits, i.e. $t$ errors occurred. 
This is reminiscent of  the situation in classical secret sharing schemes:
We should divide the ``secret'', i.e. the logical qubit, among many 
parties (i.e. physical qupits) such that no $t$ parties share any 
information about the secret. 
This analogy between secret sharing and quantum codes suggests 
that secret sharing schemes might prove useful in quantum coding theory. 
Here, we adopt the scheme suggested by Ben-Or, Goldwasser 
and Wigderson\cite{bgw}, who suggested to use random polynomials, evaluated 
at different points in a field of $p$ elements, as  a way to divide 
a secret  among a few parties. 
A random polynomial of degree $d$ is chosen, 
and then each party gets the evaluation of the polynomial at a different 
point in the field $F_p$. The secret is the value of the polynomial 
at $0$. To adopt this scheme to the quantum setting, we simply replace the 
random polynomial by a
 superposition of all polynomials, to get a quantum code. 
It turns out that we get a special case of the CSS codes.
Polynomial codes are useful from many aspects. 
First, they have a very nice algebraic structure, 
which allows to manipulate them easily. 
This allows us to apply fault tolerant  operations on states encoded 
by polynomial codes in a simple and systematic way,  as 
we will see later. 
Polynomial codes might also have  
nice applications to other areas in quantum information theory, 
as was demonstrated  recently by
 Gottesman {\it et al}\cite{gottesman9} who used polynomial codes for   
 quantum secret sharing.  
We will use both polynomial codes and general CSS codes in our fault tolerant scheme.

In order to protect a quantum computation against faults, one 
can try to compute not on the states themselves, but on states encoded by 
 quantum error correcting codes.
Each gate in the original computation will be replaced by a ``procedure'', 
which applies the encoded gate on the encoded state. 
Naturally, in order to prevent accumulation of errors, 
error corrections should be applied frequently. 
Unfortunately, computing on encoded quantum states does not
automatically provide protection against faults, even if error
corrections are applied after every procedure.  The problem lies in the 
fact that during the
computation of the procedure, faults can {\it propagate} to ``correct'' qubits.
This can happen if a gate operates on a damaged qubit and some 
``correct'' qubits- in general, this can cause all the qubits that 
participate in the gate to be damaged.
The procedures should thus be designed carefully, in such a way that 
 a fault during the operation of the procedure 
  can only effect a small number of qubits in each block. 
We refer to such procedures as fault tolerant procedures. 
It turns out that many gates can be applied {\it bitwise}, meaning that 
applying the gate on each qubit separately, achieves the desired 
gate on the encoded state. Unfortunately, not all gates can be applied fault tolerantly 
in such a simple way, and sometimes a very complicated procedure is needed. 
We are actually looking for a pair consisting of a
 quantum code which can correct $d$ errors, 
and a corresponding {\it universal} set of gates, such that
their procedures, with respect to the code,
 allow one fault to spread to at most $d$ qubits.
Since the error corrections, encoding, and decoding procedures
 are also subjected to faults, we need them to
be fault tolerant too.

In \cite{shor3}, Shor introduced a universal set of gates, which 
we denote by ${\cal G}_1$, and  showed how to apply the gates fault tolerantly on
 encoded states.
 It turns out that almost
all of the gates in ${\cal G}_1$ can be applied bitwise. 
The complicated procedures are the fault tolerant error correction and 
encoding, and  the Toffoli gate. 
Shor used noiseless measurements and classical computation extensively in these 
procedures.  
Here, we modify Shor's procedures so that they can  be performed within the 
framework of our model, i.e. all operations are quantum operations, subjected to noise.  
 Basically, we follow Shor's constructions, with some 
additional tricks. 
We remark here that in this paper we make an effort to show that 
fault tolerance can be achieved without using  classical operations or measurements. 
This fact is desirable because of two reasons. 
One is purely theoretical: One would like to know that 
measurements and classical operations are not essential, 
and that the quantum model is complete in the sense that 
it can be made fault tolerant within itself.  
The other reason is practical: In some suggestions for physical realizations
of quantum computers, such as the NMR computer, it might be very 
hard to incorporate measurements and classical computations 
during the quantum computation. 
Thus, we require that all procedures are done using only quantum computation.

For polynomial codes, we introduce a new universal set of gates which
 we denote by ${\cal G}_2$, and show how to perform these gates 
fault tolerantly on states encoded with polynomial codes. 
Here is where the advantage of polynomial codes come into play. 
Unlike the construction of the fault tolerant Toffoli gate  for CSS codes, 
(and other fault tolerant constructions) which   is a long and complicated 
 sequence of tricks, the fault tolerant procedures for polynomial codes 
all have a common relatively  simple structure, 
which uses the algebraic properties of the polynomials.  
To perform a  gates in  ${\cal G}_2$, we  apply two steps. 
First the gate  is applied pit-wise. This always achieves the correct operation of the gate, 
but sometimes the final state is encoded by polynomial codes with degree
which is twice the original degree  $d$. The second step is therefore 
a degree reduction, which can be applied using interpolation techniques, 
as the quantum analogue of the degree reductions used in \cite{bgw}. 
Thus, the fault tolerant procedures for polynomial codes have a 
simple systematic structure.

Next, we prove the universality of the two sets of gates 
which we use, ${\cal G}_1$ for the CSS codes and 
 ${\cal G}_2$ for the polynomial codes. 
Universality of a set of gates  means that 
 the subgroup  generated by the set of gates  is dense in 
the group of unitary operations on $n$ qubits, $U(2^n)$,  (for qupits the group is $U(p^n)$). 
Kitaev\cite{kitaev0} and Solovay\cite{solovay} showed, by a beautiful proof which uses 
Lie algebra and Lie groups,  that universality of a set ${\cal G}$ guarantees 
 that any quantum computation can be approximated by a circuit which uses ${\cal G}$
with only polylogarithmic cost.  
Our proof of universality of  ${\cal G}_1$ is a simple reduction to  a 
 set of gates, which 
was shown to be universal by Kitaev\cite{kitaev0}. 
The proof that  ${\cal G}_2$ is universal is more involved. 
The idea is first to generate, using gates from ${\cal G}_2$, 
  matrices with eigenvalues which are not integer roots of unity, a fact which is proved using
field theoretical arguments. This enables us to show that certain  $U(2)$
subgroups are contained in the subgroup generated by ${\cal G}_2$,
 where these $U(2)$ operate on non orthogonal subspaces,  
which together span the entire space.  
  To complete the proof, we prove a geometrical lemma 
 which asserts that  if $A$ and $B$ 
are non orthogonal then   $U(A)\cup U(B)$ generate the whole group  $U(A\oplus B)$. 
Thus we have that both sets of gates which are used
 in our scheme are universal.

We proceed to explain how a quantum code accompanied with fault tolerant 
procedures for a universal set of gates can be used to enhance the reliability 
of a quantum circuit. 
We start with our original quantum circuit,  $M_0$.
The circuit $M_1$ which will compute on encoded states
is defined as a simulation of $M_0$. 
A qubit in $M_0$  transforms  to a block of $m$ qubits
 in $M_1$, and each gate in $M_0$ transforms to a
 fault tolerant procedure in $M_1$ applied on the corresponding blocks.  
Note that procedures might take  more than  one time step to perform.  
Thus, a time step in $M_0$ is mapped to a time interval 
in  $M_1$, which is called a {\it working period}. 
In order to prevent accumulation of errors, at the end of each working period an
 error correction is applied  on each block.
The working period now consists of two stages: a computation stage
and a correction stage.
The idea is therefore to apply alternately computation stages and
 correction stages, hoping that during the computation stage 
the damage that accumulated is still small enough so that 
the corrections are still able to correct it.

We now want to show that the reliability of the simulating circuit
 is larger than that of the original circuit. 
In the original circuit, the occurrence of one fault may cause the 
failure of the  
computation. In contrast, the simulating circuit can tolerate
a number of errors, say $k$, in each procedure, since they are immediately 
corrected in the following error correcting stage.  
The effective noise rate of $M_1$
 is thus the probability for more than $k$ faults in
a procedure. This effective error rate will be  smaller than the actual error rate $\eta$, 
if $\eta$ is smaller than a certain threshold, $\eta_c$, which 
depends on the size of the procedures, and on the number of errors which the 
error correcting code can correct. If $\eta< \eta_c$, $M_1$ will be more reliable than 
$M_0$.  
 For fault tolerance, 
we need the effective error rate to be polynomially small, to ensure that 
with constant probability no effective error occurred.
In \cite{shor3} Shor uses quantum CSS codes which encode one qubit on polylog$(n)$
to  achieve fault tolerance with the error rate $\eta$ being polylogarithmically small.

The main goal of this paper is to improve this result and to show that 
quantum computation can be robust even in the presence of constant 
error rate.  Such an improvement 
from  a constant error rate $\eta$
to polynomially small effective noise rate
is impossible to achieve in the above method.
Instead, reliability in the presence of constant error rate 
 is achieved by applying the above scheme recursively. 
The idea is simple: as long as the error rate is below the threshold, 
 the effective noise rate of the simulating circuit 
can be decreased by simulating it again, and so on
 for several levels, say $r$,  to get the final circuit $M_r$.  
It will suffice that
 each level of simulation improves only slightly
 the effective noise rate,   since the 
improvement is exponential in the number of levels. 
Such a small improvement can be achieved when using a code of constant 
block size.

The final circuit, $M_r$, has an hierarchical structure. 
 Each qubit in the original circuit transforms to 
a block of qubits in the next level, and they in their turn transform to 
a block of blocks in the second simulation and so on.
A gate in the original circuit transforms to a 
procedure in the next level, which transforms to a larger procedure
containing smaller procedures in the next level and so on.
 The final circuit computes in all the levels: 
The largest procedures,  computing on the largest (highest level) blocks, 
 correspond to operations on qubits in
 the original circuit.
The smaller procedures,  operating on smaller blocks,
correspond to computation in lower levels.
Note, that each level simulates the error corrections in the 
previous level, and adds error corrections in the current level.
The final circuit, thus, includes error corrections of all the levels,
where during the computation of error corrections of larger blocks 
 smaller blocks of lower levels are being 
corrected.
The lower the level, the more often 
error corrections of this level are applied, which is in correspondence with 
the fact that smaller blocks
 are more likely to be quickly damaged.

The analysis of the scheme turns out to be quite complicated. 
To do this, we  generalizes the works of Tsirelson \cite{tsirelson1} and
 G\'ac's \cite{gacs} to the quantum case.  
First, we distinguish between 
two notions: The error in the state of the qubits, i.e. the set 
of qubits which have errors, and the 
actual faults which took place during the computation, namely the fault path, 
which is a list of points in time and place where faults had occurred (in  
a specific run of the computation.)
We then define the notion of sparse errors and sparse fault paths. 
Naturally, due to the hierarchical structure of this scheme, these notions 
are defined recursively. 
A block in the lowest level
 is said to be
 close to it's correct state  if it does not have too many errors,  
and a higher level block is ``close'' to it's correct state if
it does not contain too many blocks of the previous level which are 
far from their correct state. If all the blocks of the highest level are close to being correct, 
 we say that the set of errors is {\it sparse}.
Which set of faults does not cause the state to be too far from correct in the
 above metric? The answer is recursive too:
A computation of the
 lowest level procedure is said to be undamaged if not too many 
faults occurred in it. Computation of higher level procedures
 are not damaged if they do
not contain too many lower level procedures which are damaged.
A fault path will be called {\it sparse}, if the computations
 of all the
 highest level procedures are not damaged.
The proof of the threshold result is done by showing that the probability 
for  ``bad'' faults, i.e.  the probability for the set of faults 
not to be sparse decays exponentially   with the number of levels, $r$.
Thus, bad faults are negligible, if the error rate is below the threshold. 
 This is the easy part. 
The more complicated task is to show that the ``good'' faults are indeed good: 
i.e. if the set of faults is sparse
enough, then the set of errors is also kept sparse until the end of the computation. 
 This part is done using induction on the number 
of levels $r$, and is quite involved.

So far, we have dealt only with probabilistic faults. 
Actually, the  circuit generated by 
the above recursive scheme is robust also against 
 general local noise.
We prove this by writing the noise operator on each qubit as the identity plus 
a small error term. Expanding the  error terms in powers of $\eta$, 
we get a sum of terms, each corresponding to a different fault path. 
The threshold result for general noise 
is again proved by dividing the faults to good and bad parts.   First, we show that 
the bad part is negligible  i.e. the  norm of the 
sum of all terms which correspond to non-sparse fault paths 
is small. The proof  that the good faults are indeed good, 
i.e. that the error in the terms corresponding to sparse fault paths
is sparse, is based on the proof for probabilistic errors, 
 together with some linearity considerations. 

So far, what we have shown is that fault tolerant computation can be achieved
using one of the two sets of gates, ${\cal G}_1,{\cal G}_2$. In fact, 
fault tolerant computation can be performed using any universal set of gates. 
This last generalization is shown by first designing a circuit which uses 
gates from $\calG_1$ or $\calG_2$, and then approximate 
 each gate in the final circuit, up to constant accuracy, 
 by a constant number of gates from the desired set of gates
which we want to work with.  
This completes the proof of the threshold result in full generality.

The threshold result is shown to hold also for  one dimensional quantum circuits. 
These are systems of $n$ qubits arranged in a line, where  gates are applied only 
on nearest neighbor qubits, and all qubits are present from the beginning. 
In this model, one cannot insert a new qubit between two existing qubits,
in order to preserve the geometry of the arrangements.  However, the non unitary 
operation which  initializes a qubit in the middle of the computation 
is allowed. To apply  fault tolerant procedures in this case, 
we use swap gates in order to bring far away qubits together. 
We show that these swap gates do not cause the error to propagate too much.

Finally, it is left to estimate the exact threshold in all the above cases. 
We give a general simple formula to calculate the threshold in the error rate, 
 given the fault tolerant procedures. 
This formula depends only on two parameters of these procedures: 
 the size of the largest procedure used in our scheme, and 
the number of errors which the code corrects. 
The size of the procedures depend strongly on many variables, 
such as the geometry of the system, 
the exact universal set of gates which is used, 
and  whether  noiseless classical computation can be 
used during the computation. 
We give here an estimation of the threshold in one case,
in which no geometry is imposed on the system, 
error free 
classical operations and measurements  
can be performed during the computation, 
and the codes which are used are polynomial codes. 
The threshold which is achieved in this case is 
$\simeq 10^{-6}$.
We did not attempt to optimize the threshold, 
and probably many improvements can be made which can save a lot of 
gates in our suggested procedures. 

Some of the assumptions we made in this work cannot be released. 
First, the choice to work with circuits which allow parallelism is necessary,
 because sequential quantum computation, such as quantum  Turing machines,  
cannot be noise resistant\cite{aharonov2}, since error corrections cannot be applied 
 fast enough to prevent accumulation of errors. It is also crucial that 
 we allow 
qubits to be input at different times to the circuit unlike (Yao\cite{yao}), because
this provides  a way to discard entropy which 
is accumulating in the circuit. Without this assumption, i.e. 
under the restriction that all qubits
are initialized at time $0$, it is impossible to compute 
fault tolerantly without an exponential blowup in the size of
 the circuit\cite{aharonov4}.
An interesting question is whether other assumptions, such as the 
locality assumption, can be relaxed. A partial positive answer in this direction was given 
by Chuang {\it et. al.}\cite{lidar}. In connection with the assumptions made 
for Fault tolerance, see also \cite{aharonov0,steane7}. 

A substantial part of the 
results in this paper were published in a proceeding version in 
\cite{aharonov1}.  
The threshold result was 
 independently described also by  Knill, Laflamme and Zurek\cite{knill1}, 
and by Kitaev\cite{kitaev2,kitaev0}. This was done  using different
 quantum correcting codes,
 and required 
measurements in the middle of the quantum computation. 
In this work we provide a complete proof to all the
details of the fault tolerant quantum computation codes construction 
and in particular we rigorously show that the recursive scheme works. 
Kitaev has shown another, very elegant way to achieve fault tolerance in a different model 
of quantum computation, which works with anyons\cite{kitaev3}.  
After
 the results were first published, it was shown by Gottesman\cite{gottesman2} 
that fault tolerant procedures for the universal set of gates
introduced by Shor can be constructed for any  
stabilizer code, using basically the same technique as presented by Shor 
for the CSS codes, a result which was generalized to qupits\cite{gottesman7}.
Fault tolerant quantum computation 
in low dimensions is also discussed by Gottesman\cite{gottesman8}. 
Non binary codes were defined independently also 
by Chuang\cite{chuang1} and Knill\cite{knill5}.
For a survey of this area see Preskill\cite{preskill2}.

%The result also holds for gates which are not accurate,
%and for weak fault interactions, but this is not described in this paper.
%which after choosing (each time step?) which qubits and gates decohere,
%an adversary can choose the transformation that they all undergo.
%Another fault model which the same method also deals with
%is the following model:
%weak interactions of $k$ qubits with the environment,
% the strength of  which decays exponentially in
%$K$, and their type might differ.

%
%We now turn to the other proof, to show that if all the qubits are
%initialized together, there is no way to compute fault tolerantly
%for more then $O(log(n))$ time steps.
%To prove this, we use the notions of the  Von Newman entropy defined 
%on density matrices, and the information in this density matrix which 
%is just the number of qubits minus the entropy.
%The information in the density matrix can not increase by any computation
%operation, since the information is invariant under unitary operations
%and we show that it only decreases by measurements.
%However, choosing a specific model of single qubit faults,
%we are able to show that this information decreases by a constant factor
%each time step, and therefore approaches zero in an exponential rate.

{\bf Organization of paper:} In section \ref{sec2} we define the model of noisy 
quantum circuits. 
Section \ref{sec3} is devoted to CSS and polynomial codes.
In section \ref{sec4} we describe how to apply fault tolerant 
operations on states encoded by CSS codes. 
In section \ref{sec5} we describe fault tolerant procedures for polynomial codes. 
Section \ref{sec6} proves the universality of the two sets of gates used 
in the previous two sections. 
In section \ref{sec7} we present recursive simulations and the threshold result 
for probabilistic errors. 
In section \ref{sec8} we generalize the result to general local noise, 
and in section \ref{sec9} we generalize to circuits which use any universal set of gates. 
In section \ref{robust} we explain how to deal with exponentially decaying correlations
in the noise. 
Section \ref{sec10} proves the threshold result for one dimensional quantum circuit, 
and in  \ref{sec11} we conclude with possible implications to physics, and 
some open questions.

\section{The Model of Noisy Quantum Circuits}\label{sec2}

\thispagestyle{empty}

In this section we
 recall the definitions of quantum circuits
 with mixed states as defined by Aharonov, Kitaev and Nisan\cite{aharonov3},
 and define noisy quantum circuits. The model is based 
on basic quantum physics, and good references are Cohen-Tanudji\cite{cohentan}, 
 Sakurai \cite{saqurai}, and Peres\cite{peres}.

\subsection{ Pure States}
A quantum physical system in a {\it pure state}
is described by a unit vector in a Hilbert space, 
i.e a vector space with an inner product.
In the {\em Dirac} notation a pure state is denoted by
$|\alpha\rangle$.
The physical system which corresponds to a quantum circuit
 consists of
 $n$ quantum two-state particles, and the
 Hilbert space of such a system is
 ${\cal H}_{2}^{n}= {\cal C}^{\{0,1\}^{n}}$
i.e. a $2^{n}$ dimensional complex vector space.
${\cal H}_{2}^{n}$ is viewed as a tensor product of $n$ Hilbert spaces
of one two-state particle: 
${\cal H}_{2}^{n}={\cal H}_{2}\otimes{\cal H}_{2}\otimes\ldots
\otimes{\cal H}_{2}$.
The $k$'th copy of ${\cal H}_{2}$ will be referred to as the  $k$'th
{\bf qubit}.
We choose a special basis for   ${\cal H}_{2}^{n}$,
which is called the computational basis.
It consists of the $2^{n}$ orthogonal states:
 $|i\rangle,0\le i< 2^{n}$, where $i$ is in binary representation.
$|i\rangle$ can be seen as a tensor product of states in
${\cal H}_{2}$: 
\(|i\rangle=|i_{1}\rangle|i_{2}\rangle....|i_{n}\rangle 
= |i_{1}i_{2}..i_{n}\rangle\),
where each $i_{j}$ gets 0 or 1.
Such a state, $|i\rangle$, corresponds to the j'th particle being in the
state    $|i_{j}\rangle$.
A pure state  $|\alpha\rangle\in{\cal H}_{2}^{n}$ 
is generally a {\em superposition}
 of the basis states:  
$|\alpha\rangle = \sum_{i=1}^{2^{n}} c_{i}|i\rangle$, 
 with $\sum_{i=1}^{2^{n}} |c_{i}|^{2}=1.$
A vector in  ${\cal H}_{2}^{n}$, 
 $v_{\alpha}=(c_{1},c_{2},...,c_{2^{n}})$, written in the computational
basis
representation, with \( \sum_{i=1}^{2^{n}} |c_{i}|^{2}=1\),  corresponds
to the pure state: 
\(|\alpha\rangle = \sum_{i=1}^{2^{n}} c_{i}|i\rangle\). 
 $v_{\alpha}^{\dagger}$, the transposed-complex conjugate of  $v_{\alpha}$,
is denoted  $\langle\alpha|$.
The inner product between $|\alpha\rangle$ and $|\beta\rangle$
is denoted $\langle\alpha|\beta\rangle=
 (v_{\alpha}^{\dagger},v_{\beta})$.
The matrix  $v_{\alpha}^{\dagger}v_{\beta}$
 is denoted as  $|\alpha\rangle\langle\beta|$.

\subsection{ Mixed States}
In general, a quantum system is not in a pure state.
This may be attributed to the fact that
we have only partial knowledge about the system, 
or that the system is not isolated from the rest of 
the universe.
We say that the system  
 is in a {\it mixed state}, and assign with the system
 a probability distribution, or {\em mixture} of 
pure states, denoted by $\{\alpha\}=\{p_{k},|\alpha_{k}\rangle\}$. 
This means that the system is with probability $p_{k}$ in the pure 
state $|\alpha_{k}\rangle$. 
 This description is not unique, as different
mixtures might represent the same physical system.
As an alternative description, physicists use the notion of 
{\bf density matrices}, which is an equivalent 
description but has many advantages.
 A density matrix $\rho$ on 
${\cal H}_{2}^{n}$ is an hermitian (i.e. $\rho=\rho^{\dagger}$)
semi positive definite matrix of dimension $2^n\times 2^n$ 
with trace $\Tr(\rho)=1$.
A pure state $|\alpha\rangle=\sum_{i} c_{i}|i\rangle$ 
is represented by the density matrix:
 \(\rho_{|\alpha\rangle} = |\alpha\rangle\langle\alpha|,\)\, i.e.  
\(\rho_{|\alpha\rangle}(i,j)= c_{i}c_{j}^{*}.\) (By definition, 
$\rho(i,j)=\langle i|\rho|j\rangle$).
A mixture $\{\alpha\}=\{p_{l},|\alpha_{l}\rangle\}$,
is associated with the density matrix\,
\(\rho_{\{\alpha\}} = \sum_{l} p_{l} |\alpha_l\rangle\langle\alpha_l|.\)
This association is not one-to-one, but it is {\bf onto} the
density matrices, because any density matrix describes
the mixture of it's eigenvectors, with the probabilities being 
the corresponding eigenvalues.
 Note that diagonal density matrices correspond to 
probability distributions over classical states.
 Density matrices are linear operators on their Hilbert spaces.

\subsection{Quantum Operators and Quantum Gates}\label{op}
Transformations of density matrices are linear operators on operators
(sometimes called {\em super-operators}). 
$\LL(\calN)$ denotes the
 set of all linear operators on 
$\calN$, a finite dimensional
Hilbert space.
The most general quantum operator is any operator 
  $T:\,\LL(\calN)\to\LL(\calM)$ which sends density matrices to density matrices.
A super-operator is called {\em positive} if it sends positive semi-definite
Hermitian matrices to positive semi-definite Hermitian matrices, 
so $T$ must be positive.  
Super-operators can be extended to operate on larger spaces by
 taking tensor product with the identity operator:
$T:\LL(\calN)\to\LL(\calM)$ will be extended to 
$T\otimes I:\LL(\calN \otimes \calR)
\to\LL(\calM \otimes \calR)$, and it  is also 
required that the extended operator remains positive, and such 
 $T$ is called a {\em completely positive map}.
Clearly,  $T$ should also be trace-preserving. 
It turns out that any super operator
which satisfies these conditions indeed take 
density matrices to density matrices\cite{hellwig}. 
Thus quantum super-operators are defined as follows:

\begin{deff}\label{defgate}
A {\bf physical super-operator}, $T$, from $k$ to $l$ qubits 
is 
a trace preserving, completely 
positive, linear map from density matrices on $k$ qubits to density matrices
on $l$ qubits.
% Its action on the density matrices is denoted as follows: 
%\(\rho\longmapsto g\circ\rho\). 
%(The ``$\circ$'' sign is used for clarity and could be omitted).
\end{deff}

Linear operations on mixed states 
preserve the probabilistic interpretation of the mixture, because
\(T\circ\rho = T\circ
(\sum_{l} p_{l} |\alpha_l\rangle\langle\alpha_l|)=
\sum_{l} p_{l} T\circ( |\alpha_l\rangle\langle\alpha_l|).\)

A very important example of a quantum super-operator 
is the  super-operator
 corresponding to the standard  unitary transformation
on a Hilbert space  
$\calN$,  $|\alpha\ra \mapsto U|\alpha\ra$,\, which sends a quantum state
$\rho=|\alpha\rangle\langle\alpha|$ to the state $U \rho U^{\dagger}$.
Another important  super-operator
 is discarding a set of qubits. 
 A density matrix of $n$ qubits can be reduced to a subset,
 $A$, of $m<n$ qubits. We say that 
the rest of the system, represented by the Hilbert space $\calF=\C^{2^{n-m}}$,
is {\em traced out}, and denote the new matrix by 
$\rho|_A=\Tr_{\calF}\rho$, where 
 $\Tr_{\calF}:\,\LL(\calN\otimes\calF)\to\LL(\calN)$ is defined by: 
\( \rho|_{A}(i,j)= \sum_{k=1}^{2^{n-m}} \rho(ik,jk)\), and  $k$ runs over a basis for $\calF$.
 In words, it means averaging over $\calF$.
If the state of the qubits which are traced out is in tensor product 
with the state of the other qubits, then discarding the qubits 
means simply erasing their state. 
However, if the state of the discarded qubits is not in tensor product 
with the rest, the reduced density matrix is always a mixed state. 
In this paper, a discarding qubit gate will 
 always be applied on  qubits which are in tensor product 
with the rest of the system. 
 Any quantum operation
which does not operate on $\calF$ commutes with this tracing out. 
One more useful quantum super-operator 
 is the one which  adds a blank qubit to the system,
 $V_0:\, |\xi\rangle\mapsto|\xi\rangle\otimes|0\rangle\,: \,\C^{2^n}\to\C^{2^{n+1}}$.\,
This is described by 
 the super-operator
$T: \rho\mapsto \rho\otimes |0\ra\la 0|$.  

A lemma by Choi\cite{choi}, and Hellwig and Kraus\cite{hellwig} asserts that any
physically allowed super-operator from $k$ to $l$ qubits 
can be described as a combination of the 
above three operations: add $k+l$ blank qubits, 
apply a  unitary 
transformation on the $2k+l$ qubits and then trace out
$2k$ qubits. 
One can think of this as the definition of a quantum super-operator. 
We define a quantum gate to be the most general quantum operator.

\begin{deff}
A quantum gate of order $(k,l)$ is a super-operator 
from $k$ to $l$ qubits. 
\end{deff}

In this paper we will only use three types of 
quantum gates:  discard or add 
qubits, and of course unitary gates. 
However, quantum noise will be allowed to be a general 
physical operator. 

\subsection{Measurements}
A quantum system can be {\bf measured}, or observed.
Let us consider a set of positive semi-definite Hermitian operators 
$\{P_m\}$, such that $\sum_m P_m=I$. The measurement is a process which 
yields a probabilistic classical output. For a given density matrix $\rho$,
the output is $m$ with probability $\PR(m)=\Tr(P_m\rho)$. 
Any measurement thus defines a probability distribution over 
the possible outputs. 
In this paper we will only need 
 a basic measurement of $r$ qubits. In this case,
$P_m$ (with $1\le i\le 2^r$) are the projection on the subspace
$S_m$, which is the subspace spanned by basic vectors on which the measured 
qubits have the values corresponding to the string $m$:
$S_m={\rm span}\{\,|m,j\rangle,\  j=1,\ldots,2^{n-r}\}$.
This process corresponds to measuring the value of $r$ qubits,
in the basic basis, where here, for simplicity, we considered
measuring the first $r$ qubits.

To describe a measurement as a super-operator, we replace the 
 classical result of a measurement, $m$,  by the density 
matrix $|m\rangle\langle m|$ in an appropriate Hilbert space $\calM$. The 
state of the quantum system after a projection measurement is also defined; 
it is equal to $\PR(m)^{-1}P_m\rho P_m$. (It has the same meaning as a 
conditional probability distribution). Thus, the projection measurement can be
described by a super-operator $T$ which maps quantum states on the space 
$\calN$ to quantum states on the space $\calN\otimes\calM$, the result being 
diagonal with respect to the second variable:
\beq
  T\rho\ =\ \sum_m\, (P_m\rho P_m)\otimes\Bigl(|m\rangle\langle m|\Bigr)
\enq

\subsection{The Trace Metric on Density Operators}
A useful metric on density matrices, the {\it trace} metric, 
was  defined in \cite{aharonov3}. 
It is induced from  the {\it trace norm} for general Hermitian operators.  
The trace norm on Hermitian matrices is the sum of the absolute values of the eigenvalues.  
\(\|\rho\|=\sum_i |\lambda_i|\)
It was shown in \cite{aharonov3} that 
the distance between two matrices in this metric
$\|\rho_1-\rho_2\|$ is  
 exactly the measurable distance between them: 
It equals to the maximal total variation distance between all
possible outcome distributions 
of the same measurement done on the two matrices.

\subsection { Quantum Circuits with Mixed States}

We now define a quantum circuit:

\begin{deff}
Let $\cal{G}$ be a family of quantum gates.
{\bf A Quantum circuit that uses gates from $\cal{G}$} 
is a directed acyclic graph. 
Each node $v$ in the graph is labeled by a gate $g_{v}\in\cal{G}$ of order
$(k_v,l_v)$. 
The in-degree and out-degree of $v$ are equal $k_v$ and $l_v$, respectively.
\end{deff}

 Here is a schematic example of such a circuit. 

\begin{picture}(150,130)(-30,30)
\put(10,23){\vector(0,1){17}}
\put(10,140){\vector(0,1){15}}
%\put(30,105){\vector(0,1){45}}
%\put(50,105){\line(0,1){80}}
%\put(30,170){\line(0,1){15}}
\put(30,60){\vector(0,1){25}}
\put(30,23){\vector(0,1){17}}
%\put(50,23){\line(0,1){17}}
\put(50,60){\vector(0,1){25}}
\put(2,40){\framebox(56,20){$g_{1}$}}
\put(10,60){\vector(0,1){60}}
\put(2,120){\framebox(16,20){$g_{5}$}}
\put(70,23){\vector(0,1){62}}
%\put(62,50){\framebox(16,20){$g_{3}$}}
\put(70,105){\vector(0,1){15}}
%\put(70,110){\line(0,1){10}}
%\put(70,140){\line(0,1){45}}
\put(90,23){\vector(0,1){27}}
\put(90,70){\vector(0,1){50}}
\put(90,140){\vector(0,1){15}}
\put(110,23){\vector(0,1){27}}
%\put(110,55){\line(0,1){25}}
\put(110,70){\vector(0,1){85}}
%\put(110,170){\line(0,1){15}}

\put(82,50){\framebox(36,20){$g_{3}$}}
\put(62,120){\framebox(36,20){$g_{4}$}}
%\put(130,20){\line(0,1){80}}
%\put(130,120){\line(0,1){30}}
%\put(130,170){\line(0,1){15}}

%\put(150,20){\line(0,1){30}}
%\put(150,70){\line(0,1){80}}
%\put(150,170){\line(0,1){15}}

%\put(142,50){\framebox(16,20){$g_{5}$}}
%\put(102,150){\framebox(56,20){$g_{6}$}}
%\put(122,100){\framebox(16,20){$g_{7}$}}
%\put(22,150){\framebox(16,20){$g_{8}$}}
\put(22,85){\framebox(56,20){$g_{2}$}}

\end{picture}

{~}

\noindent
The circuit operates on  a density matrix as follows:
\begin{deff} {\bf final density matrix}:
Let Q be a quantum circuit.
 Choose a topological sort
for Q: $ g_{t}...g_{1}$.
Where $g_{j}$ are the  gates used in the circuit.
The  final density matrix for an initial density matrix 
$\rho$ is $Q\circ \rho =
g_{t}\circ...\circ g_{2}\circ g_{1}\circ \rho$.
\end{deff}

\noindent $Q\circ\rho$ is well defined and does not depend on the 
 topological sort of the circuit \cite{aharonov3}. 
At the end of the computation, the $r$ output qubits are measured, 
and the classical outcome, which is a string of $r$ bits,  
is the output of the circuit. 
For an input string $i$, the probability for an output string $j$
is 
$ \langle j|\Bigl(Q\circ|i\rangle\langle i|\Bigr)|j\rangle$.

\begin{deff}\label{func}{\bf Computed function:}
Let Q be a quantum circuit, with $n$ inputs and $r$  outputs.
$Q$ is said to compute a function $f$:
\( \{ 0,1 \}^{n} \longmapsto{\{0,1\}^{r}}, \) with accuracy $\epsilon$  if 
for any input $i\in  \{ 0,1 \}^{n}$,
 the probability for the output to be $f(i)$ is greater than $ 1-\epsilon$. 
\end{deff}

\subsection{Quantum Circuits with probabilistic Errors. }\label{loc}
To implement noise in the quantum circuits, we  divide the 
  quantum circuit to  time steps, and the gates are applied at integer time steps.
We permit that qubits are input and output at different
time steps. 
 Faults occur in between time steps, in the {\it locations} of the circuit.
\begin{deff}
A set $(q_1,q_2,..q_l,t)$ is  a {\it location} in the quantum circuit $Q$
if the qubits  $q_1,...q_l$ participated in the same gate in $Q$, 
in time step $t$. and no other qubit participated in that gate. 
For this matter, the absence of a gate is viewed as the identity
gate on one qubit, so if a qubit $q$ did not participate in any gate 
at time $t$, then $(q,t)$ is a location in $Q$ as well. 
\end{deff}

Each location in the circuit exhibits a fault with 
 with independent probability $\eta$.

\begin{deff} 
The list of times and places where faults had occurred, (in a specific run of the 
computation)  is called a fault path. 
\end{deff}

Each fault path, $\calF$,  is assigned a probability $Pr(\calF)$, 
which is a function of the number 
of faults in it. 
A fault at time $t$ at a certain location means that 
 an (arbitrary) 
 general  quantum operation is applied  on the faulty qubits after time $t$. 
Given a certain fault path, we still have to determine the 
 quantum operations that were applied on the faulty qubits.
Denote by $\calE(\calF)$ the choice of faults for the fault path,  say 
 $\calE_t, ... \calE_2,\calE_1$. The
final density matrix for this given choice  $\calE(\calF)$ is easily defined by applying 
the faults in between the gates:
 $Q (\calF)\circ \rho =
\calE_t\circ g_{t}\circ...\calE_2\circ g_{2}\circ\calE_1\circ g_{1}\circ \rho$.
 The final density matrix, $Q\circ \rho$
 of the circuit is defined as a weighted average 
over the final density matrices 
for each fault path:
\beq
Q \circ \rho= \sum_{\calF} Pr(\calF) 
Q (\calF)\circ \rho.
\enq
This final density matrix depends on the exact choice of quantum operations 
applied on the faulty gates. 
However, a noisy quantum computer is said to compute a function $f$ 
with confidence $\epsilon$ 
in the presence of error rate 
$\eta$,  if  the circuit outputs the function $f$ 
with accuracy $\epsilon$ 
 regardless of the exact specification of the quantum operations
 which occurred during the faults. 

%
%We will need notions of development in time of the density matrix
%describing the circuit.
%For a circuit of $n$ qubits, an {\it evolution} is a possible such
%  development, i.e. a general list 
%$\{E(t)\}_{t=0}^{T}$ of density matrices of $n$ qubits. 
%There are special evolutions, which develop exactly according to
% the computation.
%An initial density matrix will evolve, if no faults occur, 
%through such an evolution, called a ``trajectory''.
%We are interested in special trajectories, which correspond to 
%classical inputs.
%The trajectory which corresponds to the input $i$ will be referred to as 
%$i$'s trajectory.

%The {\it space-time} of the noisy quantum circuit is a two dimensional array,
% consisting of all 
% the pairs $(q,t)$, 
%of a qubit $q$ and time $t$, where the qubit $q$ exists at time $t$.
%$V(M)$, the volume of the circuit $M$,  is the number of points in it's
% space-time.

\subsection{Quantum Circuits with General Local Noise}\label{normproperties}

In most of the paper we will use only the probabilistic noise model 
described above. However, our result will be generalized in section \ref{noi} to 
a much more general noise model. 
In the general noise model, we replace the probabilistic faults 
which occur with probability $\eta$ in each location 
 by applying in each location in the quantum circuit an
arbitrary  general quantum operator, on which the only restriction is that 
it within $\eta$ distance  to the identity, in a certain norm of super-operators. 
Thus, in between time steps,  a {\it noise operator} operates on all the qubits. 
For the time step $t$ the noise operator is of the form 
  
\beq \label{localcond1}
\calE(t)=\calE_{A_{1,t}}(t)\otimes \calE_{A_{2,t}}(t)\otimes\cdots\otimes
\calE_{A_{l_t,t}}(t). 
\enq 
 $A_{i,t}$ runs over all possible 
 locations at time $t$, and for each one of them, 
\beq\label{localcond2}
\|{\cal E}_{A_{i,t}}(t)-I\|\le \eta
\enq

\noindent The exact definition of the norm on super-operators
is not used in this paper. 
We merely need the following properties:

\begin{enumerate}
\item\label{oper} \( \|T\rho\| \le \|T\|\|\rho\| \)
\item\label{prod} \(  \|TR\| \le\ \|T\|\|R\|\)
\item\label{tens} \( \|T\otimes R\| =\ \|T\|\|R\|\)
\item\label{unit}
 The norm of any physically allowed super-operator $T$ is equal to $1$. 
\end{enumerate}
A reasonable  norm on super-operators 
 should satisfy these properties. 
One example is the diamond norm  suggested by Kitaev\cite{kitaev0, aharonov3}.

 \subsection{Connection to Specific Noise Processes}
Independent and quasi independent 
probabilistic errors, amplitude  and phase damping, decoherence,   and 
systematic inaccuracies in gates, are all special cases of the general noise model 
defined above. 
For independent  probabilistic errors, where an error ${\cal E}$ occurs with 
probability $\eta$, we can write 
for each one of the super-operators in the product\ref{localcond1}:
\beq
{\cal E}_{A_{i,t}}(t)=(1-\eta) I_{A_{i,t}}+\eta {\cal E'}_{A_{i,t}}(t)
\enq
 where ${\cal E'}_{A_{i,t}}(t)$ is any physical super operator. 
This satisfies the conditions of  local noise \ref{localcond1},\ref{localcond2} with error rate 
$2\eta$.
This is true, in particular, also for probabilistic collapses
 of the wave function, 
a process in which each qubit randomly 
collapses to one of its basis states, for a given basis
(not necessarily the computational basis $|0\ra, |1\ra$).  
${\cal E'}_{A_{i,t}}(t)$ is then the physical super-operator 
which measures the qubit in the chosen basis. 
Depolarization, in which  a qubit is randomly 
 replaced by  a completely random qubit, i.e. a qubit 
in the identity density matrix, is also a special case of probabilistic noise, 
in which  the super-operator  ${\cal E'}_{A_{i,t}}(t)$ 
 replaces all the qubits 
in $A_{i,t}$ by completely random qubits. 
Decoherence\cite{decoherence}, amplitude damping and phase damping\cite{}
are all special cases of 
 interactions between the qubits and the environments,
and as such they can be described in the above models 
by definition, as long as the process is local.  
Systematic inaccuracies in a given gate are described by applying 
the same noise operator on all locations where that gate operated.
The generalizations described in the last subsection describe how to implement 
quasi independent probabilistic errors.

\subsection{Adding correlations to the Noise Model}\label{subaddcor}
Two very important assumptions were made when introducing
our general  model of noise. 
\begin{itemize}
\item Locality: No correlations between environments of different qubits, 
except through the gates. 
\item The Markovian assumption: The environment is renewed each time step, 
and hence no correlations between the environments at different time steps.  
\end{itemize}

Both of these assumptions can be slightly released,
to allow exponentially decaying correlations in both space and time, 
while the results of this paper still hold. 

To add exponentially decaying correlations to the 
 probabilistic noise model, we generalize it in the following way. 
Instead of considering independent probabilities for error in each location, 
we require that
 the probability for a fault path which contains 
$k$ locations is bounded by some constant times the probability
for the same fault path in the independent errors model:
\beq\label{expdecnoi0}
Pr(fault~path~with~k~errors)\le c\eta^k (1-\eta)^{v-k}.
\enq
where $v$ is the number of locations in the circuit.   
Then, an adversary chooses a noise operator which  operates on all 
the qubits in the fault path, without any restrictions. 
The most general case is as follows. 
The adversary adds some blank qubits, which 
are called the environment, at the beginning of the computation. 
Each time step, the adversary can operate a general operator 
on the environment and the set of qubits in the fault path at that 
time step. 
This model allows correlations in space, since the noise operator
need not be of the form of a tensor product of operators on
 different locations. Correlations in  time appear because the environment 
that the adversary added in not renewed each time step, so 
noise operators of different time steps are correlated.  
Note that the independent probabilistic noise process is a special 
case of this process.

% We now add exponentially decaying  correlations to the 
% general noise model.  
% To release the locality constraint, we write 
% the noise operator as a sum of terms operating on all possible unions 
% of locations. We require that the norm of a term in the sum 
% which applies a non trivial operator on $k$ locations and an identity 
% on the other locations  at one time step 
% is:
% \beq
% \|\calL\|\le c\eta^k(1-\eta)^{v-k}.
% \enq 
% To understand how to add correlations in time,  
% we have to go back to the description of quantum operators using the 
% environment, since correlations in time 
% occur due to the system interacting with the 
% same environment, instead of the environment being renewed each time step. 
% We add a set of qubits representing the environment. 
%  We write the final density matrix as a sum 
% over all possible fault paths. For each fault path, at each time step, 
% we allow an adversary to choose a 
%  noise operator to operate on all qubits in the fault path at that 
% time step, and the environment.
% We require that the product (over time) of the norms of the 
%  terms in  a 
% fault path with  exactly  $k$ locations, 
% is bounded by
% \beq\label{expdecnoi}
% \|\calL\|\le c\eta^k(1-\eta)^{v-k}
% \enq 
% where $v$ is the number of locations in the circuit. 
% We note here that the general noise model, in general does not satisfy 
% this condition, and that we restrict our selves to general noise models 
% which do satisfy this condition, when  dealing with non-probabilistic noise 
% processes with correlations.  

\section{Quantum Error Correcting Codes}\label{sec3}
Here we recall the definition of quantum codes, and of 
 Calderbank-Shor-Steane (CSS)
 codes\cite{calshor,steane1}.
These codes were first presented for qubits, that is over the field $F_2$. 
We give here a proof for the fact that CSS codes are quantum codes, 
which is significantly simpler than the original proof\cite{calshor}, 
and generalize it  to $F_p$ for 
any $p>2$.
On our way, we use the fact that 
it suffices to correct phase flips and bit flips, in order
to correct general errors. 
This was first proved by Bennett {\it et. al.}\cite{bennett14}, using
  a beautiful argument  
which involved the notion of group symmetry and a quantum operation called
``twirl''.  
We give here a simple proof of this theorem, 
which uses merely linearity of quantum operators, and is based on an idea by 
Steane\cite{steane1}. 

We then define a new class of quantum codes, 
which we call polynomial quantum codes.  These are 
a special case of CSS codes over $F_p$. 
These codes are based on ideas by Ben-Or {\it et. al.}\cite{bgw} 
who used random polynomials for distributed fault tolerant classical computation. 
This section assumes basic knowledge in  classical error correcting codes, which can be found 
in \cite{lint}.

\subsection{Quantum Codes}
\begin{deff}
A $[k,m]$-quantum code is a subspace of dimension $k$  in the
Hilbert space of $m$ qubits.
\end{deff} 
A word in the code is any  vector in the subspace, 
or more generally any density matrix $\rho$ supported on the subspace.  
We say that the word had $d$ errors in qubits $q_1,..q_d$ if a general physical operator 
on these $d$ qubits, $\calE(q_1,..q_d)$ was applied on $\rho$.
A procedure $\calR$, i.e. a sequence of gates, is said to correct an error 
$\calE(q_1,..q_d)$ on $\rho$ if 

\beq  \calR\circ
\calE(q_1,..q_d)\circ\rho=\rho
\enq
\begin{deff}
Let $C$ be a $[k,m]$ quantum code.
Let $\{|\alpha_i\ra\}_{i=1}^{k}$ be a basis for this  quantum code. 
$C$  is said to correct
 $d$ errors if there exists a 
sequence of quantum gates $\calR$, called an error correction procedure, 
 which corrects any $d$ errors on any $d$ qubits, on any word in the code, i.e 
for any $i,j$ and $\calE(q_1,..q_d)$
 \[ \calR\circ 
\calE(q_1,..q_d)\circ |\alpha_i\ra \la \alpha_j|=|\alpha_i\ra \la \alpha_j|\]
\end{deff}
Note that during the operation of 
$\calR$, it is allowed to use extra qubits, but at the 
end all ancillary qubits are discarded, and the original state 
should be recovered. 

\subsection{From General Errors to Bit Flips and Phase Flips}
A crucial and basic fact\cite{bennett14} in quantum error corrections 
is  that it suffices to correct two basic errors, called  bit flips and 
phase flips,  in a small number of qubits, 
in order to correct a general error on these qubits.
Our proof of this fact  uses merely linearity of quantum operators.
Let us define basic errors as those in which one of the Pauli unitary operators 
occur:  
\begin{equation}
{\cal I} =\left(\begin{array}{cc}1& 0 \\0 & 1\end{array}\right),
 \sigma_x=\left(\begin{array}{cc}0 & 1 \\1 & 0\end{array}\right),
\sigma_y=\left(\begin{array}{cc}0 & 1 \\-1 & 0\end{array}\right),
\sigma_z=\left(\begin{array}{cc}1 & 0 \\0 & -1\end{array}\right).
\end{equation}
$\sigma_x$ is called   a bit flip, since it takes $|0\ra$ to $|1\ra$ and 
 $|1\ra$ to $|0\ra$.
$\sigma_z$ is called a phase flip, as it multiplies $|1\ra$
by a minus sign, while leaving $|0\ra$ as is. 
$\sigma_y$ is a combination of the two, since $\sigma_y=\sigma_z \sigma_x$. 
A basic error on $d$ qubits is obtained by 
taking tensor products of $d$ of the above Pauli matrices. 
More formally, consider all  possible strings $e$ of length $d$
 in the alphabet $I,x,y,z$. 
The set of $4^d$ matrices:  
\beq
\sigma_{e}= \sigma_{e_d}
\otimes\cdots \sigma_{e_1},~~~e\in\{1,x,y,z\}^d\enq
 is a basis
 for operators  on the Hilbert space of 
 $d$ qubits. 
%We will use a convenient notation in which the basic error is tensored with identity 
%on the rest of the qubits, which gives  $\sigma_e$ with
% $e\in\{1,x,y,z\}^m$. $e$ 
%has at most $d$ coordinates different than $1$, and will be called the error vector. 
Before we show that it suffices to correct basic errors, 
 we need two definitions:
\begin{deff}
We say that  $\calR$ corrects basic errors of length $d$ in $C$ 
if for any two basis states of the code, $ |\alpha\ra$ and  $| \alpha'\ra$, 
and any basic error of length $d$, $\sigma_e$ we have 
\end{deff}
 \beq
\calR\circ 
\sigma_e|\alpha\ra \la \alpha'|\sigma^\dagger_e=
|\alpha\ra \la \alpha'|
\enq
\begin{deff}
We say that  $\calR$  {\it detects}   basic errors of length $d$ in $C$ 
if  for any two basis states of the code, $ |\alpha\ra$ and  $| \alpha'\ra$, 
and any two different basic error of length $d$, 
$\sigma_e\not=\sigma_{e'}$, 
\end{deff}
  \beq
\calR\circ 
\sigma_e |\alpha\ra \la \alpha'|\sigma^\dagger_{e'}=0
\enq
The reason for the fact that such $\calR$ is said to detect errors is 
that indeed, if at the end of the procedure 
$\calR$  the error $e$ is written down on an ancilla, 
 $\calR\circ\sigma_e |\alpha\ra=|\alpha\ra\otimes |e\ra$, then 
for two different errors the two ancilla states are orthogonal, and thus 
when discarding the ancilla qubits we get zero. 
In the rest of the paper,
 we will only use error correcting procedures which detect the 
errors. 
We can now prove the following:
\begin{theo}\label{bitphase}
Let $C$ be a quantum code, and let $\calR$ correct and detect any 
basic error  of length $d$ in $C$. 
Then $R$ corrects any general error on any $d$ qubits in $C$. 
\end{theo}

\noindent {\bf Proof:}
 We recall that any noise operator ${\cal L}$ from $d$ to $d'$ qubits 
can be written as a combination of three operators:  adding $d+d'$ blank qubits, 
applying a unitary transformation on the $2d+d'$ and discarding $2d$ qubits. 
The last operator which discards qubits will be denoted  as $\calT$. 
Combining the first two operations, we have in the most general case
the following  operation on $d$ qubits. For all $0\le i\le 2^d-1$, 
\beq
|i\ra \longmapsto |0^{d+d'}\ra|i\ra \longmapsto \sum_{j=0}^{2^d-1}|A_i^j\ra\otimes |j\ra
\enq
 where all the coefficients 
are put into the vectors $|A_i^j\ra$ which are not unit vectors. 
We now use the fact that  the basic errors,$\sigma_e$, on $d$ qubits form  an
 orthonormal basis with respect to the inner product 
 $(V,U)= \frac{1}{2^d}tr(VU^{\dagger})$ for $2^d\times 2^d$ matrices.
We define $A$,  a vector of matrices, or a matrix  
 with entries which are vectors, by  
 $A_{i,j}=|A_i^j\ra$.
For any basic error vector $e$ of length $d$,  define 
$|A_{e}\ra =(A,\sigma_{e})=
 \frac{1}{2}tr(A\cdot\sigma_{e}^{\dagger})$, 
where $A$ should be thought of as a vector of matrices, and 
thus $A\sigma_e^{\dagger}$ is also a vector of matrices, the trace of which 
is a vector. Due to the fact that 
the basic errors are orthonormal, we have 
that  $\sum_j
|A_i^j\ra\otimes |j\ra= \sum_e|A_e\ra\otimes
 \sigma_e|i\ra$, or equivalently
the two first steps of the noise operator can be written as: 
\beq
 \calL=
\sum_{e,e'}|A_e\ra\la A_{e'}|
 \otimes \sigma_e\cdot\sigma_{e'}^\dagger
\enq
Now, $\calR$  will indeed 
correct $\calE=\calT\circ\calL$, by the following: 
\beq
\calR \circ\calE\circ|\alpha\ra\la \alpha'|=\calR\circ
\calT\circ \sum_{e,e'}|A_e\ra\la A_{e'}|
 \otimes \sigma_e|\alpha\ra\la \alpha'|\sigma_{e'}^\dagger=
(\calT\circ \sum_e|A_e\ra\la A_e|)
 \otimes|\alpha\ra\la \alpha'|=|\alpha\ra\la \alpha'|
\enq
where in the second equality we have used the fact that  ${\cal T}$ commutes with $\calR$
because they operate on different qubits, and that  $\calR$ corrects and detects errors. 
In the last equality we used the fact that $ \calT\circ \sum_e|A_e\ra
\la A_e|=c$ does not depend on $|\alpha\ra, |\alpha'\ra$. 
Choosing  $|\alpha\ra=|\alpha'\ra$, $\Tr(\calR \circ\calE\circ|\alpha\ra\la \alpha|)=
\Tr(c|\alpha\ra\la \alpha|)=c$,
but on the other hand  $\calR \circ\calE$ is trace preserving so $c=1$.  $\Box$

\subsection{Calderbank-Shor-Steane Codes}
We give here the definition of CSS codes, which is slightly modified 
from the definition in \cite{calshor}. 
A linear code of length $m$ and dimension $k$
 is a subspace of dimension $k$
in $F_2^m$, where  $F_2^m$ is the $m$ dimensional 
vector space over the field of $F_2$ of two elements. 
Let $C_1,C_2$ be linear  code such that
 $\{0\}\subset C_2 \subset C_1 \subset F_2^m$, and let 
us define a quantum code by 
taking the superpositions of all words in a coset of $C_2$ in $C_1$ to be 
one basis word in the code. 
We have: 

\beq
\forall a\in C_1/C_2: |S_a\ra=\frac{1}{\sqrt{2^{dim(C_2)}}}\sum_{w\in C_2}|w+a\ra.
\enq
Note that $|S_a\ra$ is well defined and does not depend on 
the representative of the coset since 
 if $(a1-a2)\in C_2$ then $|S_{a1}\ra=|S_{a2}\ra$. 
Also, for different cosets the vectors are orthogonal, because if  
$(a1-a2)\not\in C_2$, then 
 $\la S_{a1}|S_{a2}\ra=0.$
Thus, this defines a basis for a subspace of 
 dimension $2^{dim(C_1)-dim(C_2)}$, which is 
 our quantum code.
Note that the support of $|S_a\ra$ are words in the code $C_1$. 
We will see that 
 bit flips can be corrected using  classical error correction techniques
for the code $C_1$.
Before we discuss how to  correct phase flips,  let us define 
a very important quantum gate on one qubit, called the Hadamard gate, 
 or the Fourier transform over $F_2$:
\begin{equation}\label{hadamard}
H=H^{-1}=\left(\begin{array}{ll}
\frac{1}{\sqrt{2}}&\frac{1}{\sqrt{2}}\\
\frac{1}{\sqrt{2}}&-\frac{1}{\sqrt{2}}
\end{array}
\right)
\end{equation}
Observe that 
\beq\label{had} 
H\sigma_z H^{-1} =\sigma_x
\enq
meaning that a phase flip transforms to a bit flip in the 
Fourier transform basis. 
Applying the Hadamard gate on each qubit in  $|S_a\ra$, 
we get the state:
\beq
~~~~~~~~~|C_a\ra=H\otimes H\otimes\cdots \otimes H |S_a\ra=\frac{1}{\sqrt{2^{m+dim(C_2)}}}\sum_{b=0}^{2^m-1}
\sum_{w\in C_2}(-1)^{(w+a)\cdot b}|b\ra=\frac{1}{\sqrt{2^{m-dim(C_2)}}}\sum_{b\in C_2^\perp}
(-1)^{a\cdot b}|b\ra
\enq
which is a superposition of words in $C_2^\perp$, and so to correct phase flips, 
one transforms to the Fourier basis and corrects bit flips in the code 
 $C_2^\perp$. 
These ideas lead to the following theorem by Calderbank and Shor\cite{calshor}. 
We give here a simple proof of this theorem, based on theorem \ref{bitphase}.

\begin{theo}\label{cash}
Let $C_1$ and $C_2^\perp$ be linear codes over $F_2$, of length $m$, 
 such that ${0}\subset C_2\subset C_1\subset F_2^m$, and such that 
$C^\perp_2, C_1$ can correct $t$ errors. 
 Then the subspace spanned by  $|S_a\ra$
for all $a\in C_1/C_2$ is a
$[2^{dim(c_1)-dim(C_2)},m]$ quantum code which can correct $t$ errors. 
The error correction procedure, $\calR$, is constructed by   
 correcting pit flips with respect to $C_1$ in the $S$-basis, 
 rotating to the $C$-basis by applying Fourier transform bit-wise, 
correcting with respect to $C_2^\perp$ and rotating back to the $S$-basis. 
\end{theo}

{\bf Proof:}
We define the procedure $\calR_{C_1}$ to be
a unitary embedding of $m$ qubits to $2m$ qubits, by 
\beq
\calR_{C_1}|i\ra=|w\ra\otimes |e\ra
\enq
for each $i\in F_2^m$,  
 where $w\in C_1$ is a string of minimal distance to  $i$,
and $e\in \{0,1\}^m$  satisfies $w+e=i$. 
Since this is a one to one transformation, it is a unitary embedding, 
and is a possible quantum operator. 
Let $e_b\in \{0,1\}^m$ have at most $t$ $1'$s.
Let $\calE_b$ be an error operator which is a tensor product  of bit flips
($\sigma_x$)
where $e_b$ is one and identity on the other coordinates. 
Then for any $|\alpha\ra, |\alpha'\ra$ supported on 
$C_1$ we have
\beq\label{xx}
R_{C_1}\circ \calE_b\circ|\alpha\ra\la\alpha'|=
|\alpha\ra\la\alpha'|\otimes |e_b\ra\la e_b|
\enq
$\calR_{C_2^\perp}$ is defined similarly:
 \beq
\calR_{C_2^\perp}|j\ra=|w\ra\otimes |e\ra
\enq
  where $w\in C_2^\perp$ is a string of minimal distance to  $j$,
and $e\in \{0,1\}^m$  satisfies $w+e=j$.
 Let $e_f\in \{0,1\}^m$ have at most $t$ $1'$s ($e_f$ for phase flips.). 
Let $\calE_f$ be an error operator which is a tensor product  of phase flips
($\sigma_z$)
where $e_f$ is one and identity on the other coordinates. 
Then for any $|\beta\ra, |\beta'\ra$ supported on 
$C_2^\perp$ we have
\beq\label{yy}
R_{C_2^\perp}\circ \calE_f\circ|\beta\ra\la\beta'|=
|\beta\ra\la\beta'|\otimes |e_f\ra\la e_f|
\enq
 Denote by $\calH$ the operator of applying $H$ 
on every qubit: $\calH=H\otimes H\otimes\cdots \otimes H$.
We claim that the operator 
\beq\label{zz}
\calR=\calT_f\circ\calT_b\circ \calH\circ\calR_{C_2^\perp}\circ\calH\circ\calR_{C_1}
\enq
is the desired error correcting procedures. 
 $\calT_f,\calT_b$ are the operators discarding the 
qubits added for $\calR_{C_2^\perp},\calR_{C_1}$, respectively. 

By theorem \ref{bitphase}, it is enough to show that this procedure 
corrects and detects $d$ basic errors. 
To show that it corrects  $d$ basic errors, write 
the error vector $e$ in two parts, $e_b$ and $e_f$ as follows: 
$e_b$ will be $1$ in the coordinates where $\sigma_x$ or $\sigma_y$ occurred, 
and $0$ elsewhere. $e_f$ will be $1$ in the coordinates where $\sigma_z$
 or $\sigma_y$ occurred, 
and $0$ elsewhere. We can therefore write the
  error operator $\calE$  as 
  a product of two operators, $\calE=\calE_b \calE_f$.
Now, using, in the following order,   equation \ref{zz},  the fact that 
 $\calR_{C_1}$ commutes with $\calE_f$, equation 
 \ref{xx},\ref{had}  and \ref{yy}, we have the desired result:
\begin{eqnarray}\label{bitandphase}
\calR\circ\calE\circ|\alpha\ra\la\alpha'|&=&
\calT_f\circ\calT_b\circ\calH\circ\calR_{C_2^\perp}\circ\calH\circ\calE_f\circ\calR_{C_1}\circ\calE_b\circ
|\alpha\ra\la\alpha'|=\\\nonumber&&
\calT_f\circ\calT_b
\circ\calH\circ\calR_{C_2^\perp}\circ\calH\circ\calE_f\circ|\alpha\ra\la\alpha'|\otimes
|e_b\ra\la e_b|=\\\nonumber&&
\calT_f
\circ\calH\circ\calR_{C_2^\perp}\circ\calH\circ\calE_f\circ|\alpha\ra\la\alpha'|=\\\nonumber&&
\calT_f\circ\calH\circ\calR_{C_2^\perp}\circ(\calH\circ\calE_f\circ\calH)\circ(\calH\circ
|\alpha\ra\la\alpha'|)=
\\\nonumber&&
\calT_f\circ\calH\circ(\calH\circ|\alpha\ra\la\alpha'|)\otimes |e_f\ra\la e_f|=
|\alpha\ra\la\alpha'|. 
\end{eqnarray}
It is left to show that $\calR$ also detects $d$ basic errors.
 This follows from the fact
that if $e\not=e'$, then either $e_f\not= e_f'$ or $e_b\not= e_b'$. 
Now reducing two orthogonal vectors gives zero, i.e. if $e_b\not= e_b'$
$\calT_b\circ |e_b\ra \la e_b'|=0$ and  if $e_f\not= e_f'$
$\calT_f\circ |e_f\ra \la e_f'|=0$. $\Box$.

%(1) For $u\in C1$and $e$ with weight not 
%larger than $t$, correct a state $|u+e>$
%using an ancilla $|A>$, to the state $|u>$ so that the ancilla 
%transforms to some state $|A_e>$ which depends only on $e$.
%(2) Rotate the code from the  $S-$basis  to the $C-$basis
%and apply the same unitary correction operation as before,

\subsection{CSS codes  over $F_p$}
The theory of quantum error corrections can be generalized to 
quantum computers which are composed of quantum particles of  $p>2$ 
states, called qupits. 
To generalize  the notion of bit flips and phase flips to qupits
define  the following two matrices:
\begin{itemize}
\item \(B: ~~~~B|a\ra=|(a+1)\rm{mod}~p\ra \)
\item \(P: ~~~~P|a\ra= w^{a} |a\ra  \)
\end{itemize}
where  $w=e^{\frac{2\pi i}{p}}$.
We will consider combinations of powers of these matrices, 
i.e. the  $p^2$ matrices, 
\beq
B^cP^{c'},~~~ \forall c,c'\in F_p. 
\enq
This set can be easily seen to be an 
 orthonormal basis for  the 
set of $p\times p$ complex matrices, with the inner product
 $(U,V)= \frac{1}{p}tr(UB^{\dagger})$.
Like in the case of qubits, errors of type $B$ transform to errors of type 
$P$ and vice versa, via a Fourier transform, which is defined to be
 
\beq
W: |a\ra\longmapsto\frac{1}{\sqrt{p}}\sum_{b\in F}w^{ab}|b\ra.
\enq

And it can be easily checked that 
 \beq
\forall c \in F_p, ~~
W P^c W^{-1}  = B^{c}~~, ~~W B^c W^{-1}
P^{c}
\enq

We can therefore define CSS codes over $F_p$. The statements and proofs 
of theorems \ref{bitphase} 
and  \ref{cash} are generalized to $F_p$  
using the above definition of bit flips and phase flips, 
where  $F_2^m$ is replaced by  $F_p^m$ 
and Hadamard gate is replaced by the Fourier transform $W$ over $F_p$
everywhere. 

\subsection{Polynomial Quantum Codes}
We define polynomial codes over $F_p$. 
Set $d$ to be the degree of the polynomials we are going to use, 
and set $m$ to be the length of the code. 
$p>m$ will be the number of elements in the field $F_p$
we will be working with. 
Set $\alpha_1,...\alpha_m$ to be $m$ distinct non zero elements of the field
$F_p$. Define the linear codes:
\begin{eqnarray}
C_1 &=&\{(f(\alpha_1)\cdot f(\alpha_m))| f(x)\in F(x), deg(f(x))\le d\}\subset F_p^m\\\nonumber
C_2 &=&\{(f(\alpha_1)\cdot f(\alpha_m))| f(x)\in F(x), deg(f(x))\le d, f(0)=0\}\subset C_1
\end{eqnarray}

We can now define the quantum code:
\beq
\forall a\in C_1 /C_2,~~  |S_a\ra=\frac{1}{\sqrt{p^d}}\sum_{f\in V_1, f(0)=a} |f(\alpha_1)\cdots f(\alpha_m)\ra
\enq
Clearly, $C_2$ has $p$ different cosets in $C_1$, and so 
$C^\perp_1$ has $p$ disjoint cosets in $C_2^\perp$. 
Thus the dimension of the code is $p$, and the code encodes 
exactly one qupit.
Note that this code is a special case of CSS codes, which will be used to prove the following
theorem:
 
\begin{theo}\label{pol}
A polynomial code of degree $d$ with length $m$ over $F_p$ is a $[p,m]$ quantum code which
 corrects $\min\{\lfloor \frac{m-d-1}{2} \rfloor,\lfloor \frac{d}{2}\rfloor\}$
 errors. 
\end{theo}

{\bf Proof:}
Two different words in $C_1$ agree on at most $d$ coordinates, and thus $C_1$ is a linear code of distance $m-d$. It can thus correct and detect
$\lfloor (m-d-1)/2 \rfloor$ errors. 
$C_2^\perp$ is a linear code of minimal distance $\ge d+1$. 
This is true since the projection 
on  any $d$ coordinates of the code $C$ contains all possible strings 
of length $d$, and therefore the only vector of length $d$ orthogonal 
to all the vectors is the $0$ vector.  
Thus, $C_2^\perp$ corrects and detects $\lfloor d/2 \rfloor$ errors. 
 Theorem \ref{pol} follows from theorem \ref{cash}. $\Box$

\section{Computing on States Encoded by CSS codes}\label{explicit}\label{sec4}
CSS codes will be used to perform  computations fault tolerantly:
The idea is to compute on quantum  states encoded by CSS codes.  
Each gate is replaced by a procedure which imitates the operation of
 the gate on the encoded states.
In order for the computation to be fault tolerant, 
the procedures  have to be designed in such a way so that  a small number of errors 
during the procedure  cannot propagate to too many errors
at the end of the procedure,  
before error correction can be applied.
The basic definitions are given next.
We then  show how to perform fault tolerant operations on states encoded 
by CSS codes. 
We follow  Shor's constructions very closely, with
some  additional tricks 
and modifications mainly in the construction of the ancilla state 
required for the Toffoli gate, and 
in the decoding and error correction procedures. 
These modifications are done in order to avoid 
 measurements and classical operations 
during the computation.  In fact,  
this is the only difference between the results 
presented in this section, and the results derived by Shor in \cite{}.

\subsection{Fault Tolerant Procedures on Encoded States- General Definitions}
Say we have a unitary gate $g$ which was applied on the state $|\alpha\ra$ 
in the original circuit. 
We now want to apply a sequence of gates, or a ``procedure'', 
 $P(g)$,  on the state  encoding $|\alpha\ra$, 
such that $P(g)$ will take the encoded $|\alpha\ra$ to the state encoding 
$g|\alpha\ra$.
\begin{deff}
A sequence of gates $P(g)$ is said to encode a gate $g$ for the code $C$ 
if for any superposition   $|\alpha\ra$, 
 $P(g)|S_{|\alpha\ra}\ra=|S_{g|\alpha\ra}\ra$.
\end{deff}

We will want $P(g)$ to be such that an error 
occurring at one gate or qubit during the procedure will not 
affect too many qubits at the end of the procedure, so that 
the error can be corrected. 
A location  $(q_1,..q_l,t)$ effects a  
qubit $q'$ at time $t'>t$ if  there is a path in the circuit
from $(q_1,..q_l,t)$ to $(q',t')$.
\begin{deff}
The ``spread'' of a procedure is the maximal number of qubits
in one block in the output of the procedure, which are effected by
one location in this procedure.
\end{deff}

If we use only procedures with small spread,
the error corrections will still be able  to 
 correct the damage using the undamaged qubits, provided that
 not too many errors happened during the procedure.

The notion of reduced density matrices is useful here:
At the end of a fault tolerant procedure  which operates on a 
state $|\alpha\ra$, the result will be ``correct'' on all qubits except 
those effected by the fault. 
This means that the reduced density matrix on all 
the qubits except those which where effected, is the same as it would have been 
 if no fault occurred at all. 

\subsection{Questions of Ancilla Qubits} 
In some of  the procedures, we will use ancilla qubits, as extra working space. 
At the end of the procedure these qubits will
 be discarded, in order to get exactly the state we need. 
As was explained in subsection \ref{op}, 
we will always discard qubits which are in tensor product with the rest of the system, 
so the operation of discarding means simply erasing their state, and the resulting 
state is a pure state. This is necessary if we want to 
operate unitary operations on the encoded states.  
We will describe a procedure by specifying what it does to 
basic states of the code. 
It is easy to see that if for any input basis state, 
 the state of the ancilla qubits at the end of the procedure 
are in a tensor product with the rest of the qubits, and their state 
 does not depend on 
the input basis state, i.e. \beq
|S_a\ra \longmapsto |S_{g(a)}\ra \otimes |A\ra
\enq
where $A$ is independent of $a$, then 
 for any input superposition for the procedure 
 the ancilla qubits will be in tensor product with
 the rest of the qubits, and thus they can be discarded 
simply by erasing them.

This requirement of independence of the ancilla 
qubits on the basis state can be released in two cases. 
An encoding procedure, is the procedure which takes a block of input bits, $0^m$ 
to the state encoding $0$, $|S_0\ra$, and likewise for $1$. 
For  the encoding procedure, we know that the procedure always gets as an input 
a basis state. Thus, we release the above requirement, 
and demand only that 
the ancilla state is in a tensor product with the computational qubits 
 when the input to the procedure 
is a basis state. (The ancilla might therefore depend on the input, and indeed it will, 
but we are allowed do discard it anyway.)
The requirement of tensor product with the ancilla 
 can also be released in the decoding procedure, since the state will 
not be used any more, and we should only check that it gives the correct 
answer when measured. 

It suffices to check 
the independence of the ancilla  in the case of 
no fault in the procedure. The fact that the
 procedures are fault tolerant guarantees that even if a fault occurred, 
the reduced density matrix of the non-effected  qubits is 
just like in the case in which no fault occurred at all. 

\subsection{Some Restrictions on the CSS Codes}

%In this section we first present a general way to perform initialization, reading
%and correction procedure, fault tolerantly. The presented procedures are
%good whenever we are dealing with the class of quantum codes presented by
%Calderbank and Shor\cite{}. These procedures are fault tolerant inherently- 
%this does not depend on the set of quantum gates, except that we need that 
%this set contains the
% controlled not gate, and the rotation around the x axis gate.

In the following, we will put some restrictions on  the CSS codes
which we will use. 
This is done 
in order to be able to apply  several gates fault tolerantly bitwise, 
as will be seen shortly. 
We require that $C_1$ is a punctured doubly even self
 dual code, and that $C_2=C_1^\perp$. 
A punctured code is a code which is obtained 
 by deleting one coordinate from
a code $C'$, and a punctured self dual code means that we require in addition that $C'$ is  
self dual, i.e. $C'=C'^\perp$.
 We also require that $C'$ is a doubly even code, i.e.
the weight of each word in the code is divisible by $4$.
To see that $C_2=C_1^\perp\subset C_1$, as in the definitions of CSS codes, 
observe that if $v\perp C_1$, then $v0\perp C'$ so $v0 \in C'^\perp=C'$, 
so $v\in C_1$. 
We will denote $C_1=C$, and $C_2=C^\perp$. 

We now claim that there are only two cosets of $C^\perp$ in $C$. 
If the length of $C$ is $m$, then $dim(C'^\perp)=dim(C')=(m+1)/2$. 
Hence $dim(C)=(m+1)/2$ as well, since $|C|=|C'|$, because 
no two words in $C'$ are mapped to the same word in $C$ 
by the punctuation.
Hence,  $dim(C^\perp)=m-(m+1)/2=(m-1)/2$
and so $dim(C)-dim(C^\perp)=1$. 
Observe that $C$ includes the all one vector: $\vec{1}\in C$.
This is true since $1^{m+1}\in C'^\perp$, because $C'$ is even, 
and since $C'$ is self dual, $1^{m+1}\in C'$. Hence $1^m\in C$. 
Observe also that the length $m$ must be odd due to the above considerations. 
This implies that 
$\vec{1}\not\in C^\perp$. 
The two code words in our quantum code can thus be written as 
\begin{eqnarray}
|S_{\vec{0}}\ra&=& \sum_{w\in C^\perp} |w\ra \\\nonumber
|S_{\vec{1}}\ra&=&\sum_{w\in C^\perp} |w+\vec{1}\ra
\end{eqnarray}

$|S_{\vec{0}}\ra$,$|S_{\vec{1}}\ra$,  can be thought of 
as encoding $|0\ra$, $|1\ra$ respectively.
We will make use of the fact that 
$\vec{a}\cdot \vec{b}~\rm{ mod}~2=ab$, for $a,b\in {0,1}$, and that 
$\vec{a}+ \vec{b}=\overrightarrow{a+b}$. 
This fact allows us to  shift easily between 
operations on vectors, and operations  on the bits they represent. 
We will therefore usually omit the vectors in the notations of 
$|S_{\vec{0}}\ra$ and $|S_{\vec{1}}\ra$, unless there is ambiguity.

\subsection{The Set of Gates for CSS codes}
We work with the following set of gates, which we denote 
by ${\cal G}_1$:

\begin{enumerate} 
\item Not: $|a\ra\longmapsto |1-a\ra$, 
\item Controlled not: $|a,b\ra\longmapsto |a,a+b\ra$,
\item Phase rotation: $|a\ra\longmapsto i^a|a\ra$,
\item Controlled Phase rotation: 
$|a\ra|b\ra \longmapsto (-1)^{ab} |a\ra|b\ra$,
\item  Hadamard: $|a\ra\longmapsto\frac{1}{\sqrt{2}}\sum_{b}
 (-1)^{ab}|b\ra$,
\item  Toffoli gate: $|a,b,c\ra\longmapsto |a,b,c+ab\ra$,
\item Swap $|a\ra|b\ra \longmapsto |b\ra|a\ra$,
\item Adding a qubit in the state $|0\ra$,
\item Discarding a qubit.
\end{enumerate}

where all the addition and multiplication are in $F_2$(i.e. mod $2$).
We will show later on that this set of gates is universal. 
We remark here, that ${\cal G}_1$  is by no means
 a minimal universal set of gates, but the fault tolerant procedures become simpler
and shorter if we have a larger repertoire of fault tolerant
 gates that we can use. 
The following theorem  shows how to perform gates from ${\cal G}_1$ 
on encoded states fault tolerantly. 

\begin{theo}\label{procedures-css}
There exists fault tolerant procedures which simulate the operations of 
all the gates from ${\cal G}_1$, on states encoded by 
punctured self dual doubly even CSS codes such that 
one error in a qubit or a gate effects at most four
 qubits in each block at the end of the procedure. 
There exist such procedures also for encoding, decoding and error 
correction. Moreover, all these procedures use only gates from ${\cal G}_1$, 
and in particular do not use measurements.
\end{theo}

To prove the theorem, we first show this for 
the gates which can be applied bit-wise, then for 
the encoding, decoding and error correction procedures, and 
then for the rest of the gates. 

\subsection{Bitwise Fault Tolerant Procedures}
Let $g$ be a gate on $k$ qubits. 
A bitwise procedure of this gate is 
defined by
labeling the qubits in each one of $k$  blocks from $1$ to $m$, and 
then applying the gate $m$ times, each time on all qubits 
with the same label.
Obviously, an error in this procedure can effect only one qubit. 
All the gates in $G_1$ can be applied bitwise, 
except the Toffoli gate and the gate which adds a blank qubit.  
These are the
NOT, Controlled not, Phase rotation, Controlled Phase rotation,
 Hadamard, Swap and Discarding a qubit.
This is trivial for the SWAP gate, and the gate 
which discards a qubit. 
Simple calculations show this also for the other gates.
In the following we omit overall normalization factors, 
since all vectors are known to be unit vectors. 
We also set  $a,b\in C/C^\perp$. 
\beq
 |S_a\ra=\sum_{w\in C^\perp}|a_1+w_1\ra\otimes...|a_m+w_m\ra\longmapsto
\sum_{w\in C^\perp}|a_1+1+w_1\ra\otimes...|a_m+1+w_m\ra=|S_{a+\vec{1}}\ra
\enq
For CNOT, 
\begin{eqnarray}
 |S_a\ra|S_b\ra&=&\sum_{w\in C^\perp}|a_1+w_1\ra
\otimes...|a_m+w_m\ra\sum_{w'\in C^\perp}|b_1+w'_1\ra
\otimes...|b_m+w'_m\ra\longmapsto\\\nonumber
&&\sum_{w\in C^\perp}|a_1+w_1\ra\otimes...
|a_m+w_m\ra\sum_{w'\in C^\perp}|a_1+b_1+w_1+w'_1\ra
\otimes...|a_m+b_m+w_m+w'_m\ra\\\nonumber
&&=|S_a\ra|S_{a+b}\ra
\end{eqnarray}
where the last equality follows from the fact that $C^\perp$ is a
 linear subspace, 
and therefore summing over $w+w'$ for a fixed $w$ in the code 
is equivalent to summing over $w'$. 
For the Phase gate, apply the gate 
$|a\ra \longmapsto i^a|a\ra$ three times on each coordinate.
This gives
\beq
 |S_a\ra=\sum_{w\in C^\perp}|a_1+w_1\ra\otimes...|a_m+w_m\ra\longmapsto
\sum_{w\in C^\perp}i^{3(\sum_k a_k+w_k)}
|a_1+w_1\ra\otimes...|a_m+w_m\ra.
\enq
This is the desired result, because $C$ is a punctured doubly even self dual code, 
and it is easy to check that 
 all words in $C^\perp$ have weight which is divisible by $4$, 
and 
all words in $C$ but not in $C^\perp$ have weight which is $3$ mod $4$. 

%$\sum_k w_k=0$ mod $4$ for any $w\in C^\perp$, and  $\sum_k a_k=3$ mod $4$
%for $a\in C- C^\perp$.
%To see that this is indeed the case, recall that 
% $C$ is a punctured doubly even self dual code. 
%Let $C'$ be the doubly even self dual code from which $C$ is 
%constructed by deleting one coordinate.
%Consider the set $V'_0\subset C'$ which consists  of all
% words which have $0$ on the deleted coordinate, 
%and denote by $V_0\subset C$ the set of words which we get by erasing 
%this coordinate. 
%Let $V'_1\subset C',V_1\subset C$,
%  be the sets defined similarly for words which 
%have $1$ in the deleted coordinate. 
%Of course, $C=V_0\cap V_1$. 
%We claim that $V_0=C^\perp$. To see this, note first that $V'_0\perp C'$, 
%so $V_0\perp  C$, and hence $V_0\subseteq C^\perp.$ To see the other 
%direction,  let $w\perp C$. 
%Then $w0\perp C'$, and hence $w0\in C$ since $C$ is self dual, so 
%$w\in V_0$.  
For the encoded controlled phase gate, 
\beq 
|S_{\vec{a}}\ra|S_{\vec{b}}\ra=
\sum_{w,w'\in C^\perp}|\vec{a}+w\ra|\vec{b}+w'\ra \longmapsto
\sum_{w,w'\in C^\perp}(-1)^{(\vec{a}+w)\cdot (\vec{b}+w')}|\vec{a}+w\ra|\vec{b}+w'\ra
\enq

Now, 
$(\vec{a}+w)\cdot (\vec{b}+w')=\vec{a}\cdot \vec{b}$ mod $2$. 
This is true since $\vec{a}\in C$ and $w'\in C^\perp$ so 
 $\vec{a}\cdot w'=0$ mod $2$, and likewise   $w\cdot \vec{b}=w\cdot w'=0$
 mod $2$. 
Moreover, $\vec{a}\cdot \vec{b}$ mod $2$ is equal to $ab$, and so 
the final state is indeed the desired state. 
Finally, for the Hadamard gate, 
\begin{eqnarray}
 |S_a\ra&=&\sum_{w\in C^\perp}|a_1+w_1\ra\otimes...|a_m+w_m\ra\longmapsto
\sum_{x\in F^m_2}\sum_{w\in C^\perp}(-1)^{(a+w)\cdot x}|x\ra=\\\nonumber
&=&\sum_{x\in C}(-1)^{a\cdot x}|x\ra=
\sum_{b\in C/C^\perp}\sum_{w\in C^\perp}(-1)^{a\cdot (b+w)}|b+w\ra=
\sum_{b}(-1)^{a\cdot b}|S_{b}\ra.
\end{eqnarray}

\subsection{Fault Tolerant Error Correction Procedure for CSS Codes}
So far, we have described how to apply those procedures 
which can be obtained by bitwise operations. 
Before we continue to the Toffoli gate and the gate which adds 
a blank qubit, we first show how to apply encoding, decoding and error 
correction fault tolerantly. 
These procedures will be used later in the remaining of the gates.

We now construct the error correction procedure. 
It is composed from detecting and correcting bit-flips, 
using classical error correction techniques for the code $C$,
rotating by Hadamard gate to the $C-$basis, and correcting bit-flips
using again classical error correction techniques. 
Finally we rotate back by applying Hadamard bit-wise. 

We now describe how to correct bit-flips. 
For this, we recall that for linear classical codes over $F_2$, 
one can define the {\it parity check matrix} $H$ of the code $C$. 
The kernel of $H$ is exactly the words in $C$. 
If an error occurred in a word $w$, the 
new word can be written $w+e$, where $e$ is called the error vector. 
$H(w+e)=He=s$ is called the syndrome of the error, 
and given the syndrome $s$, one can find the error vector of minimal weight 
which gives $s$.  The error correction is therefore done by first finding 
the syndrome, and then deducing the error vector from it. 

We now want to compute the $j'th$ bit of the syndrome fault tolerantly.
This is simply the inner product of the $j'$th row of $H$ with the 
word.  
If out computation was noiseless, we could do the following: 
 we apply CNOT from 
the block we are correcting to one blank qubit, 
 only on the coordinates which are $1$ in the $j'$th row
of the parity check matrix
of the code. 
This will compute the inner product of the word we are correcting with the 
$j'$th row of the parity check matrix, and the extra blank qubit will thus 
contain the $j'th$ bit of the syndrome. 
However, this is not a fault tolerant procedure, since 
an error in one of the CNOT gates causes an error in the 
extra qubit, which propagates through the other CNOT gates to 
other qubits in the block. 

To avoid this problem, we 
 will use an ancillary state  on $l$ qubits:
\beq
\frac{1}{2^{\frac{(l-1)}{2}}} \sum_{b;b\cdot \vec{1}} |b\ra
\enq
 where $l$ 
is the number of $1'$s in the  in the $j'th$ raw of $H$, the 
parity check matrix of $C^{\perp}$. 
This state is actually the superposition of all states with an even number
 of $1$'s.  It can be easily constructed by 
starting with $|0^l\ra$, applying Hadamard 
on the first qubit and then CNOT 
from this qubit to all the other $l-1$ qubits, to get
the ``cat state'' on $l$ qubits, 
\beq\label{cat}  
|cat_l\ra=\frac{1}{\sqrt{2}}(|0^l\ra+|1^l\ra).
\enq
and then  rotating each qubit of the cat state to get the desired state.

We would now like to apply CNOT bitwise from
the block we are initializing to the 
 cat state only on the coordinates which are $1$ in the $j'$th row 
of the parity check matrix. 
Then we apply a classical computation, using only gates from ${\cal G}_1$, 
 on the ancilla state, 
to find out the parity of the strings in the ancilla state,
 and write this parity on another 
qubit, initialized in the state $|0\ra$. 
The resulting state  will be the   $j'$th syndrome bit.

The only problem in this scheme is that the generation of the cat state 
is not done fault tolerantly, and thus one error can cause the entire 
cat state to be ruined; The CNOT gates will then propagate this error to 
many qubits in the block. 
The solution to this problem is to verify that the cat state is indeed 
a superposition of the two states 
$|0^l\ra$ and $|1^l\ra$,   before continuing.
The relative phase between the two states can still be mistaken, but 
this will not cause any error to propagate to the state
by the CNOT gates,   but only cause the syndrome to be mistaken.
To see this, suppose we have instead of the cat state, the state
$c_0|0\ra+c_1|1\ra$. 
This state transforms after the Hadamard gates to the state:
\beq
c_0\sum_i |i\ra +c_1 \sum_i (-1)^{i\cdot \vec{1}}|i\ra
\enq
Now, consider 
applying CNOT's on all the $1$ coordinates in the parity check matrix 
from a  correct  encoded state to $\sum_i |i\ra$. 
This is simply the identity operator:
\beq
(c_0|S_0\ra+c_1|S_1\ra)\sum_i |i\ra\longmapsto 
(c_0|S_0\ra+c_1|S_1\ra)\sum_i |i\ra
\enq
and the same for the other term $ \sum_i (-1)^{i\cdot \vec{1}}|i\ra$. 
This means that in spite of the fact that 
the cat state is not correct, this does not cause any error 
in the original state.

This verification  
 was done by Shor\cite{shor3} using measurements and assuming noiseless 
classical computation, by measuring the XOR of pairs of the qubits.
If we want to avoid measurements, we proceed as follows. 
We now want to compute whether the bits in the cat state are all equal. 
For this, note that checking whether two bits are equal, and writing the result 
on an extra blank qubit, can be easily done by a small circuit  
which uses Toffoli and NOT gates. We denote this circuit by $S$, and also  
denote the qubits in the cat state by $1,...l$. 
We add
 $l-1$ extra blank qubits, 
and apply the circuit $S$ first   from each even pair of qubits
 (e.g. the pair of qubits $(1,2), (3,4)...$),  to one of the blank qubits;
Then apply $S$ from each odd pair of qubits
 (e.g. the pairs  $(2,3), (4,5)...$) to one of the remaining blank qubits. 
We get $l-1$ qubits which are all $1$, if no error occurred, indicating that all the 
qubits are equal. 
We then apply a classical circuit on all these qubits, which checks whether they are all 
$1$, and write the result on an extra blank qubit, which 
is our check bit, and indicates that all the bits in the cat state are equal. 
We now want to use the cat state, but  condition all the operations 
 on the fact that the check bit is indeed $1$. 
However, if we do this, an error in the check bit,
 might propagate to all qubits in the state we are trying to correct. 
Hence, to keep the procedure fault tolerant, 
we construct $m$ different check bits, one for each qubit in the 
state we are correcting.  This is done using $m(l-1)$  blank qubits,
and   applying 
 all the operations above, where each operation is repeated  $m$ times.  
to $m$ different target qubits.  
Thus, we can verify that the cat state is of the form $c_0|0\ra+c_1|1\ra$, 
fault tolerantly. 
We can now condition all the operations done in the syndrome measurement 
involving the $i'$th qubit, on the $i'$th check bit. 

From the syndrome  $s$ we can compute the error vector $e$
which is the vector of minimal weight which gives $He=s$. 
This can be done  by classical 
operations, and need not be done fault tolerantly. 
Applying a CNOT from the result bit $e_i$ to the $i$'th bit
will correct a bit flip in the $i'$th bit.
Now, we do not correct all qubits according to the error vector 
computed from one copy of the syndrome, since an error during
 the  computation of the error vector might result in the entire error 
vector being mistaken, and the procedure will not be fault tolerant.  
Instead, we compute the syndrome independently $m$ times, 
and from the $i'$th copy of the syndrome we compute $e_i$, from 
which we apply a CNOT to the $i'$th bit of the state which 
we are correcting. 
The fact that for each bit we compute the syndrome independently 
ensures us that  the procedure is fault tolerant. 

% first bit has an error, as follows: 
%the vector space can be divided to non-intersecting cosets of the 
%subspace $C^{\perp}$. Each coset can be written as 
%$C^{\perp}+e$ where $e$ is a
%vector. Each such $e$ gives one possible syndrome.($He=s$).
%Given the syndrome, we compute the {\bf table} $s ->e $,
%and decide whether a qubit is wrong by asking whether it 
%is in the support of $e$,
%meaning that the corresponding  coordinate in $e$ is $1$. 
%Finally apply a controlled not from the result to the first qubit.

To apply error correction, we first apply error corrections of bit flips, 
according to the code $C$.
We then apply a Hadamard gate to transform to the $C$-basis, 
and correct bit flips again, according to the code $C$. 
Then, we apply a Hadamard gate again, to get back to the $S$-basis.

In all the procedures we have described so far, 
 one fault causes at most one error in each block. 
Unfortunately, the propagation of errors in the  error correction procedure
is much worth. 
One fault in one of the circuits $S$, checking whether two qubits in the cat state are the
 same, might cause four qubits in this cat state to be contaminated,
 where their 
check bits might still indicate a green light for following procedures. 
Hence, during the error corrections, each one of the four errors can propagate
 to the one qubit which is conditioned upon it. 
Apart from this bottle neck, all other errors have exactly one spread.

\subsection{Fault Tolerant Encoding Procedure for CSS codes}

An encoding procedure  takes
\begin{eqnarray}
|0^{m}\ra &\longmapsto &|0^m\ra |S_0\ra\\\nonumber
|1^{m}\ra &\longmapsto &|1^m\ra |S_1\ra
\end{eqnarray}
To construct this procedure fault tolerantly, it is enough to 
  generate 
 a state $|S_0\ra$ fault tolerantly. 
  This suffices since  given a string of bits,
$|0^m\ra$ or $|1^m\ra$, we first generate $|S_0\ra$  and then 
apply  a  controlled not bitwise from the 
given input string to the encoded state  $|S_0\ra$.

To generate the state $|S_0\ra$, 
start with $m$ blank qubits,  $|0^m\ra$,  then  
 apply an error  correction procedure with respect to a code 
which consists of one word: $|S_0\ra$ alone.
This is done by first  applying Hadamard gates bitwise on  
  $|0^m\ra$, and then correcting bit flips
according to the  code $C^{\perp}$.
To see that we indeed get the state $|S_0\ra$, note that
after the first rotations we have
 $\sum_{i=0}^{2^m-1}|i\ra$. The corrections will then take 
this state to a uniform distribution over all the 
basic states in $C^{\perp}$,
 due to the linearity of the code.

\subsection{Error Correction which Projects any Word into the Code}

%(Note that the only effect of the CNOT from the 
%state which we want to correct to the ancilla block, is 
%by shifting the states of the ancilla state.  ) 

We would like to insert here one important modification,
which is not necessary for the fault tolerant error correction, 
 but will become 
crucial when applying the error corrections in the recursive scheme. 
This is the requirement that the error correction takes any word to 
some word in the code, regardless of the number of faults.
Roughly, this is done by checking whether too many errors 
occurred, and if so replacing the entire block by another block
which is initialized in the state $|S_0\ra$. 
However, we should be careful to keep the procedure fault tolerant. 
We do this in the following way: 
 Before starting the correction procedure, 
we generate  another state $|S_0\ra$ on ancilla qubits, 
as in the encoding procedure.
When computing from the $i'$th copy of a 
syndrome whether the $i'$th qubit is wrong, also compute whether the number
of faults according to the syndrome is smaller than $d$, and
 write the answer on another qubit.
The CNOT which checks if the $i'$th bit is wrong and 
if so applies  $NOT$ on the $i'th$ qubit,  is replaced by
a Toffoli gate 
 which also checks if the number of faults is smaller than $d$.
We then swap the $i$'th qubit 
with the $i'$th qubit of the state 
  $|S_0\ra$, conditioned that 
 the number of faults is indeed larger that $d$. 
To achieve a swap between the second and third qubit conditioned on the first 
qubit, apply three Toffoli gates one after the other, where
 the target of the first gate is the third qubit, 
the target of the second gate is the second qubit and the 
target of the third gate is again  the third qubit, i.e. 
\beq
Controlled(a)SWAP(b,c)=T(a,b,c) T(a,c,b) T(a,b,c)
\enq  
which is easy to check. 

To see that this procedure indeed  
takes any word to some word in the code, observe that 
this is true if no error occurs in the procedure itself, and 
since the procedure is fault tolerant, the final state will 
differ from a word in the code only in the qubits effected by an error.

\subsection{A Fault Tolerant Decoding Procedure for CSS codes}
A decoding procedure applies
\beq
|S_a\ra \longmapsto |A_a\ra|\vec{a}\ra
\enq
where the state $|A_a\ra$ is an ancillary state which depends on $a$. 
(We can discard this state at the end of the procedure; 
we will see that whenever the decoding procedure is applied, the state 
encodes  a well defined logical bit $a$. )  
 To apply this transformation, we compute 
$a$ independently $m$ times from the state $|S_a\ra$. 
To do this, we add  $m^2$ blank qubits, and copy each qubit from 
 $|S_a\ra$ $m$ times to $m$ different blank qubits, using $m$ CNOT gates.
We get $m$ ``copies'' of $|S_a\ra$.
 (These of course are not really copies of $|S_a\ra$, since they are entangled. However, each word in the classical code is copied $m$ times.)
  On each copy of the word  we apply the quantum analog of the classical 
computation that computes what is the logical bit that 
the word encodes. 
 The answer, which is $a$ if no error occurred, 
is written on another blank qubit. 
  For this computation we use 
Toffoli, CNOT and NOT gates. We might need some 
extra blank qubits as working space.  
The  computation of $a$ is done is the shortest way possible, regardless 
of whether it is fault tolerant; 
An error in this computation can effect only the one copy of $a$ 
which it computes. 
A fault in the first stage of copying the qubits $m$ times can only
effect one qubit in each of the copies, and if the number of errors
in $S_a$ plus number of faults in the first stage is smaller than 
the  number of errors 
 correctable by the code, these faults have no effect. 
 One fault in the second stage of the   procedure,
during the computation of one of the $a$'s,  can 
effect only the correctness of that $a$. 
% bit which 
%is computed by  at most one error in the bit which  is smaller than  is one  spread is $1$, as long as the sum of damage in the block 
% and number of faults in the 
%first stage is smaller than half the minimal distance of the code.  

\subsection{Toffoli Gate}

To apply the Toffoli gate, we  roughly follow
 Shor's scheme, where we construct an ancillary state and use it 
to obtain the Toffoli gate. 
The main difference from Shor's scheme is in the  construction of the
 ancillary state, which is not completely
 straightforward if one wants to avoid 
using measurements. 
The following construction involves many details, and has no 
underlying structure, as far as we can see. 
As we will see in the next section, the Toffoli gate 
on polynomial codes can be applied in a much easier and shorter way. 
However, for completeness, we also give the details of 
the Toffoli procedure in the case of CSS codes.

\subsubsection{ Construction of the Ancilla state $|A\ra$}

We would now like to show how to construct  
the state 
\beq
 |A\ra =\frac{1}{2} |S_0S_0S_0\ra +|S_0S_1S_0\ra+|S_1S_0S_0\ra+|S_1S_1S_1\ra.
\enq

This will be done using the help of the encoded cat states. 
We will need the definition of the state
\beq
 |B\ra =\frac{1}{2} |S_0S_0S_1\ra +|S_0S_1S_1\ra+|S_1S_0S_1\ra+|S_1S_1S_0\ra.
\enq
which is easily convertible to $|A\ra $ by
 applying an encoded NOT on the third block.

The idea is that 
\beq
 \frac{1}{\sqrt{2}} (|A\ra+|B\ra)
\enq
 is actually very easy to construct, because 
it is equal to 
\beq
 \frac{1}{2\sqrt{2}}(|S_0\ra+|S_1\ra)(|S_0\ra+|S_1\ra)(|S_0\ra+|S_1\ra)
\enq
which can be constructed easily from $3$ block of $m$ blank qubits
by applying 
an encoding procedure and then an encoded Hadamard gate on each block. 
In order to convert this state to $|A\ra$, we 
use encoded cat states.  
An encoded cat state,
\beq
|S_{cat}\ra=\frac{1}{\sqrt{2}}(|S_0\ra^m+|S_1\ra^m)
\enq
is also easy to construct fault tolerantly, in the following way:
 Generate $m$ $|S_0\ra$ states,
$|S_0\ra^m$, fault tolerantly, using the encoding procedure
on $m$ blocks each containing $m$ blank qubits.  
Then,  apply an encoded Hadamard gate on the first block, 
 and then copy this block to all the other blocks 
by applying encoded CNOT from the first block to all other blocks. 
This results in an encoded cat state. 
We now apply fault tolerant error corrections on each block.

Now, consider the transformation
\begin{eqnarray}\label{69}
|S_0\ra^m|A\ra \longmapsto |S_0\ra^m|A\ra\\ \nonumber
|S_1\ra^m|A\ra \longmapsto |S_1\ra^m|A\ra\\ \nonumber
|S_0\ra^m|B\ra \longmapsto |S_0\ra^m|B\ra\\ \nonumber
|S_1\ra^m|B\ra \longmapsto -|S_1\ra^m|B\ra\\ \nonumber
\end{eqnarray}
If we can apply this transformation fault tolerantly, 
we could start with the state 
\beq
\frac{1}{2}(|S_0\ra^m+|S_1\ra^m)(|A\ra+|B\ra)
\enq
and apply the transformation \ref{69}
to get

\beq
\frac{1}{2}(|S_0\ra^m+|S_1\ra^m)|A\ra+
\frac{1}{2}(|S_0\ra^m-|S_1\ra^m)|B\ra
\enq
We might then be able to compute fault tolerantly in which 
of the cat states $\frac{1}{2}(|S_0\ra^m\pm|S_1\ra^m)$
 the ancilla is, and apply a NOT on the third 
block of $|B\ra$, conditioned that the result is the second state
$\frac{1}{2}(|S_0\ra^m-|S_1\ra^m)$. 
Let us refer to this process as a ``measurement'' of the cat state, 
though no measurement will be involved. 

The transformation \ref{69} can be done by 
 applying
\beq\label{mix}
|S_a\ra|b\ra|c\ra|d\ra \longmapsto (-1)^{a(bc+d)}|S_a\ra|b\ra|c\ra|d\ra
\enq
on the $i$'th block in the encoded cat state and the $i$'th bit
 in each of the three 
blocks of $|A\ra+|B\ra$. 
Note that transformation \ref{mix} need not be fault tolerant, 
and it might ruin the entire block $|S_a\ra$. 
Applying transformation \ref{mix} for $1\le i\le m$
it is easy to check that we get
\beq
|S_a\ra^m|S_b\ra|S_c\ra|S_d\ra \longmapsto (-1)^{a(bc+d)}|S_a\ra^m|S_b\ra|S_c\ra|S_d\ra
\enq
which exactly achieves transformation \ref{69}.

We now run into the problem of how to measure the encoded cat state. 
To solve this problem, we use three encoded cat states, instead of one, 
and repeat everything that we did on the first encoded cat state also 
for these states. 
The resulting state is

\beq\label{28}
\frac{1}{2}(|S_0\ra^m+|S_1\ra^m)^3|A\ra+
\frac{1}{2}(|S_0\ra^m-|S_1\ra^m)^3|B\ra
\enq

Now it becomes possible to measure the triple cat state, 
 using majority. 
This can be done in several ways. One way is to  apply
 encoded  Hadamard gates on all 
blocks of the encoded cat states, which 
 takes the encoded states into superpositions of encoded states, 
with certain parity: 
\begin{eqnarray}
\frac{1}{\sqrt{2}}(|S_0\ra^m+|S_1\ra^m)\longmapsto \frac{1}{\sqrt{2^{m-1}}}
\sum_{i=0, i\cdot \vec{1}=0}^{2^m} |S_{i_1}\ra|S_{i_2}\ra... |S_{i_m}\ra,\\\nonumber
\frac{1}{\sqrt{2}}(|S_0\ra^m-|S_1\ra^m)\longmapsto \frac{1}{\sqrt{2^{m-1}}}
\sum_{i=0, i\cdot \vec{1}=1}^{2^m} |S_{i_1}\ra|S_{i_2}\ra... |S_{i_m}\ra\nonumber
\end{eqnarray}

To compute the parity fault tolerantly, 
we compute from each block the bit it represents,
decoding it fault tolerantly.   Then independently compute 
 the parity of the $m$  $i'$th bits in the $m$ blocks.  
The parity can be computed using Toffoli, CNOT and NOT gates.
We now have $m$ parity bits, for each encoded cat state. 
We compare them bitwise by applying on the
 three $i$'th parity bits and an ancilla bit 
 a majority vote 
\beq
|a,b,c,d\ra \longmapsto |a,b,c, d+maj(a,b,c)\ra 
\enq
which can be constructed from  
 Toffoli, CNOT and NOT gates, since they are universal for
 classical computations.
Now,  apply CNOT bitwise from the resulting majority votes of the parity bits
 to the qubits in the third block of $|A\ra$.

%For each encoded cat state, we apply encoded CNOT 
%from the first block to all other blocks, 
%and then an encoded Hadamard gate. 
%This takes the state to 
%\beq
%\frac{1}{2}(|S_0\ra|S_0\ra^{m-1})^3|A\ra+
%\frac{1}{2}(|S_1\ra|S_0\ra^{m-1})^3|B\ra
%\enq%

%Now apply the fault tolerant decoding procedure 
%on the first block of each encoded cat state, 
% and then apply a majority vote among the 
%$i'$th bits of the three decoded blocks.
%A  majority operation can be easily constructed using  
%the classical gates: Toffoli, NOT and CNOT.  
%Finally, apply CNOT's to the third block 
%of $A\ra$, conditioned on the results. 

It is left to see that one error in this procedure effects 
only one qubit in $|A\ra$. 
We will consider errors in different stages of the procedure. 
The construction of an 
encodes cat state id fault tolerant since 
it is composed of the fault tolerant encoding procedure, 
and bitwise CNOT's. 
The construction of the state $\frac{1}{\sqrt{2}}(|A\ra+|B\ra)$
 is fault tolerant 
because we only used the fault tolerant encoding procedure 
and the fault tolerant encoded Hadamard gate.
A problem arises in transformation 
\ref{69}. 
During this transformation, an error can cause the entire block 
$|S_a\ra$ to be ruined. 
Hence, an error during this transformation can cause 
a whole block in one of the cat states in the state (\ref{28})
 to be effected, together 
with one qubit in each one of the last $3$ blocks.  
However, one such block can ruin the 
parity bits of only one encoded cat state, 
and as long as the other two are still fine, 
the majority vote will still give the correct parity. 
An error in the parity computations or in the majority vote 
cannot effect more than one qubit in $|A\ra$, since they are done bitwise. 
This concludes the fault tolerant construction of the state $|A\ra$. 

Note, that the above procedure is fault tolerant if one error occurred, 
but if we are unlucky and two errors occurred during transformation 
(\ref{28}) when applied on two different encoded states, this can 
effect the parity bits of two encoded cat states, and thus 
the majority vote will fail here. 
In order to tolerate more than one error in a procedure, we will need to 
use more encoded cat states. Using $2k+1$ encoded cat states, we can 
still say that one error effects only four qubits in each block,
 as long as the number 
of errors is less than $k$.

\subsubsection{Construction of Toffoli Gate Given $|A\ra$}

The construction of the Toffoli gate given $|A\ra$ follows Shor's scheme
almost exactly, with minor changes 
due to the fact that measurements are replaced by CNOT gates to additional 
blank qubit. Also, classical conditioning 
on the results of the measurements are replaced 
by classical unitary gates (i.e. permutation matrices)
 on the computation qubits and extra blank qubits. 
Here is a short description of how this is done:
We try to generate the transformation 
\beq
|\alpha\ra=|S_a\ra |S_b\ra|S_c\ra \longmapsto |S_a\ra |S_b\ra|S_{c+ab}\ra
\enq
For this, we will first generate ``half'' a Toffoli gate: 
\begin{eqnarray}\label{half}
|S_0\ra |S_0\ra |S_0\ra   \longmapsto |S_0S_0S_0\ra\\\nonumber
|S_0\ra |S_1\ra |S_0\ra  \longmapsto |S_0S_1S_0\ra\\\nonumber
|S_1\ra |S_0\ra |S_0\ra \longmapsto |S_1S_0S_0\ra\\\nonumber
|S_1\ra |S_1\ra |S_0\ra  \longmapsto |S_1S_1S_1\ra\\\nonumber
\end{eqnarray}
A Toffoli gate can be generated from transformation (\ref{half}) in the
 following way. 
We start with the three blocks on which we want to apply Toffoli, 
$|S_a\ra |S_b\ra|S_c\ra$.
Then we generate $|S_0\ra$ on an extra block, using the encoding procedure. 
We then apply transformation (\ref{half}) on the first two 
blocks and the newly generated $|S_0\ra$. 
Then, we apply an encoded CNOT from our original  third block
$|S_c\ra$ to the new block. 
We finally apply an encoded Hadamard on the  original  third block. 
This gives the overall transformation:

\begin{eqnarray}\label{almost}
|S_0\ra |S_0\ra |S_0\ra |S_0\ra  \longmapsto 
\frac{1}{\sqrt{2}}|S_0S_0S_0\ra(|S_0\ra+|S_1\ra)\\\nonumber
|S_0\ra |S_1\ra |S_0\ra |S_0\ra  \longmapsto 
\frac{1}{\sqrt{2}}|S_0S_1S_0\ra(|S_0\ra+|S_1\ra)\\\nonumber
|S_1\ra |S_0\ra |S_0\ra |S_0\ra  \longmapsto 
\frac{1}{\sqrt{2}}|S_1S_0S_0\ra(|S_0\ra+|S_1\ra)\\\nonumber
|S_1\ra |S_1\ra |S_0\ra |S_0\ra  \longmapsto 
\frac{1}{\sqrt{2}}|S_1S_1S_1\ra(|S_0\ra+|S_1\ra)\\\nonumber
|S_0\ra |S_0\ra |S_1\ra |S_0\ra  \longmapsto 
\frac{1}{\sqrt{2}}|S_0S_0S_1\ra(|S_0\ra-|S_1\ra)\\\nonumber
|S_0\ra |S_1\ra |S_1\ra |S_0\ra  \longmapsto 
\frac{1}{\sqrt{2}}|S_0S_1S_1\ra(|S_0\ra-|S_1\ra)\\\nonumber
|S_1\ra |S_0\ra |S_1\ra |S_0\ra  \longmapsto 
\frac{1}{\sqrt{2}}|S_1S_0S_1\ra(|S_0\ra-|S_1\ra)\\\nonumber
|S_1\ra |S_1\ra |S_1\ra |S_0\ra  \longmapsto 
\frac{1}{\sqrt{2}}|S_1S_1S_0\ra(|S_0\ra-|S_1\ra)\\\nonumber
\end{eqnarray}

Note that if the fourth block was not there, we are done, because 
the operation on the first three blocks is exactly Toffoli.
We next decode the fourth block, and apply some operations conditioned 
on the qubits containing the results of the decoding.
Note that if the decoding is $\vec{0}$, there is no phase to correct, 
since the gate which has been performed is exactly the Toffoli gate. 
Hence, we would like to apply the operation
\beq
|S_a\ra |S_b\ra |S_c\ra \longmapsto (-1)^{ab+c}|S_a\ra |S_b\ra |S_c\ra
\enq
on the first three blocks, conditioned that the decoded qubits are $1$. 
This can be applied bitwise, in the following way. 
First, apply bitwise a controlled phase shift on $|S_c\ra$ and the decoded 
qubits. 
This will give the factor   $(-1)^{c}$. 
To apply $|S_a\ra |S_b\ra\longmapsto (-1)^{ab}|S_a\ra |S_b\ra$
 conditioned on the decoded qubits, 
apply bitwise the operation:
$|a\ra |b\ra|d\ra\longmapsto (-1)^{abd}|a\ra |b\ra|d\ra$ where $|d\ra$
is a
decoded qubit. 
This operation can be achieved by adding a blank qubit, 
$|a\ra |b\ra|0\ra |d\ra$
applying a Toffoli gate on the first three qubits, 
followed by a controlled shift on the last two qubits, and 
then by a Toffoli gate again on the first three qubits.

It is left to show how to construct transformation (\ref{half}). 
This is done by  
 generating the ancilla state $|A\ra$ as before,
where one of the blocks of $|A\ra$ is the third block. 
and apply encoded CNOT from the first block of $|A\ra$ to the first block 
of $|\alpha\ra$ and from the second block of $|A\ra$ to the second block 
of $|\alpha\ra$.
This achieves the transformation:
\begin{eqnarray}
|S_0\ra |S_0\ra |A\ra \longmapsto \frac{1}{2}(|S_0S_0S_0S_0S_0\ra+
|S_0S_1S_0S_1S_0\ra+
|S_1S_0S_1S_0S_0\ra+|S_1S_1S_1S_1S_1\ra)\\\nonumber
|S_0\ra |S_1\ra |A\ra \longmapsto \frac{1}{2}(|S_0S_0S_0S_1S_0\ra+
|S_0S_1S_0S_0S_0\ra+|S_1S_0S_1S_1S_1\ra+
|S_1S_1S_1S_0S_0\ra)\\\nonumber
|S_1\ra |S_0\ra |A\ra \longmapsto \frac{1}{2}(|S_0S_0S_1S_0S_0\ra+
|S_0S_1S_1S_1S_1\ra+|S_1S_0S_0S_0S_0\ra+
|S_1S_1S_0S_1S_0\ra
)\\\nonumber
|S_1\ra |S_1\ra |A\ra \longmapsto \frac{1}{2}(|S_0S_0S_1S_1S_1\ra+
|S_0S_1S_1S_0S_0\ra+|S_1S_0S_0S_1S_0\ra+|S_1S_1S_0S_0S_0\ra)\nonumber
\end{eqnarray}
We note that the last three blocks have strong connection to the 
Toffoli gate. 
More precisely, we note that projecting the above states 
on the subspace where the  first two blocks
is $|S_0\ra|S_0\ra$, the gate which was achieved is exactly the Toffoli gate. 
The projection on the subspace in which the two blocks are 
$|S_0\ra|S_1\ra$ is not a Toffoli gate, but can be converted to a Toffoli gate
by applying a NOT on the second block and a CNOT from the first
 block to the second block. Projecting to the other two possible subspaces,
 we again get a Toffoli gate by simple corrections:
for $|S_1\ra|S_0\ra$ we should apply $NOT$ on first block and CNOT 
from second to third block, 
while for $|S_1\ra|S_1\ra$ the gate can be corrected to Toffoli by 
applying a CNOT from the second block to the third, then a CNOT from the 
first to the third and then a NOT on the last black.

We now have to 
 make these adjustments, conditioned on the state of the first two 
blocks,   without measurements. 
This is done by  decoding 
the first two blocks, and applying the adjustments conditioned on the state 
of the decoded qubits, bitwise, meaning that 
we apply the following operation on five qubits:

\begin{eqnarray}
|0,0\ra |a,b,c\ra&\longmapsto& |0,0\ra|a,b,c\ra\\\nonumber
|0,1\ra |a,b,c\ra &\longmapsto& |0,1\ra NOT(2)CNOT(1,2)|a,b,c\ra\\\nonumber
|1,0\ra |a,b,c\ra &\longmapsto& |1,0\ra NOT(1)CNOT(2,3)|a,b,c\ra\\\nonumber
|1,1\ra |a,b,c\ra &\longmapsto& |1,1\ra NOT(3)CNOT(1,3)CNOT(2,3)|a,b,c\ra\\\nonumber
\end{eqnarray}
This operation is reversible, and is classical. 
Therefore it can be constructed out of the classical gates 
which are in the set ${\cal G}_1$.  
It is easy to check that indeed it achieves the correct transformation.

\section{Computing on States Encoded by Polynomial Codes} \label{sec5}
In the last section it was shown how to implement fault tolerantly 
the set of gates ${\cal G}_1$. 
It turned out that all the gates were very easy to implement, except for 
the last gate: The Toffoli gate. 
This turned out to involve many tricks, and seems not to have an underlying 
theoretical structure.
In this section we show that when transforming to qupits, 
and using the polynomial codes, one can use the algebraic 
properties of these codes to implement 
the gates fault tolerantly in a manner which seems logical and easy to 
explain. 
Moreover, the procedures are shorter, and require less operations. 
Thus, as we will see later, the threshold which we achieve
using polynomial codes is significantly better than 
that achieved using CSS codes. 
The error correction, decoding and encoding procedures will be 
similar to those used for the CSS codes.

We define a set of gates ${\cal G}_2$, which 
will be shown to be universal for qupits in the next section.
We show how to apply all gates from  ${\cal G}_2$
fault tolerantly on states encoded by polynomial codes. 
It turns out that all these procedures can be applied 
using the same basic underlying idea. 
This idea is that when applying the gates bitwise, we always get the 
desired result, except for one problem: 
Instead of getting the final state as a superposition of polynomials 
with degree $d$,   we sometimes end up with the correct 
logical pit, encoded with polynomials with degree $2d$
instead of $d$.  
To get back into the code which uses polynomials of degree $d$, 
 one has to apply degree reduction. We use techniques
which were introduced by Ben-Or {\it et al} \cite{bgw} for fault tolerant 
classical distributed 
 computation,  on a secret which is shared by several parties. 
In order to be able to use these degree reduction techniques,
 it is required that 
after the degree has increased to $2d$, the state is still 
a polynomial code, so that errors can still be corrected. 
For this, we will work with  codes of length 
$m=3d+1$.

\subsection{The Set of Gates for Polynomial Codes}
We work with the following set of gates, which we denote 
by ${\cal G}_2$:

\begin{enumerate}

 \item Generalized NOT: $\forall~c\in F$, $|a\ra\longmapsto|a+c\ra$,
\item Generalized CNOT: $|a,b\ra\longmapsto |a,a+b\ra$,

\item Swap $|a\ra|b\ra \longmapsto |b\ra|a\ra$,

\item Multiplication gate: $0\ne c\in F$: $|a\ra\longmapsto|ac\ra$, 

\item generalized Phase Rotation:
$\forall c\in F$ $|a\ra\longmapsto w^{ca}|a\ra$, 

\item generalized Hadamard (Fourier Transform)
 $|a\ra\longmapsto\frac{1}{\sqrt{p}}\sum_{b\in F}w^{rab}|b\ra, \forall 0<r<p$.

%\item Generalized Controlled 
%Phase rotation: $|a\ra|b\ra \longmapsto (w)^{ab} |a\ra|b\ra$,

\item generalized Toffoli: $|a\ra|b\ra|c\ra\longmapsto |a\ra|b\ra|c+ab\ra$, 

\end{enumerate}

where all the addition and multiplication are in $F_p$(i.e. mod $p$).
The fact that this set of gates is universal will be shown later, 
in section \ref{universal}.  
The following theorem  shows how to perform gates from ${\cal G}_2$ 
on encoded states fault tolerantly. 
Like ${\cal G}_1$,  ${\cal G}_2$  is by no means
 a minimal universal set of gates, and in fact we make it as large 
as possible so that 
 the fault tolerant procedures become more 
 efficient. 

\begin{theo}\label{procedures-poly}
There exists fault tolerant procedures which simulate the operations of 
all the gates from ${\cal G}_2$, on states encoded by polynomial codes, 
 such that 
one error in a qubit or a gate effects only one
 qubit in each block at the end of the procedure. 
There exist such procedures also for encoding, decoding and error 
correction. Moreover, all these procedures use only gates from ${\cal G}_2$, 
and in particular do not use measurements.
\end{theo}

{\bf Proof:}
It turns out that all the gates can be applied pit-wise, 
and then applying degree reduction. 
The proof of this theorem, like that of theorem \ref{procedures-css}
is done by first showing how to perform the encoded gates in the cases
in which pit-wise applications of the gate will do (without the need 
in reducing the degree of the polynomials.)
 Then we show how to apply
 the encoding, decoding
and correction procedures. Finally, we show a procedure  to
reduce the degree of the polynomials, which enables us to   
construct  the Fourier transform and the Toffoli gate. 
Both of these gates can be achieved by 
 applying them pit-wise, and then applying degree reductions.

\subsubsection{Pit-wise Fault Tolerant Procedures}

Applying a generalized NOT gate on each one of the qupits in a block 
gives an generalized NOT on the entire block, as can be easily checked. 
Applying a generalized CNOT from  the $i'$th  qupit in the first block to 
the $i'$th of the second block, for all $1\le i\le m$s, gives an encoded
 generalized 
CNOT from the first block to the second block. 
In the same way, applying the  SWAP gate and the 
 multiplication gate  pit-wise, 
achieves an encoded SWAP and an encoded multiplication gate respectively.

\subsubsection{Fault tolerant Procedure of 
General Phase Rotation}

Define   $c_l$ as  the interpolation coefficients such that 
\beq
\forall~ f\in F[x],~ deg(f)\le m-1, f(0)=\sum_{i=1}^{m}c_if(\alpha_i).
\enq
To achieve a rotation by $w^{ca}$, we 
 apply on the $l'th$ qupit
the gate $|a\ra\longmapsto w^{c_la}|a\ra$.
This achieves the desired operation because:
\begin{eqnarray}
|S_a\ra=&\sum_{f\in V,f(0)=a}|f(\alpha_1),...,f(\alpha_m)\ra
&\longmapsto
\\\nonumber
&\sum_{f\in V1,f(0)=a}\Pi_{i=1}^{m}w^{c_lf(\alpha_{l})}
|f(\alpha_1),...,f(\alpha_m)\ra&=\\\nonumber
&\sum_{f\in V,f(0)=a}w^a
|f(\alpha_1),...,f(\alpha_m)\ra&.
\end{eqnarray}

\subsubsection{Fault Tolerant Error Correction Procedure}\label{errorcor}
These procedures can be applied exactly as was done in the case of the 
CSS codes, except that everything is replaced by the natural analogue over the field $F_p$. 

For the error correction procedure, we first generate 
the general cat state, which is 
a  superposition of all strings of equal pits:
\beq
|cat^p_l\ra=\sum_{a=0}^{p-1} |a^m\ra
\enq
This is done by applying the generalized Fourier transform on a blank qupit, 
and then copying it using generalized CNOT gates to more blank qupits. 
We now apply a generalized Fourier transform on each of the coordinates, to
 achieve the state 

\beq\label{par}
\sum_{j,j\cdot \vec{1}=0~mod ~p} |j\ra,
\enq
which is simply an equal superposition of all 
strings $j$  which satisfy 
$\sum_k j_k=0 ~mod ~p$, as a natural generalization of the corresponding state 
for qubits which was the superposition of all states with parity zero. 
The calculations of the syndrome, like in the CSS case, 
are done  by computing the inner product of the corresponding 
rows in the parity check matrix of the code. 
To compute the inner product for the $i$'th row of the check matrix, 
$h_{i,l}$, 
with a vector $a_l$,  we need to sum $\sum_l h_{i,l} a_l $. 
We thus add $h_{i,l} a_l$ to the $l'$th coordinate of the 
ancillary state. 
If no error occurred, the inner product is supposed to be zero, 
so the ancillary state does not change, 
since we have added to the ancilla 
 a vector of which  the coordinates sum up to $0$ mod $p$. 
The rest of the error correction transforms smoothly to the case of 
computing over the field $F_p$. 
As a general rule,  the gates $NOT$, $CNOT$ and Hadamard are replaced 
by their generalized version, i.e. generalized NOT, generalized CNOT and 
Fourier Transform over $F_p$, and this achieves the desired transformation.

\subsection{Fault Tolerant Decoding and Encoding Procedures}
The encoding procedure is achieved exactly like in the case of CSS 
codes: We generate $|S_0\ra$ by correcting according to the code 
$C_2$, and then we add the input pits pit-wise. 
The decoding is also done exactly in the same way. 

\subsubsection{ Fault Tolerant Procedure of the Fourier Transform Gate}\label{ftpoly}

The desired transformation is:
 
\beq
|S_a\ra\longmapsto\frac{1}{\sqrt{p}}\sum_{b\in F}w^{ab}|S_b\ra.
\enq
To achieve this,  first fix  \(c_1,...,c_m\) to be the interpolation 
coefficients, i.e.  
for any polynomial $f(x)$ over $F_p$ with $deg(f)\le m-1$,
the zero coefficient of $f$ satisfies 
$f_0=\sum_{i}c_if(\alpha_i)$.
%(Such $e_i$ exist, since 
% interpolation via $\alpha_1,...\alpha_m$ is a linear 
%functional: 
% if $A$ is the Vandermonde matrix $A_{i,j}=\alpha_i^j$, 
%then the vector of coefficients of $f$, $f_0,...f_{m-1}$ is 
% $A^{-1}$ times the vector $f(\alpha_1),...f(\alpha_m).$)
Denote by  $w_l=w^{c_l}, l=1,...,m$.
and recall that in our notation
 \beq W(w_l): |a\ra\longmapsto \frac{1}{\sqrt{p}}\sum_{b\in F_p}w_l^{ab}|b\ra.\enq
We now apply  $W(w_l)$ to the $l'$th qupit
for all $1\le l\le m$.

\beq
|S_a\ra\longmapsto  W(w_1)\otimes W(w_2)\otimes\cdots\otimes W(w_m) |S_a\ra.
\enq

 The transformation takes us from $|S_a\ra$ to a state which we  denote
 by $|\alpha\ra$: 
\begin{eqnarray}\label{super}|S_a\ra&=&
\frac{1}{\sqrt{p^d}}\sum_{f\in V,f(0)=a}|f(\alpha_1),...,f(\alpha_m)\ra
\longmapsto\\\nonumber
|\alpha\ra&=&\frac{1}{\sqrt{p^{d+m}}}\sum_{b1,b2,..bm\in F}\sum_{f\in V,f(0)=a}
w^{\sum_{l=1}^{m}e_lf(\alpha_l)b_l}|b1,..bm\ra.
\end{eqnarray}

For each string  $b_1,...b_m\in F_p$, associate 
the unique polynomial $b(x)$ which satisfies
 $b(\alpha_l)=b_l$, and has  degree $deg(b)\le m-1$.
The exponent of $w$ in equation 
\ref{super} can be written in a much simpler form 
when  $b(x)$ is of degree  $deg(b)\le m-d-1$. 
For such $b(x)$, the polynomial $h(x)=b(x)f(x)$ is of degree 
 $deg(h)\le m-1$ so:
\beq
\sum_{l=1}^{m}e_lf(\alpha_l)b(\alpha_l)=\sum_{l=1}^{m}e_l
h(\alpha_l)= h(0)=f(0)b(0)
\enq
Hence, the sum over all  $b$ with  
 $deg(b)\le m-d-1$ in equation \ref{super} gives:

\begin{eqnarray}
\frac{1}{\sqrt{p^{d+m}}}\sum_{b1,b2,..bm\in F,deg~ b(x)\le m-d-1}
~\sum_{f\in V,f(0)=a}
w^{b(0)f(0)}|b1,..bm\ra=\\\nonumber
\frac{1}{\sqrt{p^{m-d}}}\sum_{b1,b2,..bm\in F,deg~ b(x)\le m-d-1}
w^{b(0)a}|b1,..bm\ra=\\\nonumber
\frac{1}{\sqrt{p}}\sum_{b\in F_p} w^{ab}\frac{1}{\sqrt{p^{m-d-1}}}
\sum_{b1,b2,..bm\in F,deg~ b(x)\le m-d-1, b(0)=b}
 |b1,..bm\ra=\\\nonumber \frac{1}{\sqrt{p}}w^{ab}\sum_{b\in F_p}|S'_b\ra 
\end{eqnarray} 
Where $|S'_b\ra $ is the code word of the polynomial code 
when the degree of the polynomials is at most 
$m-d-1$, instead of $d$. 

Now, we claim that the sum over the rest of the $b'$s must vanish. 
The reason is that the norm of the above vector  is $1$.  
Now $|\alpha\ra$ can be written as a sum 
of two vectors: The contribution from  $b$'s with $deg(b) \le m-d-1$,
 and from the rest of the $b$'s. The two are orthogonal, 
since different $b'$s are orthogonal. Hence,  
the squared norm of  $|\alpha\ra$, which is $1$, (because 
the operation is unitary and we started with a norm one vector)
  is the sum of 
the  squared norms of the contribution of $deg(b) \le m-d-1$,
  which is also $1$, and 
the norm of the orthogonal vector. Thus, the norm 
of the sum over $b'$s with  $deg(b) > m-d-1$ must vanish.

Now, since $m=3d+1$, 
 the degree of the polynomials in $|S'_b\ra$ is larger 
than that of $|S_b\ra$. To fix this, we apply 
a degree reduction, which we will show shortly. 
The degree reduction  takes the state  
$|S'_b\ra$ to $|S'_b\ra|S_b\ra$. 
To complete the Hadamard transformation, 
 we have to erase $|S'_b\ra$.
 This is done by applying pit-wise subtraction 
of the second state from the first state. 
 $|S'_b\ra|S_b\ra$ will then be taken to $|S'_0\ra|S_b\ra$. 
Then, we can discard the first register, and this completes 
the generalized Fourier transform.

Note that 
to achieve a generalized Fourier transform with $w^r$ instead of $w$,  
we should simply  replace 
$w$ by $w^r$ in all places where $w$ appears in the above procedure.

\subsubsection{Fault Tolerant Generalized Toffoli Gate }

%We will first construct the transformations
%\beq\label{toftof}
%|S_a\ra|S_b\ra|S_c\ra|S_d\ra \longmapsto |S_a\ra|S_b\ra|S_c\ra|S_{ab+c+d}\ra
%\enq
%and 
%\beq\label{toftof2}
%|S_a\ra|S_b\ra|S_c\ra|S_d\ra \longmapsto |S_a\ra|S_b\ra|S_c\ra|S_{ab+c-d}\ra
%\enq
%It is easy to get the Toffoli gate from these two transformations,
% in the following way:
%\beq
%|a,b,c,0,0\ra\longmapsto |a,b,c,0,ab\ra\longmapsto |a,b,c+ab,0,ab\ra\longmapsto |a,b,c+ab,0,0\ra
%\enq
%by applying the transformation \ref{toftof} on qupits $1,2,4,5$ then 
%applying generalized $CNOT$ from qupit $5$ to $3$ and then  
%  transformation \ref{toftof2} on qupits $1,2,4,5$.
%To apply the two transformations \ref{toftof} and \ref{toftof2},
%To get the transformation \ref{toftof2} we simply replace in the
% final interpolation stage of the degree reduction 
%addition by subtraction. 

To apply the generalized Toffoli gate on $|S_a\ra|S_b\ra|S_c\ra$, 
 we will again use degree reduction. 
We first add another state,  $|S'_0\ra$ 
which is an encoded $0$  using the same polynomial code except the 
 degree is $2d$. 
Applying  the general
 Toffoli gate pit-wise  on the $m$ coordinates, from the first two states 
$|S_a\ra|S_b\ra$ to the extra state 
 gives 
 $|S_a\ra|S_b\ra|S'_{ab}\ra|S_c\ra$, as is easy to check. 
We now apply degree reduction on the third register, 
using the fourth register as the target of the interpolation
as is explained in the next section.  
This gives $|S_a\ra|S_b\ra|S'_{ab}\ra|S_{ab+c}\ra$. 
Applying the reverse of the generalized Toffoli gate,  pit-wise
on the first three registers gives 
  $|S_a\ra|S_b\ra|S'_{0}\ra|S_{ab+c}\ra$, 
and we can discard the extra $|S'_{0}\ra$
to obtain the desired state  $|S_a\ra|S_b\ra|S_{ab+c}\ra$.

%To obtain the state $|S'_c\ra$ from $|S_c\ra$, 
%(which is the opposite of degree reduction), 
% add $|S_c\ra$ to this state by applying bitwise generalized
%CNOT gates from $|S_c\ra$ to $|S'_0\ra$. This gives 
%$|S_c\ra|S'_c\ra$.  
%Then we subtract the second register from the first, 
%by applying pit-wise subtraction (using again generalized CNOT gates.)
%This gives $|S'_0\ra|S'_c\ra$, and we can discard the first register. 

%\begin{eqnarray}
%|S_a\ra|S_b\ra|S_0\ra &=&\\\nonumber &&
%\frac{1}{\sqrt{p^{3d}}}\sum_{f,g,h\in V_1,f(0)=a,g(0)=b,h(0)=0}
%|f(\alpha_1),...,f(\alpha_m)\ra|g(\alpha_1),...,g(\alpha_m)\ra |h(\alpha_1),...,h(\alpha_m)\ra\longmapsto \\\nonumber &&
%\frac{1}{\sqrt{p^{3d}}}\sum_{f,g,h\in V_1,f(0)=a,g(0)=b,h(0)=0}\\\nonumber &&
%|f(\alpha_1),...,f(\alpha_m)\ra|g(\alpha_1),...,g(\alpha_m)\ra |f(\alpha_1)g(\alpha_1)+h(\alpha_1),...,f(\alpha_m)g(\alpha_m)+h(\alpha_m)\ra
%\end{eqnarray}
%On the third block we obtain the values of the polynomial 
%the polynomial $D(x)=f(x)g(x)+h(x)$ which satisfies  $D(0)=ab$,
%and its   degree satisfies $deg(D)\le 2d$.
%where instead of the state $|S_0\ra$ on which we interpolate, 
%we use the original state $|S_c\ra$.  
%This gives the final state 
%the above state tensored with $|S_{c+ab}\ra$. 
%We then  apply the reverse of the Toffoli gate pit-wise, to get back 
% $|S_a\ra|S_b\ra|S_0\ra$ on the first three blocks. 
%This gives us the transformation \ref{toftof}.

\subsection{Fault Tolerant Degree Reduction}
The term {\it degree reduction} means the following. 
Let $C'$ and $C$ be two quantum polynomials codes of the same
 length $m$, 
which use polynomials of degree $d'$ and $d$ respectively, 
such that $d'>d$, where the polynomials are evaluated at the same $m$
 points in the field.   
A degree reduction procedure takes  a word in the first code
 $|S'_a\ra$ to the word $|S_a\ra$ in the second code 
which encodes the same logical pit.
The restriction on our procedure 
 is that it is fault tolerant. 
 This means that we cannot simply decode $a$ and encode 
it again to get $|S_a\ra$,
because then even one error cannot be corrected, since the 
 state of the  environment will depend on the encoded pit.  

The following solution, which 
generalizes  classical degree reduction techniques 
by Ben-Or {\it et al}\cite{bgw},
is best  illustrated in the following figure: 

\begin{figure}[h!]
\begin{center}
\vbox{\epsfxsize=4in\epsfbox{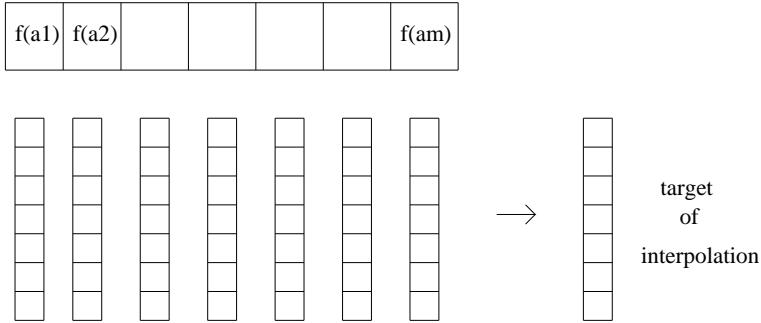}}
\caption{Scheme for degree reduction.}
\end{center}
\end{figure}

%\begin{figure}[h!]
%\centerline{\psfig{figure=degreereduction.eps,height=5cm,width=10cm}}
%\caption{Scheme for degree reduction.}
%\end{figure}

The idea goes  as follows. 
Let 
\beq
\sum_j c_j f(\alpha_j)=f(0)
\enq
so $c_j$ are the interpolation coefficients for polynomials of degree 
less than $m$. 
We would have wanted to apply interpolation on the coordinates 
of the encoded state, so as to get $a$, and then to encode $a$
by the code $C$. 
However, this is not fault tolerant. 
The trick is to first  encode each coordinate in 
the original state  $|S'_a\ra$, so each coordinate  is {\it opened} to an encoded 
word in the code $C$. 
 This is done by first  adding
 $m$ encoded states $|S_0\ra$, 
and then, from each coordinate, say the $i'$th coordinate, 
 apply $m$ generalized CNOT gates to all the coordinates 
 in the $i'$th state $|S_0\ra$, so that 
the extra state now encodes the coordinate. 
  This is illustrated in the figure by an $m\times m$  matrix, 
the columns of which are the 
 encoded words, each encoding one coordinate 
in the original encoded word. 
We can now apply interpolation pit-wise from the opened coordinates to 
an extra  state $|S_0\ra$, by computing along the rows of the matrix. 
If the matrix elements are symbolically denoted by $a_{i,j}$,
We apply the sum 

\beq
\sum_j c_j a_{i,j} 
\enq
over the elements in the $i'$th row of the matrix, and write it on
the $i'$th coordinate of the extra state  $|S_0\ra$. 
Note that this computation is done separately on each row, 
or pit-wise,  on the encoded opened words. 
It is easy to check that 
if no errors occurred, this indeed achieves the correct state
$|S_a\ra$ on the extra state which was the target of the interpolation. 
 However,  the state of the 
ancilla pits certainly depend on the original encoded pit. 
To take care of this, we ``wrap'' the encoded states by 
applying the reverse of the operations which we applied when we opened
 the states; To wrap the  $i'$th coordinate, 
 apply $m$ generalized CNOT gates to subtract 
 this coordinate from all the coordinates 
 in the $i'$th opened state.
If no error occurred, we get back  the original $m$ states $|S_0\ra$, 
which can be discarded.

We would now like to take care of errors that might occur during this 
procedure, and this will force us to change the above procedure a little, 
and add error corrections. 
Note that an error during the opening of each coordinate 
 might cause the entire encoded state to 
be wrong, since this part was not done 
fault tolerantly. Note that the interpolation 
is completely ruined, if even one of the 
columns of the matrix encodes a wrong pit.
Thus, we must apply error corrections before we apply the interpolation. 
We first apply an error correction on each of the encoded 
states, so that it is a word in the code.
Then, we apply an error correction  according to the code $C'$, 
{\it encoded} by the code $C$, to the overall word.
We will immediately show how to do this. 
Given this transformation, 
 it is easy to see 
 that everything we have done is 
fault tolerant: We first open each coordinate, then apply error 
corrections, then apply the interpolation, and finally 
close each encoded state back to one coordinate, and discard the ancilla
qubits. 
This achieves
the transformation of $|S'_a\ra$ to $|S'_a\ra|S_a\ra$ 
fault tolerantly, and 
 completes the description of the degree reduction.

%
% or even a superposition of 
%such. We consider an encoded pit-flip, and an encoded phase flip, and both, 
%on these encoded states. An encoded phase flip does not change the 
%result of the state which is the target of the interpolation. 
%On the other hand, a 

In the same way, one can achieve the transformation 
$|S'_a\ra|S_b\ra$ to $|S'_a\ra|S_{a+b}\ra$ 
by using the state $|S_b\ra$ as the target of the interpolation, instead of 
$|S_b\ra$. This will be useful in the construction of the Toffoli gate.

%In fact, one can apply the above technique, except the final part of 
%erasing the original state, 
% to any state in the classical code. 
%It is easy to check that starting with  a state of the form
%$|f(\alpha_1)....f(\alpha_m)\ra$ where $deg(f)\le d' $ and $f(0)=a$. 
%and applying the degree reduction to this  state we end up with
%$|f(\alpha_1)....f(\alpha_m)\ra|S_a\ra$. 
%This will be used shortly in the Toffoli gate. 

\subsection{Fault Tolerant Encoded Error Correction}
We show how to apply an encoded 
error correction on a doubly encoded word. 
This means the following. 
The state which we operate on is generated by encoding a state 
using the code $C$, and then encoding the resulting state 
again, using the code $C'$. 
We want to apply on the entire state an encoded error correction 
procedure. This means that the procedure is exactly the 
error correction procedure according to the code  $C$,
but each gate in the procedure is encoded by the code $C'$.
We thus want to correct $C'-$encoded pit flips, which means 
the operator on words in $C'$ 
\beq
\calE_{S_b(C')}: |S_a(C')\ra\longmapsto |S_{a+1}(C')\ra
\enq
and $C'-$encoded phase flips, operating on $C'$ by 
   \beq
\calE_{S_f(C')}: |S_a(C')\ra\longmapsto w^{a}|S_{a}(C')\ra
\enq
The natural way to correct these errors on words in the code $C$
encoded by $C'$ is  simply to apply the error 
correction procedure of the code $C$, where each gate is replaced
by a fault tolerant procedure according to the code $C'$. 
However, 
one must be careful, in order not to make a circular 
argument here. The
 problem is that  we want to apply an encoded error correction
in spite of the fact that we not yet have  the fault tolerant procedures
for the entire set
 of gates used in the error correction, and in particular, we 
cannot apply encoded Fourier transform, because in this procedure 
we use
degree reduction, which uses encoded error correction. 
It turns out that this does not pose any difficulty since
 in this case we can omit the degree reduction everywhere. 
We proceed as follows.

For an encoded error correction, 
we need cat states as in equation \ref{cat}, encoded by the code $C'$. 
Instead of generating cat states of length $l$ as we do in the error 
correction procedure,  we generate an encoded cat states
on $ml$ qupits. 
We first generate  a state $|S_0(C'')\ra$ in the polynomials code 
with degree $m-d-1$, which we denote by $C''$, 
using an encoding procedure according to this 
code. We then apply pit-wise generalized Fourier transforms, 
to get 
\beq
\frac{1}{\sqrt{p}}\sum_{i=0}^{p-1}|S_a(C')\ra
\enq
 as is shown  
in section \ref{ftpoly}.
We then copy it on $l-1$ ancilla states $|S_0(C')\ra $  
using generalized CNOT gates pit-wise. 
This generates the state 
\beq
|S_{cat_l}\ra=\frac{1}{\sqrt{p}} \sum_{i=1}^{p-1} |S_i(C')\ra|S_i(C')\ra\cdots|S_i(C')\ra
\enq
(It might seem as if generating encoded cat states is much simpler than 
generating regular cat states, as we do in section \ref{errorcor},
because we avoid here the main problem:  
verifying that the cat states are correct fault tolerantly. 
In fact, these verifications are not avoided here, 
 but actually hidden in the generation of 
the encoded states, $|S'_0\ra$ and $|S_0\ra$, which are done using error 
corrections.)

From these cat states, it is easy to generate the encoded 
version of the state 
\ref{par}, using generalized Fourier transforms pit-wise. 
We thus have: 

\beq\label{parencoded}
\sum_{j,j\cdot \vec{1}=0~mod ~p} |S_{j_1}(C')\ra\cdots |S_{j_l}(C')\ra
\enq

We now want to compute the $j'$th  syndrome pit. 
Recall that this was done by computing the inner product of 
the string $(f(\alpha_1)...f(\alpha_m))$ 
with the  $j'$th parity check matrix of the code. 
The un-encoded version of this operation is done by adding
$h_jf(\alpha_j)$ to the $j$'th coordinate in the un-encoded 
version of the state \ref{parencoded}. 
The natural encoded version of this is done by 
adding $h_ja_{i,j}$ to the $(mi+j)$'th coordinate in the state 
\ref{parencoded}, 
where   $a_{i,j}$ 
is the $i'$th coordinate in the state encoding $f(\alpha_j)$. 
We now apply a classical computation on the ancilla state, 
computing the sum of the encoded coordinates mod $p$, to give the 
corresponding syndrome bit. In this way we can compute all the 
pits in the syndrome, where like in the error correction procedure, 
everything 
 is done all over again for each syndrome pit, to avoid 
propagation of errors.  
The syndrome is computed in this way for each one of the 
$m^2$ pits,  and from the $(i,j)'$th copy we compute whether the 
$j$'th encoded coordinate suffered a pit-flip.  If so we apply the 
reverse operation on  
the  $(i,j)'$th pit. 

A problem arises when we want to transform to the encoded 
$C-$basis, since the encoded Fourier transform gate 
 is not yet in our repertoire. 
As was shown in the subsection \ref{ftpoly}, applying
generalized Fourier transforms pit-wise (without 
degree reduction) takes us to the $C-basis$ of the polynomial code 
 with  degree $m-d-1$, if $C'$ is of degree $d$. 
And so after the rotation to the $C-$basis we should actually 
correct pit-flips as in the first step, except we should 
replace $C'$ everywhere by $C''$. 
At the end of this correction stage, we apply 
the inverse of the generalized Fourier transforms, pit-wise.

To show that this procedure indeed corrects general encoded errors, 
we first note that it suffices to correct encoded pit flips and encoded 
phase flips, just as in theorem \ref{bitphase}. 
The proof of theorem \ref{cash} can be slightly modified to show that 
the procedure 
 corrects and detects $C'-$encoded pit flips and $C'-$encoded phase flips. 
Denote by $\calH$ the tensor product of these 
generalized Fourier transforms applied pit wise. 
The crucial fact is that a $C'-$encoded phase flip 
transforms by $\calH$ to a $C''$-encoded pit flip: 
\begin{eqnarray}
~~~~~~~\calH \circ \calE_{S_f(C')}|S_a(C')\ra= \calH (-1)^a |S_a(C')\ra=(-1)^a|C_a(C'')\ra=
 \calE_{S_b(C'')}|C_a(C'')\ra=
 \calE_{S_b(C'')}\circ\calH |S_a(C')\ra\\\nonumber
\end{eqnarray}
Hence, the sequence of equalities  \ref{bitandphase}
holds also in our case, which shows that the above procedure 
corrects general encoded errors, if their number is smaller than 
that which the code $C$ can correct.

\section{Universality of the Sets of Gates ${\cal G}_1$ and
 ${\cal G}_2$.}\label{sec6}\label{universal}
The sets of gates ${\cal G}_1$ and ${\cal G}_2$ were not chosen arbitrarily; 
They were chosen so that any quantum computation can be expressed using these
sets. 
In other words, these are universal sets of gates.
A proof of this fact for ${\cal G}_1$ is missing in Shor's paper introducing 
the set ${\cal G}_1$, and we provide here a detailed 
proof, relying on a universality proof by Kitaev. 
A similar result was achieved independently by Boykin {\it et al.}\cite{boykin}. 
The proof that the set of gates  ${\cal G}_2$ is universal 
is more complicated, 
and is based on  geometrical arguments on the groups $U(n)$, together 
with some basic facts from field theory. 
We start with a discussion of the issue of 
universality, and  then show that our sets of gates 
${\cal G}_1$ and  ${\cal G}_2$ are indeed universal.

\subsection{Universal Sets of Gates}
We consider a set of gates, and ask which unitary operations can be 
constructed when gates from this set are applied on the qubits,
in any order.

\begin{deff}
Let $p\ge 2$. 
A set of gates ${\cal G}$ on $k>1$ qupits is said to be universal
if ${\cal G} \cup \{e^{i 2 \pi \theta}I\}_{real~\theta}$ generates a dense subset 
in $U(p^k)$.
\end{deff}

Note that in quantum computation, one can 
multiply a gate by an overall scalar 
of absolute value $1$ (which is sometimes called a phase factor)
and the density matrices will remain the same. 
Therefore we can add to ${\cal G}$ all phase factors, 
without changing the set of operations on the quantum states.  
%Similarly, we can multiply all gates in ${\cal G}$
%by an overall phase such that their determinants are $1$,
%and ask what is the subgroup which is generated by these gates in 
%the special unitary group $SU(n)$. 
This definition indeed captures the notion of universality because 
it turns out that  
 an algorithm using any set of gates 
 can be translated to an algorithm which uses only gates from 
a universal set of gates ${\cal G}$, 
such that 
$a)$, the new circuit computes 
a function which  
approximates to any given 
 accuracy the original function (in total variation distance)  $b)$, the new circuit is only polylogarithmically 
larger and deeper than the original one, and $c)$, the design of the new circuit can be efficiently computed given the design of the original circuit. 
This fact is based on two known results. 
The first is that  a universal set of gates can be used to 
approximate matrices exponentially fast, and the sequence of gates from the universal set
can be found efficiently: 

\begin{theo} Kitaev, Solovay,:

Consider a universal 
 set of gates ${\cal G}$ over $U(m)$, for some integer $m$.  
Then there exists a polynomial $p_m$ 
such that  any element of $U(m)$ can be approximated 
up to $\epsilon$ by a word from $G$ which is not longer 
than    $p_m(log(1/\epsilon))$, and this word can be found 
efficiently (in time which is  $p_m(log(1/\epsilon))$)
 by a classical Turing machine. 
\end{theo}

The proof of this theorem\cite{kitaev0,solovay} uses Lie
 groups and Lie Algebras, and will not be discussed here.  
The second result which we will need to justify the choice of definition of 
universality, is that an operation on any number of qubits (qupits) can be 
achieved using gates on two qubits (qupits). 
This was proved for qubits by  DiVincenzo\cite{twobit}, and simplified 
by Barenco {\it et. al.}\cite{barenco4}. 
We give here a proof for the general case of qupits.

\begin{theo}\label{twobit}
Let  ${\cal G}$ be a universal set of gates on $k\ge 2$ qupits, for $p\ge 2$.
Consider the Hilbert space of $m>k$ qupits. 
Then the gates in the set ${\cal G}$, extended to $m$ qupits, generate 
 a dense subset of $U(p^m)$. 
\end{theo}

{\bf Proof:} 
The proof generalizes ideas which were first used by
 Deutsch\cite{deutsch2} and Barenco {\it et. al}\cite{barenco4}.
We define a generalized Toffoli gate on $m$ qupits, $T_{m}(Q)$, to be 
a gate which applies $Q$ on the $m'$th qupit conditioned that the first $m-1$ 
qupits are in the state $p-1$:

{~}

\setlength{\unitlength}{0.030in}
\begin{picture}(40,40)(-80,5)
\put(0,15){\line(1,0){10}}
\put(20,15){\line(1,0){10}}
\put(0,30){\line(1,0){30}}
\put(0,45){\line(1,0){30}}
\put(15,30){\circle*{3}}
\put(15,45){\circle*{3}}
\put(15,20){\line(0,1){25}}
\put(10,10){\framebox(10,10){$Q$}}
\end{picture}

Where the number of vertical wires is $m$. 
The conditioned $Q$ can be applied  on  the $k'$th qupit, 
instead of the $m'$th one, 
in which case we denote $T_{m,k}(Q)$. 
Following Barenco {\it et. al. }\cite{barenco4} we show an explicit 
sequence of generalized Toffoli gates on $m-1$ qupits, $T_{m-1}(Q)'$s,  
 which  constructs   $T_{m}(Q)$. 
Denote $V=Q^{\frac{1}{p}}$, so:

\setlength{\unitlength}{0.030in}

\begin{picture}(40,60)(-10,0)

\put(0,15){\line(1,0){10}}
\put(20,15){\line(1,0){10}}
\put(0,30){\line(1,0){30}}
\put(0,45){\line(1,0){30}}

\put(15,30){\circle*{3}}
\put(15,45){\circle*{3}}

\put(15,20){\line(0,1){25}}

\put(10,10){\framebox(10,10){$Q$}}

\end{picture}
\begin{picture}(20,60)(0,0)

\put(0,0){\makebox(20,60){$=$}}

\end{picture}
\begin{picture}(90,60)(0,0)

\put(0,15){\line(1,0){10}}
\put(20,15){\line(1,0){20}}
\put(50,15){\line(1,0){20}}
\put(80,15){\line(1,0){10}}
\put(0,30){\line(1,0){90}}
\put(0,45){\line(1,0){90}}

\put(30,30){\circle{6}}
\put(60,30){\circle{6}}

\put(15,30){\circle*{3}}
\put(45,30){\circle*{3}}
\put(30,45){\circle*{3}}
\put(60,45){\circle*{3}}
\put(75,45){\circle*{3}}

\put(75,20){\line(0,1){25}}
\put(15,20){\line(0,1){10}}
\put(30,27){\line(0,1){18}}
\put(45,20){\line(0,1){10}}
\put(60,27){\line(0,1){18}}

\put(10,10){\framebox(10,10){$V^{p-1}$}}
\put(70,10){\framebox(10,10){$V$}}
\put(40,10){\framebox(10,10){$V^{\scriptsize \dag}$}}

\end{picture}

The picture should be interpreted as follows. 
The first wire is actually duplicated $m-2$ times.
The sign $\oplus$ is actually the pit-flip,  which adds $1$ modulo 
$p$ to every element in the field (This is a generalization of 
the NOT for $F_2$, and we sometimes call it the generalized NOT) 
\begin{equation}
Pit-flip:|a\ra \longmapsto |(a+1) \rm{mod} p\ra
\enq
The sequence of gates which consists of the  
second  and third gates, i.e. the controlled generalized $NOT$ and the 
controlled $V^\dagger$, 
 is  repeated $p-1$ times. 
It is easy to check that the above circuit indeed gives 
the desired controlled $Q$,  
by considering what happens to the basis states, 
in two cases:  all first $m-2$ qupits are equal to $p-1$, or not. 
Using the above scheme recursively we can construct a circuit
which uses two-qupit gates and applies
 $T_{m}(Q)$ for any $m>2$ and any one-qupit $Q$.
Note that the recursion starts with two-qupit gates, which is the reason 
why we require $k\ge 2$.

The gate 
$T_{m}(Q)$ can be seen as applying a general  $Q$ on the subspace 
spanned by the last $p$ 
basis vectors, while applying identity on the rest. 
The next step is to construct a generalization of the above gate, 
i.e. a gate which applies  $Q\in U(p)$ on the subspace spanned 
by any $p$ basis vectors $|i_1\ra,..,|i_p\ra$, 
 while applying identity on the rest
of the basis vectors. Denote this matrix by $T_m(Q,i_1,...,i_p)$.
We first note that   $T_m(Q)$  with $Q$ being any permutation on 
the basis vectors of the Hilbert space of one qupit, together 
with  generalized 
$NOT$ gates, 
 generate all permutation matrices
 on $m$ qupits.
Generalized $NOT$ gates are one-qupit gates,
so we are allowed to use those.   
It suffices to construct all matrices of the form
 $\tau_{i,j}$, where 
 $|i\ra,|j\ra$ are basis vectors for which the
 strings $i,j$ differ in one coordinate.
  $\tau_{i,j}$
 switches the two basis vectors  $|i\ra,|j\ra$  
and  $\tau_{i,j}|k\ra=|k\ra$ for any $k\ne i,j$.
W.l.o.g. let the coordinate on which $i,j$ disagree be the last coordinate, 
and let this coordinate be equal to  $a$ in the string $i$, and to $b$ 
in the string $j$. 
 To construct $\tau_{i,j}$, apply sufficiently many 
generalized $NOT$ gates  
on all the coordinates except the last coordinate, 
so that all these coordinates in both strings become equal to $p-1$.
Now, apply $T_m(Q)$ with $Q$ on the last coordinate 
being the matrix which permutes $|a\ra$ and $|b\ra$,  
leaving the rest of the basis vectors untouched.  
Then reverse all  the generalized  $NOT$ gates  on 
all the coordinates but the last one, which  
 gives $\tau_{i,j}$, and therefore all permutations on basis vectors. 
The general  $T_m(Q,i_1,...,i_p)$ for any $Q$  can clearly be achieved by
first permuting the basis vectors $ i_1,...,i_p$ to the last 
$p$ vectors, applying $T_m(Q)$, and then permuting back. 

The last step is to use $T_m(Q,i_1,...,i_p)$
to construct a general 
 $p^m\times p^m$ unitary matrix $U$. 
Let us denote the $p^m$ eigenvectors of $U$ by
 $|\psi_j\ra$ with corresponding
 eigenvalues
$e^{i\theta_j}$.
$U$ is specified by \(U|\psi_j\ra= e^{i\theta_j}|\psi_j\ra.\)
Define:
\begin{equation}
 U_k |\psi_j\ra = \left\{\begin{array}{ll}
|\psi_j\ra &\mbox{if $k\ne j$} \\
 e^{i\theta_k}|\psi_k\ra & \mbox{if $k= j$}\end{array}\right.
\end{equation}
Then $U=\Pi_{k=1}^{p^m} U_k$.
It is left to show how to construct $U_k$:
We first construct a transformation $R$ which 
takes  $|\psi_k\ra$ to $\lambda|(p-1)^m\ra$
where $\lambda $ is a complex number of absolute value $1$. 
We don't care what $R$ does to the rest of the vectors.  
          After applying $R$ we apply the generalized Toffoli which takes 
 $|(p-1)^m\ra \longmapsto e^{i\theta_k}|(p-1)^m\ra$  
and does nothing on the rest of the basis states. Then
 we take $\lambda|(p-1)^m\ra$ to $|\psi_k\ra$ by applying $R^{-1}$.
This indeed achieves $U_k$, as can be easily checked. 
To construct $R$, and similarly $R^{-1}$, 
we start with $|\psi_k\ra$, and we first make the coefficient in front 
of $|0^m\ra$ to be zero by 
 a rotation in the plane spanned by $|0^m\ra$ 
and $|(p-1)^m\ra$, which is a special case of the $T_m(Q,i_1,...,i_p)$
we have dealt with before. 
Thus, the weight of $|0^m\ra$ has been shifted to 
$|(p-1)^m\ra$. 
In this way, 
the weights in front of all basis vectors, one by one, 
 are shifted to $|(p-1)^m\ra$, and this achieves $R$. 
$\Box$

We proceed to  prove  a few geometrical lemmas
 which will be useful in 
proving universality of the sets of gates we use. 
The first and second lemmas were
 used by Kitaev\cite{kitaev0} for proving universality
of his set of gates. We will use these lemmas to prove 
universality of $\calG_1$ and $\calG_2$.
It seems that the method which is used here to prove universality is quite general. 
\begin{lemm}\label{one}
Let $n\ge 3$
Let $|\alpha\ra\in {\cal C}^n$.  
Let $H$ be the subgroup in $U(n)$, which fixes  $|\alpha\ra$, 
Let $V\in U(n)$ be a matrix not in $H$.  
Then, the subgroup generated by $H$ and $V$ is dense
in $U(n)$.
\end{lemm}

{\bf Proof:}
We use some ideas which were first used by Deutsch\cite{deutsch2}.
Let $G\subseteq U(n)$ be the closed subgroup generated
by $H$ and $V$.   
Let $U$ be a 
 $n\times n$ unitary matrix, and 
 denote the $n$ eigenvectors of $U$ by
 $|\psi_j\ra$ with corresponding
 eigenvalues
$e^{i\theta_j}$.
$U$ is specified by \(U|\psi_j\ra= e^{i\theta_j}|\psi_j\ra.\)
Define:
\begin{equation}\label{uk}
 U_k |\psi_j\ra = \left\{\begin{array}{ll}
|\psi_j\ra &\mbox{if $k\ne j$} \\
 e^{i\theta_k}|\psi_k\ra & \mbox{if $k= j$}\end{array}\right.
\end{equation}
Then $U=\Pi_{k=1}^{n} U_k$.
It is thus enough to construct $U_k$.
Let $|\psi_k\ra=c|\alpha\ra+d|\delta\ra$.
Let $|\gamma\ra$ be perpendicular to both $|\alpha\ra$ and $|\delta\ra$. 
Such $|\gamma\ra$ exists because the dimension is at least $3$. 
 To construct $U_k$, 
 it suffices to rotate $|\psi_k\ra$
to $f|\gamma\ra$, where $f$ is an additional phase factor, 
 apply a matrix which 
multiplies
$|\gamma\ra$  by $e^{i\theta_k}$  and does nothing to 
vectors orthogonal to $|\gamma\ra$, then rotate back  $f|\gamma\ra$ to $|\psi_k\ra$. 
The middle operation of multiplication by the phase $e^{i\theta_k}$
of $|\gamma\ra$ is clearly in $H$. 
The  transformation which
 takes  $|\psi_k\ra$
to $f|\gamma\ra$ 
can be approximated to any degree 
of accuracy as follows. 

We first find some improved versions of $V$ in $G$.
We have $V|\alpha\rangle =
a|\alpha\rangle + b|\beta\ra$. By multiplying $V$
on the left by a suitable element of $H$, we can arrange that the
resulting element $V_1$ satisfies: $V_1|\alpha\rangle =
a|\alpha\rangle + b|\delta\ra$. There is a unit vector $|\eta\rangle$
which is mapped by $V_1$ onto $\bar{b}|\alpha\ra - \bar{a}|\delta\ra$. So
$V_1|\alpha\ra$ is perpendicular to $V_1|\eta\ra$. Hence
$|\eta\rangle $ is perpendicular to $|\alpha\rangle$. So we can find
$h \in H$ such that $h|\delta\ra = |\eta\ra$. Let $V_2 = V_1\cdot
h$. Then $V_2|\alpha\ra = a|\alpha\ra + b|\delta\ra$, and $V_2$ 
maps the space spanned by
$|\alpha\ra$ and $|\delta\ra$ onto itself.
$|\gamma\ra$ and $V_2|\gamma\ra$ are
 perpendicular to both $|\alpha\ra$ and $|\delta\ra$. Hence there is an
$h_1 \in H$ that fixes both $|\alpha\ra$ and $|\delta\ra$ and 
that moves $V_2|\gamma\ra$
onto $|\gamma\ra$. Let $V_3 = h_1\cdot V_2$. Then $V_3$ fixes $|\gamma\ra$ and
$V_3|\alpha\ra= a|\alpha\ra + b|\delta\ra$.

We now construct a sequence  of elements $W_i \in G$ (by induction on
$i$) such that $W_i|\psi_k\ra$ has the form  $ca^i |\alpha\ra + 
d_i|\gamma\ra$. If we can do this,
 then for $i$ large, $ca^i$ tends to $0$ since
$|a| < 1$. Hence $|d_i|$ tends to $1$ as $i$ goes to $\infty$ and the
lemma will be proved.
For $i = 0$, we simply choose $W_0 \in H$ such that $W_0|\delta\ra =
|\gamma\ra$. Now suppose $W_i$ has been chosen to satisfy our inductive
hypothesis.
Then 
\beq
 V_3W_i|\psi_k\ra = ca^{i+1}|\alpha\ra + bca^i|\delta\ra + d_i|\gamma\ra
\enq
The vector $bca^i|\delta\ra + d_i|\gamma\ra$ 
is clearly perpendicular to $|\alpha\ra$ and
non-zero. Hence we can find $h^\star \in H$ that moves this vector to
a vector of the form $d_{i+1}|\gamma\ra$. It suffices now
to set $W_{i+1} = h^\star V_3 W_i$. $\Box$

\begin{lemm}\label{corkitaev}
Let $U_1,U_2$ be two non-commuting matrices in $SU(2)$, 
such that their eigenvalues are not integer roots of unity. 
The subgroup generated by $U_1$, $U_2$  
is dense in $SU(2)$.
\end{lemm}

{\bf Proof:}
 If $x$ is an element of $SU(2)$ not of finite order, then the
closed subgroup generated by $x$ is connected and of dimension
$1$. The closed group generated by both $U_1$, and $U_2$ 
is thus connected, and is non commutative. 
 Any connected non-commutative subgroup of $SU(2)$ is \emph{all}
of $SU(2)$. $\Box$

%$
%|\psi_k\ra=a|\alpha\ra+b|\delta\ra$
%where $|\delta\ra\in S$. 
%Then, rotate $b|\delta\ra$ to $b|\gamma\ra$
%while leaving everything else as is. 
%We now apply alternately the matrix $H$ which transforms some of the weight 
%in $|\alpha\ra$ to $|\beta\ra$, and a rotation 
%which takes all the weight in $|\beta\ra$ and puts it in $|\gamma\ra$. 
%This reduces the coefficient in front of $|\alpha\ra$ each time by a factor 
%of $a$, while transforming all the weight of the vector to $|\gamma\ra$. 

\begin{lemm}\label{bridge}
Let $A,B$ be two non orthogonal subspaces of 
$C^n$. 
Let $G_A$, $G_B$ be  dense subsets of $U(A)$,$U(B)$ 
respectively. 
Then the subgroup generated by $G_A\cup G_B$ 
 is dense 
in $U(A\oplus B)$. 
\end{lemm}

\noindent {\bf Proof:}
We will first prove the lemma for the case $dim(A)=dim(B)=1$, and 
$A\not=B$.  
Let $G$ be the group generated by $U(A)\cup U(B)$. 
We consider the natural map from $U(2)$ to $SU(2)$, 
$U\longmapsto U/\sqrt{Det(U)}$,
and look at the image of $G$ in $SU(2)$ under this map.  
We can find in $U(A)$ and in $U(B)$ two matrices such that their  
images are of infinite order and do not commute, 
and by lemma \ref{corkitaev}, the image of $G$ 
 is dense in $SU(2)$. 
Hence, $G$ contains all matrices in $SU(2)$ multiplied by a certain 
phase. Now, observe that taking the commutators of these matrices 
 erases the phases.
Using the fact that the commutator group of $SU(2)$ is $SU(2)$ itself, 
$[SU(2),SU(2)]=SU(2)$, we have that $[G,G]\subseteq G$ contains a dense
subset of $SU(2)$. 
Using $U(A)$ we can generate all phase factors, 
which proves the lemma for the case $dim(A)=dim(B)=1$.

%Let us consider the operation of $U(A)$,$U(B)$ on the subspace $A\oplus B$.   
%Let $|\alpha\ra,|\alpha^\perp\ra$ be 
%an orthonormal basis for $A\oplus B$, such that 
%$|\alpha\ra$ is a basis for $A$, and similarly 
%let $|\beta\ra,|\beta^\perp\ra$ be an orthonormal
% basis for $A\oplus B$, such that 
%$|\beta\ra$ is a basis for $B$. 
%Clearly, any matrix which is diagonal in the basis 
%$|\alpha\ra,|\alpha^\perp\ra$  is in our repertoire.
%Let $U_A\in SU(2)$ be such a matrix of infinite order. 
%Similarly, let $U_B\in SU(2)$ be a 
% matrix of infinite order  
%which is diagonal in the basis 
%$|\beta\ra,|\beta^\perp\ra$.
%$U_A$ and $U_B$ do not commute, because 
%they cannot be diagonalized simultaneously.  By lemma \ref{corkitaev}
%they generate the entire $SU(2)$. 
%Adding the phase factors, this gives $U(2)=U(A\oplus B)$, 

Let us now consider the general case. 
Let $|\alpha_1\ra,...|\alpha_m\ra$ be a basis for $A$. 
Let $|\beta_{m+1},....|\beta_s\ra$ be a basis for $B$, such that 
each vector in the basis of $B$  is not orthogonal to $A$. 
(To construct such a basis, let us start with any given basis. 
At least one basis vector is not orthogonal to $A$. 
Adding it to all the vectors in the basis which are orthogonal to $A$
gives the desired basis. ) 
Let $B_i$ be the subspace spanned by $|\beta_i\ra$, and denote by 
$A_i= A\oplus B_1\oplus B_2\oplus\cdots\oplus B_i$. 
We will show by induction on $i$ that 
the group $U(A_{i-1})$
is in the subgroup generated by $U(A)\cup U(B)$ and all phase factors
on $A\oplus B$.

For $i=0$, the claim is trivial.  
We assume for $i-1$, that the group 
 $U(A_{i-1})$ is in the subgroup generated
by  $U(A)\cup U(B)$, 
and prove for $i$.
If
$|\beta_i\ra\in A_{i-1}$, the induction step is proven. 
Hence, we assume $|\beta_i\ra\not\in A_{i-1}$. 
We have $|\beta_i\ra\in B$, so 
we have in our repertoire $U(|\beta_i\ra)$. 
We divide to two cases. 
If $dim(A_{i-1})=1,$ 
then the induction step follows from the simple case in which 
$dim(A)=dim(B)=1$. 
If $dim(A_{i-1})\ge 2$, then  $dim(A_{i-1}\oplus B_i)\ge 3$, 
and we can use lemma \ref{one}. 
Let $|\gamma_i\ra\in A_{i}$ be the (non trivial) 
 projection of $|\beta_i\ra$ 
on the subspace orthogonal to $A_{i-1}$.
 $U(A_{i-1})$ is exactly the subgroup of $U(A_{i})$ which fixes 
 $|\gamma_i\ra$. 
The matrix which multiplies  $|\beta_i\ra$ by a non 
trivial phase factor, leaving all orthogonal vectors untouched, 
does not fix  $|\gamma_i\ra$. 
Hence lemma \ref{one} can be applied, and we have $U(A_i)$. $\Box$

\subsection{Universality of the Set of Gates ${\cal G}_1$ used for CSS Codes}

The set of gates used for $CSS$ codes is shown here to be universal.  
This is done by a reduction to a proof of Kitaev\cite{kitaev0}, 
who used lemmas \ref{one} and  \ref{corkitaev}.  

\begin{theo}\label{un}
 ${\cal G}_1$ together with all phase factors 
 generate a dense subgroup in the group of special unitary 
matrices operating on five qubits, $U(2^5)$.
\end{theo}

\noindent
{\bf Proof:} 
The proof is based on a result by Kitaev\cite{kitaev0}, which asserts that
the following set of gates is universal:
 $CP$ is a two qubit gate which takes $|11\ra\longmapsto i |11\ra$ 
and applies identity on the rest, and the Hadamard gate.
Let us first show how to construct $CP$ from our set of gates:  
We will denote by $T_k$ a  generalized Toffoli on $k$ qubits, which  
applies not on the $k$'th bit conditioned that the first $k-1$ bits are $1$.
We first construct $T_4$, using five qubits. 
This can be done using $T_3'$s, as follows:
\begin{eqnarray}
&a, b, c,  d, e  &\longmapsto\\\nonumber
&a, b, c+ab, d, e &\longmapsto\\\nonumber
&a, b, c+ab, d, e+cd+abd &\longmapsto\\\nonumber
&a, b, c, d, e+cd+abd &\longmapsto\\\nonumber
&a, b, c, d, e+abd &
\end{eqnarray}
which is exactly $T_4$ from qubits $1,2,4$ to $5$.
%We first construct $T_4$, which can be done since all permutations on 
%the basis states can be achieved with $NOT$ gates and $T_3$,
%because $T_3$ is a transposition of the last two 
%basic states, and applying NOT gates we can construct 
%any transposition between two basis states of Hamming distance $1$, 
%which suffice for any permutation on basis states.  
Define  $X=P_4^3 T_4 P_4 T_4$, where $P_4$ applies the phase 
flip $P$ on the fourth qubit. $X$ takes 
$ |1110\ra \longmapsto i|1110\ra$, 
$ |1111\ra \longmapsto -i|1111\ra$ and does nothing to the other basis states. 
$X^2 $ is 
the three qubit gate   which gives $|111\ra\longmapsto -|111\ra$ 
and identity on the rest of the basis vectors, tensored with identity on 
the fourth qubit.
Now, to construct $CP$ we apply  $ P_3^3 T_3 P_3 T_3$ which  gives 
$|110\ra \longmapsto i|110\ra, |111\ra\longmapsto -i|111\ra$ and identity 
on the rest,  and applying $X^2$ we get $CP$ tensor
with identity on the third qubit.

The theorem now follows from Kitaev\cite{kitaev0}, who used the 
geometrical lemmas \ref{one} and \ref{corkitaev} to show that 
$CP$ and the Hadamard gate are universal. 
 Here is Kitaev's argument. Denote by 
\begin{eqnarray}
X_1= H_1 (CP)_{1,2} H_1\\\nonumber
 X_2=  H_2 (CP)^{-1}_{2,1} H_2
\end{eqnarray}
Define  $Y_1= X_1X_2^{-1}$ and 
$Y_2= X_2X_1^{-1}$. Note that 
$Y_1,Y_2$ both operate as the identity 
on the two states $|00\ra$, $|\eta\ra=|01\ra+|10\ra+|11\ra$.
Denote by $L$ the subspace orthogonal to  $|00\ra$ and $|\eta\ra$.
Then $Y_1,Y_2\in SU(L)$. 
$Y_1,Y_2$ do not commute, and 
their eigenvalues  are $\frac{1}{4}(1\pm \sqrt{15})$. 
Hence, by lemma \ref{corkitaev} they generate a dense subgroup in $SU(L)$. 
Now, add to $Y_1, Y_2$ also the gate $CP$ itself. 
This gate fixes $|00\ra$, but does not stabilize the space  $|\eta\ra$. 
Thus, in the space $L\oplus |\eta\ra$, $Y_1, Y_2$ generate a dense subgroup 
in the subgroup which fixes $|\eta\ra$, while $CP$ is not in this 
subgroup.
We can use lemma \ref{bridge} to show that 
$Y_1, Y_2, CP$ generate a dense subgroup in $SU(L\oplus |\eta\ra)$. 
Finally, add $H_1$ to the set $Y_1, Y_2, CP$. 
We have seen that $Y_1, Y_2, CP$ generate a dense set in the
subgroup  
that fixes $|00\ra$, while 
$H_1$ is not in this subgroup.  
Hence,   $H_1, Y_1, Y_2, CP$, (which are all gates generated by 
$H$ and $CP$), generate a dense subgroup of 
$SU(L\oplus |\eta\ra\oplus |00\ra)=SU(4)$.
Together with all phase factors, we get the unitary group on two 
qubits. The result follows from theorem \ref{twobit}. 
$\Box$

\subsection{Universality of the Set of Gates ${\cal G}_2$ for Polynomial Codes}

We would now like to show that the set of gates
for polynomial codes, $\calG_2$,  is universal.  
Note that in this case we are working with qupits, i.e over the field 
$F_p$, so universality is proved for matrices operating on qupits. 
We will show that  ${\cal G}_2$ together with all phase factors 
 generate a dense subgroup in $U(p^3)$ (theorem \ref{un2}).
We will first prove an analogue of the fact that one qubit gates 
and classical gates are universal for qubits\cite{}.

\begin{lemm}\label{cnotuniv}
The set of gates consisting of all one-qupit gates $U(C^p)$  and 
all classical two-qupit gates  generates all 
unitary matrices on two qupits, $U(C^{p^2})$. 
\end{lemm}

{\bf Proof:}
Let $G$ be the closed subgroup in $U(C^{p^2})$
generated by   all one-qupit gates $U(C^p)$ on each of the $2$ qupits, 
 and all classical two-qupit gates. 
We will use the fact that the lemma we are trying to prove
 is already known 
for the case of $p=2$. 
Let $S$ be the two dimensional subspace in $C^p$  
spanned by the first two basis vectors $|0\ra$ and $|1\ra$. 
Clearly, $U(S)$ is in $U(C^p)$. 
Let $S_l$ be the two dimensional subspace 
in the Hilbert space of the $l'$th qupit, and let 
$A=S_1\otimes S_2$. 
This subspace is isomorphic to the Hilbert space of $2$ qubits, and by 
 lemma \ref{cnotuniv} for the case of qubits, we have 
that $U(A)$ can be generated. 
In the same way, we can define all possible subspaces of the form $A$:
For each qupit, pick two basis vectors out of the $p$ possible basis 
vectors. Let $A$ be the tensor product 
of the subspaces spanned by these pairs of vectors.
 There are 
$m=\big(\begin{array}{c}p\\2\end{array}\big)^2$ subspaces of this form.  
By the same argument as before,  $U(A)$ is in $G$ for any such $A$. 

We now want to apply lemma \ref{bridge}. 
First, we claim that the subspaces of the above form can be 
ordered, $A_1,...A_m$ such that 
\beq 
\oplus_{i=1}^{j-1} A_i \not\perp A_j
\enq
as is easy to check. 
Second, 
\beq 
\oplus_{i=1}^{m} A_i =C^{p^k}
\enq
Since  $U(A_i)\subset G$, the proof follows from lemma \ref{bridge}. 
$\Box$

It is thus enough to prove that all one qupit gates are in our repertoire. 
Denote by $Q_0$ the one qupit matrix of the form:
\beq
Q_0 =\left( \begin{array}{cccc}
             w &   &  &   \\
               & 1 &  &   \\
               &   & \ddots &  \\
               &   &        & 1 \end{array}\right)
\enq
                
For $0\le i < p$, we can similarly 
denote by $Q_i$  the one qupit diagonal matrix 
which multiplies $|i\ra$ by $w$ and applies identity of the other basis 
states.

\begin{lemm}\label{Q}
 $Q_i$ and $Q_i^{-1}$ 
are in the subgroup generated by ${\cal G}_2$ on three qupits.
\end{lemm}

{\bf Proof:}
We will generate $Q_i\otimes I\otimes I$, 
which applies the following transformation:
\begin{eqnarray}
 |i\ra|a\ra|b\ra &\longmapsto &w |i\ra|a\ra|b\ra\\\nonumber
 |j\ra|a\ra|b\ra &\longmapsto & ~~ |j\ra|a\ra|b\ra~ ~~~j\not=i
\end{eqnarray}
To achieve this transformation,  we view this gate
as applying multiplication by 
$w$ of the second qupit, conditioned that the first qupit is $i$. 
 Recall that in our notation,  $P$ was a one qupit gate 
which applies  a phase shift,
$P|a\ra=w^a|a\ra$, and  $B$  is a one qupit gate 
which applies  a pit shift, $B|a\ra=|a+1\ra$.
Both $B$ and $P$ are in our repertoire. 
We now claim that the controlled $B$, which we denote 
by $CB$,  is also in our repertoire. 
$CB$ is the gate which applies $B$  on the second qupit, conditioned
that the first qupit is $i$, and applies the identity on the second qupit
if the state of the first qupit is anything but $i$.
We can generate $CB\otimes I$, since 
the generalized Toffoli gate, together with the addition gate, 
generate all permutations on basis states of three qupits. 
 Now consider the commutator
\beq
P^{-1}\cdot CB^{-1}\cdot P\cdot CB. 
\enq
This is exactly the gate we want, since 
if the first qupit is $i$, the matrix which is applied on the second 
qupit is $P^{-1}\cdot B^{-1}\cdot P\cdot B=wI$.
If the first qupit is in a basic state which is 
not $|i\ra$, the matrix which is applied on the second qupit  
is simply the identity. $\Box$

We now consider the two commutator matrices
in  $<{\cal G}_2>$:
\beq
X_i=HQ_iH^{-1}Q_i^{-1}~~,~~Y_i=HQ_i^{-1}H^{-1}Q_i
\enq
where $H$ is the generalized Fourier transform. 
We also define the two dimentional subspace $S_i$: 
\beq
S_i=span\{|i\ra, \sum_{b\in F_p, b\not = i} w^{ib}|b\ra\}.
\enq
The claim is that $X_i$ and $Y_i$ operate as the identity on the 
orthogonal subspace to $S_i$. 

\begin{lemm}\label{le2}
$X_i$ and $Y_i$
operate as the identity on $S_i^\perp$. 
\end{lemm}

{\bf Proof:}
It is easy to write down explicitly the matrix elements of
$X_i=HQ_iH^{-1}Q_i^{-1}$ and $Y_i=HQ_i^{-1}H^{-1}Q_i$. 
%As an example,  for $i=0$:
%\beq
%X_i =\left( \begin{array}{ccccc}
%             1        & w-1 & w-1   &\cdots &w-1   \\
%             1-w^{-1} & w   &       &       &      \\
%             1-w^{-1} &     & w     &       &      \\ 
%             \vdots   &     &       &\ddots &      \\
%             1-w^{-1} &     &       &       & w \end{array}\right)
%\enq
%\noindent where all the empty entries are equal to $w-1$.         
\begin{eqnarray}\label{del}
(X_i)_{ab} &=& \left\{\begin{array}{ll}
\delta_{ab}+\frac{1}{p}(w-1)w^{(a-b)i} & \mbox{if $b\not=i$} \\
\delta_{ab}+\frac{1}{p}(w-1)w^{(a-b)i}]w^{-1} & \mbox{if $b=i$}\end{array}\right.
\\\nonumber
(Y_i)_{ab} &=&\left\{\begin{array}{ll}
 \delta_{ab}+\frac{1}{p}(w^{-1}-1)w^{(a-b)i} & \mbox{if $b\not=i$} \\
  \delta_{ab}+\frac{1}{p}(w^{-1}-1)w^{(a-b)i}]w & \mbox{if $b=i$}\end{array}\right.\nonumber
\end{eqnarray}
\noindent To see that  $X_i$ and $Y_i$ operate as the identity on 
the subspace orthogonal to $S_i$, consider the matrices 
$X_i-I$ and $Y_i-I$, which satisfy equation \ref{del} where we substruct 
$\delta_{ab}$ from each term. 
   It is easy to see that 
the orthogonal vectors to $S_i$ are all in the kernel of 
$X_i-I$ and $Y_i-I$, since the vectors $v$ orthogonal to $S_i$  
satisfy:
\beq
v_i=0~~,~~~ \sum_{b,b\not=i} v_b w^{-ib}=0
\enq
\noindent 
and thus 
\begin{eqnarray}
\sum_{b\in F_p} (X_i-I)_{ab}v_b&=&
~~\frac{1}{p} (w-1)w^{ai}\sum_{b\not= i}v_b w^{-bi}=0\\\nonumber
\sum_{b\in F_p} (Y_i-I)_{ab}v_b&=&
\frac{1}{p} (w^{-1}-1)w^{ai}\sum_{b\not= i}v_b w^{-bi}=0.\nonumber~~~~ \Box
\end{eqnarray}

 We now consider the operation of $X_i$ and $Y_i$ on the subspace 
$S_i$. We claim that they generate a
 dense subgroup in the group of $2\times 2$ unitary matrices 
$U(2)$ operating on  $S_i$.
We will want to use lemma 
\ref{corkitaev}, and the main effort is to prove that 
the eigenvalues of $X_i$ and $Y_i$
are not integer roots of unity.
The proof of this fact is based on some basic results regarding 
cyclotomic fields and Galois fields, which can 
 be found in ``Introduction to Cyclotomic Fields'' 
by Washington\cite{wash}.

\begin{lemm}\label{irrational}
For $p>3$, 
the eigenvalues of $X_i$ and $Y_i$ confined to $S_i$ are not integer 
roots of unity. 
\end{lemm}

\noindent We now consider the operation of $X_i$ and $Y_i$ on the subspace 
$S_i$, which we span by the orthonornal basis  vectors $|i\ra$ and  
$|\alpha_i\ra=\frac{1}{\sqrt{p-1}}\sum_{b\in F_p, b\not = i} w^{ib}|b\ra$. 
By equation \ref{del} and a little algebra we get:
\begin{eqnarray}
X_i|i\ra= (1+\frac{1}{p}(w-1))w^{-1}|i\ra + \frac{\sqrt{p-1}}{p}(w-1)w^{-i^2-1}
|\alpha_i\ra\\\nonumber
X_i|\alpha_i\ra= (\frac{\sqrt{p-1}}{p}(w-1)w^{i^2}) |i\ra+
(1+\frac{p-1}{p}(w-1))|\alpha_i\ra \nonumber
\end{eqnarray}
\noindent and for $Y_i$ we get the transformation:
\begin{eqnarray}
Y_i|i\ra= (1+\frac{1}{p}(w^{-1}-1))w|i\ra + \frac{\sqrt{p-1}}{p}(w^{-1}+1)w^{-i^2-1}
|\alpha_i\ra\\\nonumber
Y_i|\alpha_i\ra= (\frac{\sqrt{p-1}}{p}(w^{-1}-1)w^{i^2}) |i\ra+
(1+\frac{p-1}{p}(w^{-1}-1))|\alpha_i\ra \nonumber
\end{eqnarray}

\noindent We denote by $X'_i, Y'_i$ the two matrices confined to $S_i$. 
Observe now that the determinant of $X'_i$ is $1$,  because
$|Q_i|=w, |H|=1$ and 
$|X_i|=|H|\cdot |Q_i|\cdot |H^{-1}|\cdot |Q_i^{-1}|=1$, 
and $|X'_i|=|X_i|$ since $X_i$ operates 
as the identity on the subspace orthogonal   to $S_i$.
Hence $X'_i$ and similarly $Y'_i$ have determinant $1$. 
The typical polynomial for $X'_i$ is $\lambda^2-Tr(X'_i)\lambda+Det(X'_i)$, 
which amounts to:
\beq
f(\lambda)=\lambda^2-\frac{2+(p-1)(w+w^{-1})}{p}
\lambda+1
\enq

\noindent $Y'_i$ has exactly the  same typical polynomial. 
We want to show that the roots of this polynomial are not integer
 roots of unity. 
Let us assume that one of the roots of the above polynomial 
 is a primitive $n'$th roots of unity,  denoted  by 
$\zeta_n$.  
The other solution is the complex conjugate of $\zeta_n$, and we have
\beq\label{wada}
\frac{2+(p-1)(w+w^{-1})}{p}=\zeta_n+\zeta_n^{-1}
\enq
We will first prove that 
 $n=p$. 
Denote by $Q(w), Q(\zeta_n)$ 
the Galois extentions of the field of rationals  
obtained by adjoining $w$,  $\zeta_n$,
respectively,  to the field 
of  rationals $Q$. 
Also, denote by  $Q(w)^+$, 
 the maximal real subfield of $Q(w)$, obtained by 
extending $Q$ by $w+w^{-1}$, and similarly denote the 
maximal real subfield of $Q(\zeta_n)$ by  $Q(\zeta_n)^+$.
The idea is that by equation \ref{wada}, $Q(\zeta_n)^+= Q(w)^+$. 

The degree of the extension $deg(Q(w)/Q(w)^+)$ is exactly $2$, 
 since $w$ is a root of the minimal two degree
 polynomial $x^2-(w+w^{-1})x+1$ over the field $Q(w)^+$.
Similarly, $deg(Q(\zeta_n)/Q(\zeta_n)^+)=2.$
On the other hand,   $deg(Q(w)/Q)=p-1$ and $deg(Q(\zeta_n)/Q)=\phi(n)$, 
by theorem $2.5$ in \cite{wash}. 
Now, for three fields, $F_1,F_2,F_3$ 
such that $F_3$ extends $F_2$ which extends $F_1$,  we have 
$deg(F_3/F_1)=deg(F_3/F_2)deg(F_2/F_1)$. It follows that:
\beq
deg(Q(w)^+/Q)= \frac{p-1}{2}~~~, ~~~ deg(Q(\zeta_n)^+/Q)=\frac{\phi(n)}{2}
\enq
But $Q(w)^+=Q(\zeta_n)^+$, which implies that
the degrees of extensions are equal, so $\phi(n)=p-1$. 

Now if $p>3$, then  $w+w^{-1}\not\in Q$, since 
$deg(Q(w+w^{-1})/Q)=(p-1)/2>1$.  Since $w+w^{-1}\in 
  Q(w)\cap Q(\zeta_n)$ we have that $Q\not=Q(w)\cap Q(\zeta_n)$. 
 If $p$ and $n$ 
were relatively prime we would have $Q=Q(w)\cap Q(\zeta_n)$ (by 
 proposition $2.4$ in \cite{wash}) and so $p$ must devide $n$, say 
 $n=p^rm$, with $m$ coprime to $p$. 
This implies:

\beq
\phi(n)=p^{r-1}(p-1)\phi(m)=p-1. 
\enq
This can only be satisfied if $n=p$. 
We get:

\beq
\frac{1+(p-1) cos(2 \pi/p)}{p}=  cos (k* (2 \pi/p)).
\enq

\noindent
This equation 
 is not satisfied by any integer  $k$, for $p>3$, 
since the left hand side is a convex 
combination of $cos(2 \pi/p)$ and $1$, 
 and no real part of a $p'$th root of unity lies between 
these two points. 
This shows that the eigenvalues of $X'_i$ and $Y'_i$ are not integer roots of unity.  $\Box$

We can now prove the theorem. 

\begin{theo}\label{un2}
 ${\cal G}_2$ together with all phase factors 
 generate a dense subgroup in $U(p^3)$, for $p>3$. 
\end{theo}

\noindent
{\bf Proof:} 
Since we are allowed to operate on three qupits, we can generate $Q_i$
by 
lemma \ref{Q}. 
We can thus generate $X_i$ and $Y_i$. 
 It is easy to see that  if 
$w\not = \pm 1$, or $p\not = 2$
the off diagonal terms of  $X'_iY'_i-Y'_iX'_i$ are not zero,  
so $X'_i$ and $Y'_i$ do not commute. 
 We can apply lemma  
\ref{corkitaev}, using the fact that the eigenvalues are not roots of unity, 
 by lemma \ref{irrational}. 
We thus have a dense subset of the unitary group on 
all subspaces $S_i$. 
We have that $\oplus_i S_i=C^p$, and $S_i$ is not orthogonal to 
  $\oplus_{j=0}^{i-1} S_j$. Using 
 lemma \ref{bridge} and induction on $i$, we
 show that we can generate  a dense subgroup of
 $U(C^p)$, i.e. all operations on one qupit. 
Note that apart from all one-qupit gates, 
we also have in our repertoire all classical gates on 
three qupits, and in particular all classical gates 
on two qupits which act trivially on the third qupit. 
The theorem follows from lemma \ref{cnotuniv}, which gurantees 
universality on two qupits, and theorem \ref{twobit} which 
shows that these matrices can be used to construct 
all matrices on three qupits, $U(p^3)$.  $\Box$

\section{Fault Tolerance for Probabilistic Noise}\label{sec7}
In this section we show  
 how to use the fault tolerant procedures described in the previous sections, 
in order to achieve robustness against probabilistic noise with
constant error rate.
We use qubits all along, but everything works for qupits in
 exactly the same way.  
The scheme is not specific for the 
quantum codes defined earlier  but works with any quantum code,
as long as it is a {\it quantum computation code} ( 
to be defined shortly). 
We then define the recursive simulation of the unreliable circuit, 
 using such codes. To analyze the propagation of errors, 
we define the notion of sparse errors and sparse fault paths.  
The threshold theorem is proved in two parts. 
First, we show that sparse fault paths are good, meaning that 
they cause sparse errors. Then we show that non-sparse 
fault paths are rare.  
We give the exact threshold condition on the error rate, 
which depends on the size of the fault tolerant procedures.

\subsection{Quantum Computation Codes}
In order to improve reliability of a quantum circuit, we 
need a quantum code accompanied with a universal 
 set of gates which 
can be applied fault tolerantly on states encoded by $C$. 
The code should also be accompanied 
with fault tolerant decoding, encoding and error correction procedures. 
In the last three sections we have shown that 
$CSS$ codes accompanied with the set of gates 
${\cal G}_1$, and polynomial codes accompanied with the set of gates 
${\cal G}_2$, are quantum computation  codes. 
However, any quantum computation code 
can be used in order to improve the reliability of a quantum circuit, 
provided that its procedures satisfy two simple restrictions. 
\begin{deff}
A quantum  code is called a quantum computation code 
if it  is accompanied  with a universal set of gates $G$, with fault 
tolerant procedures, and with fault tolerant encoding, decoding and correction 
procedures. Moreover, we require that  
 (1) all procedures use only gates from $G$, and 
(2) The correction procedure
 takes any density matrix to some word in the code.
\end{deff}

The first restriction allows us to use this code recursively. 
The second restriction is required for a reason which 
will become clear in the proof of lemma \ref{Theproof}.
Note  that the procedures for CSS codes and polynomial codes 
satisfy the above restrictions.  
We define the spread of the computation code.

\begin{deff}
Let $C$ be a computation code using the set of gates  $\calG$. 
Consider a fault tolerant procedure for a gate in  $\calG$
 preceded by fault tolerant error corrections 
on each block participating in the procedure.
The spread of the code is $l$ if 
one fault which occurs during this sequence of gates,
for any gate in  $\calG$,  effects at most $l$ qubits 
 in each block at the end of the procedure. 
\end{deff}

We require that the number of errors that the code can correct, $d$, 
is larger than the spread $l$, so that we can tolerate 
at least one error in a procedure preceded by error corrections. 
\beq
l\le d.
\enq

\subsection{Recursive Simulations}
From now on, fix a  quantum computation code $C$.  
It encodes one qubit on $m$ qubits, 
it corrects $d$ errors, it is accompanied with 
 a universal set of gates $\calG$, and a set of fault tolerant procedures
with spread $l$. 
We will use $m$ which is constant and does not grow with $n$. 
Let $M_0$ be a quantum circuit using gates from $\calG$.
 Then we simulate $M_0$ by a more reliable circuit
$M_1$, as follows. 
Each qubit is replaced by a block of qubits. Each time step 
in $M_0$, transforms in $M_1$ to a working period, which 
consists of two stages. In the first stage,  an error correction procedure is applied on 
each block. At the second stage,
 each gate which operated in the simulated time step in $M_0$ is 
replaced in $M_1$ by its procedure, operating on the corresponding blocks.

The input of $M_1$ is  the input to $M_0$, where each input bit is duplicated $m$ times. 
 Before any computation is done on this input, 
we apply in $M_1$  a fault tolerant encoding procedure on each block, which takes
$|0^m\ra$ to $|o^m\ra|S_0\ra$ and similarly for $1$. 
At the end of the computation  we will again use redundancy, 
for each block, we apply a decoding procedure which decodes the state to 
$m$ copies of the logical bit, and  the output 
of $M_1$  is defined as the majority of the bits in each block.
  Note that redundancy in the input and output  
is unavoidable if we want robustness to noise, because 
otherwise the probability for the input and output to be correct 
is  exponentially small. 
This is also assumed in the classical scenario\cite{neumann}.

The above mapping, denoted by $M_1=\phi(M_0)$, 
is one level of the simulation. 
$\phi$ is then applied  again, on $M_1$,  
 to give $M_2$, and we repeat this $r$ levels to get 
 $M_r=\phi^r(M_0)$,  an $r$-simulating circuit of $M_0$.
The number of levels $r$  will be 
 $O(\rm{polyloglog}(V(M_0)))$, where $V(M_0)$ 
 is the volume of $M_0$, i.e. the number of locations in $M_0$. 
The output of $M_r$ is defined  by taking 
 recursive majority on the  outputs. 
This means that first we 
take the majority in each block of size $m$, then we take the majority, 
of $m$ such majority bits, and so on for $r$ levels, to give one output bit.  

The advantage of using recursive simulations, instead of 
one step of simulation, as in Shor's scheme\cite{shor2}, 
 is that  in each  level error 
corrections to the current level are added. This means that 
 error corrections of all scales 
are applied frequently during the computation  
procedures, preventing accumulation of errors in all levels.  
This allows robustness against constant error rate, which seems 
impossible to achieve in one level of simulation.

\subsection{Blocks and Rectangles}  
The recursive simulations induce a definition of $s$-blocks:
Every qubit  transforms to a block of $m$
 qubits in 
the next level, and this block transforms to $m$ blocks of $m$ qubits and
so on.
One qubit in $M_{r-s}$ transforms to  $m^s$ qubits in $M_r$.
 This set of qubits in $M_r$ is called an $s$-block.
%This induces a division of the 
% qubits in Mr to $s$-blocks, and this division is 
%a refinement of the division to s+1 blocks.
An $0$-block in $M_r$  is simply a qubit. 
In the same way, one can define $s-$working periods. 
Each time step in $M_0$ transforms to $w$ time steps in $M_1$, 
and an $s-$working period is the time interval in $M_{r}$ which 
corresponds to one time step in $M_{r-s}$. 

Recall the definition  of a location in a quantum circuit in subsection \ref{loc}. 
 The recursive simulation induces a
 partition of the set of locations in $M_r$ to generalized rectangles. 
An $r-$rectangle in $M_r$ is 
the set of locations which originated 
 from  one location in $M_0$. This is best explained by an example:   
Consider a CNOT gate which is applied in $M_0$ at time $t$ on qubits 
$q_1,q_2$.  The location $((q_1,q_2),t)$ in $M_0$ 
 transforms in $M_1$ to  error correction procedures on 
 both blocks, followed 
by the procedure of the CNOT gate.  
The set of locations in these three procedures is the $1$-rectangle in $M_1$ which 
originated from the location $((q_1,q_2),t)$ in $M_0$. 
More generally, an $s-$rectangle 
in $M_r$ is the set of points in $M_r$ which originated from 
one location in  $M_{r-s}$. 
Note that the partition to $s$-rectangles is a refinement 
 of the partition to $(s+1)$-rectangles. 
An $0$-rectangle in $M_r$ is just one location.

\subsection{Sparse Errors and Sparse Faults}

In a noiseless scenario, 
 the state of $M_r$ at the end of each $r$-working period 
encodes the state of $M_0$ at the end of the corresponding time step. 
However, we assume that errors occur in $M_r$ and we want to
 analyze those. In order to analyze the propagation of errors
in $M_r$,  we need to distinguish between the actual faults 
that occur during the computation, 
and the errors that are caused in the state. 
First, we focus on the errors 
in the states, and define  a distance between encoded states.  
The hierarchy of blocks requires
a recursive definition. 
\begin{deff}
Let $B$ be the set of qubits in $n$ $r-$blocks.
An $(r,k)$-sparse set of qubits $A$ in $B$  is 
a set of qubits in which for every $r-$block in $B$, there are at most 
$k$ $(r-1)-$blocks such that the set $A$ in these blocks
 is not $(r-1,k)$ sparse. 
An $(0,k)-$sparse set of qubits $A$ is an empty set of qubits.
\end{deff}

Two density matrices $\rho_1,\rho_2$ of the set of qubits $B$
 are said to be
$(r,k)$-deviated if 
%$k$ is the minimum integer such that 
there exists an $(r,k)$-sparse set of qubits $A\subseteq B$,
with $\rho_1|_{B-A}=\rho_2|_{B-A}.$
The deviation satisfies the triangle inequality since the union of
 two sets which are $(r,l_1),(r,l_2)$-sparse respectively
is   $(r,l_1+l_2)$ sparse, by  induction on $r$.

%Then, $A\cap B$ is  $(r,l_1+l_2)$ sparse.
%\begin{lemm}\label{sparsemetric}:
%Let $C$ be the set of qubits in $n$ $r-$blocks.
%Let $A,B\in C$ be two  sets of qubits which are $(r,l_1),(r,l_2)$ sparse
%respectively.
%Then, $A\cap B$ is  $(r,l_1+l_2)$ sparse.
%\end{lemm}
%{\bf Proof:}
%Use induction on $r$.
%For $r=0$, the union of two empty sets is empty.
%Let us assume for $r$,
%and prove for $r+1$.
%$A$ is $(r+1,l_1)$ sparse, so there are at most $l_1$ $r-$blocks
%which are not  $(r,l_1)$ sparse.
%$B$ is $(r+1,l_2)$ sparse, so there are at most $l_2$ $r-$blocks
%which are not  $(r,l_2)$ sparse.
%In all the other $r-$blocks, the union of $A$ and $B$ is $(r,l_1+l_2)$ sparse
%by induction.
%So there are at most $l_1+l_2$ $r-$blocks in which the union is not 
%$(r,l_1+l_2)$ sparse,
%so the union is $(r+1,l_1+l_2)$ sparse.\bbox
%The idea behind the recursive simulation is to keep the density matrix
%only a little deviated, in this metric.
%During the computation, the distance in this metric will grow a little bit,
%but the error corrections will correct it.
%This ``game'', in which the deviation in this metric grows during the 
%computation and shrinks during the error corrections, happens
%{\it in all the levels} of computation simultaneously. 

A computation is successful if the error at the end of each $r-$working 
period is sparse enough.     
The question is which fault paths keep the errors sparse.  
 We will show in lemma \ref{Theproof} that 
this is guaranteed if the fault path is sparse:

\begin{deff}
A set of locations  in an $r-$rectangle 
is said to be $(r,k)$-sparse if  there are no more than $k$ 
$(r-1)-$rectangles, in which the set is not $(r-1,k)$-sparse.
An $(0,k)$-sparse set in an $0-$rectangles 
is an empty set. A fault path in $M_r$ is $(r,k)$-sparse
if in each
$r-$rectangle, the set is $(r,k)-$sparse. 
\end{deff}

\subsection{The Good Part: Sparse Fault Paths Keep the Error Sparse }

%The deviation is a metric  since it satisfies the triangle inequality,
%and it is zero only for equal density matrices. 
%\begin{deff}
%A circuit $M1$ block simulates a circuit $M0$ by the quantum code $\phi$
%with parameters $m$ and $w$ if for any trajectory $E$ of $M0$,
%the trajectory $E1$ of $M1$ which satisfies that the initial state
%of the input blocks is the  input of $M1$ duplicated $m$ times,
%satisfies that $E1(wt)=\phi(E(t))$.
%\end{deff} 
%
%^We will ask what 
% (linear) quantum code from $k$ qubits to $m$ qubits 
% which corrects $d$ errors is a subspace of
%dimension $2^k$ of  
%the Hilbert space of $m$ qubits.  
%$\phi: \cal{C}$$^{2} \longmapsto \cal{C}$$^{2^{m}}.$
%$m$ is called the size of the block in the code.
%Such a code  induces a linear function from $\cal{C}$$^{2^{n}}$
%to $\cal{C}$$^{2^{mn}}$ in the following way:
%a pure state in $\cal{C}$$^{2^{n}}$, $|\alpha\rangle=\sum_{i} c_i |i\rangle$
%will transform to 
%$|\beta\rangle=\sum_{i} c_i 
%\phi|i_{1}\rangle\phi|i_{2}\rangle...\phi|i_{n}\rangle$.
%A pure state in the image of $\phi$ is called a word in the code.
%The above definition can be extended to density matrices:
%A mixed state of $n$ qubits will be 
%encoded by the corresponding probability 
%over the encoding of the pure states.
%A mixture of words in the code is also said to be in the code.

We claim  that if the fault path is sparse enough,
then the error corrections keep the deviation small.
 The number of faults allowed in one rectangle 
is bounded so that when taking into account the spread of the fault, 
the number of qubits effected in each block 
at the end of one working period 
is not too big, so that the density matrix can  still be recovered.

\begin{lemm}\label{Theproof}
Let $C$ be a computation code that corrects $d$ errors,
with spread $l$.
 Let $M_r$ be the $r-$simulation of $M_0$ by $C$.      
Consider a computation subjected to an $(r,k)$-sparse fault path
with $kl\le d$.
At the end of each $r-$working period
the error  is $(r,d)$-sparse.
\end{lemm}

{\bf Proof:}
It is instructive to first prove this lemma for $r=1$. 
This is done by induction on the time $t$. 
For $t=0$ the deviation is zero.
Suppose that the density matrix at the end of the $t'$th working period 
 is $d$-deviated from the correct 
matrix.
If no errors occur during the $t'$th working period, the error 
corrections would have corrected the state to $\phi(\rho(t))$, and the procedures would 
have taken it to the correct state $\phi(\rho(t+1)).$
However, due to the fact that $k$ errors did occur in each rectangle, 
we have in each  block at most $kl$ qubits which  are effected 
by this error, and the deviation is at most $kl\le d$.
This proves the theorem for $r=1$. 
For general $r$, 
we  prove
 two assertions together,  using induction on  $r$.
 The first assertion implies the desired 
result. 
\begin{enumerate}
\item\label{comp}
 Consider $n$ $r-$blocks, in a density matrix $\rho_r$ which is 
$(r,d)-$deviated from $\phi^r(\rho_0)$,
where $ \rho_0$ is a density matrix of $n$ qubits.
At the end of an $r-$working period which  $r-$simulates the operation $g_0$ on  $\rho_0$,
with an $(r,k)$ sparse set of faults,
the density matrix is  $(r,d)$ deviated from $\phi^r(g_0\circ\rho_0)$.
\item\label{totalcor}
 Consider $n$ $r-$blocks, in any density matrix, $\rho_r$.
At the end of an $r$-working period which   $r-$simulates the operation $g_0$ on  $\rho_0$,
with an $(r,k)$ sparse set of faults,
the density matrix is $(r,d)$ deviated from some word $\phi^r(g_0\circ\rho_0)$,
where $ \rho_0$ is a density matrix of $n$ qubits.
\end{enumerate}

For $r=1$ the proof of the first assertion is 
as before, while the second assertion is true because 
of a similar argument, using the extra requirement that  the error correction 
takes any word to a quantum code. 
Let us now assume both  claims for $r$, and prove each of the claims 
for $r+1$.

\ref{comp}.
We consider an $(r+1)-$working period operating on $\rho_{r+1}$, and  
 $(r+1)-$simulating the operation $g_0$ on  $\rho_0$.
Let us assume for a second two wrong assumptions. 
First, that all the $r-$rectangles in the $(r+1)-$stage have $(r,k)$ sparse set of
faults. Second, that  $\rho_{r+1}$  is $(r,d)$ 
deviated from 
$\phi^{r}(\rho_1)$, where $\rho_1$ is a density matrix 
which is $(1,d)-$deviated from $\phi(\rho_0)$.
We can now use the induction assumptions. 

The $(r+1)-$simulation working period consists of two stages,
one is the stage which $r-$simulates the error correction procedures 
 in $M_1$, and the other part $r-$simulate one working period 
in  $M_1$ which $1-$simulates the operation of $g_0$ in $M_0$.  
The first part $r-$simulates a computation in $M_1$,  which, if no errors occur, 
is supposed to take  $\rho_1$ to $\phi(\rho_0)$. 
We can use the induction assumption on the first assertion,  
to prove that at the end  each one of the $r$-working periods 
in this $(r+1)$-working period, the matrix is $(r,d)$ deviated from 
what it is supposed to be. 
After $w$ applications of this induction assumption,
 (where $w$ is the number of time steps 
in the error correction procedure) we get that   the matrix at 
the end of the error correction stage in the $r-$working period, is $(r,d)$ deviated from
  $\phi^{r}(\phi(\rho_0))=\phi^{r+1}(\rho_0)$.
This is true under the two wrong assumptions. 

Now, we release the second assumption. We actually start the computation 
with a matrix which is $(r+1,d)-$deviated from  $\phi^{r+1}(\rho_0)$.
So most of the $r-$blocks are 
$(r,d)-$deviated from $\phi^{r+1}(\rho_0)$,
except maybe $d$ $r-$blocks in each $(r+1)-$block which are problematic.
By the induction stage on claim \ref{totalcor},
after the first $r-$working period which $r$-simulates the operation of 
$g_1$ in $M_1$,  
the $d$  problematic blocks in each $(r+1)-$block are 
$(r,d)-$deviated from $\phi^{r}(g_1\circ\rho'_1)$.
So after the first $r-$working period  the density matrix is 
$(r,d)-$deviated from $\phi^{r}(g_1\circ\rho_1)$, where $\rho_1$
is $(1,d)-$deviated from $\phi(\rho_0)$. 
Thus after the first $r-$working period we are in the same situation as 
if the second assumption holds, and we can proceed with the above argument. 

We now consider the computation part of the $(r+1)-$working period. 
Still under the first wrong assumption, using the induction assumption on the 
first claim,  we have that at the end of the $(r+1)-$working period, 
the state is $(r,d)-$deviated from $\phi^{r+1}(g_0\circ\rho_0)$.
Thus under the first wrong assumption, 
the final density matrix is $(r,d)-$deviated from $\phi^{r+1}(g\circ\rho_0)$.

We now relax the first wrong assumption, and 
take into account the fact that there where $k$
 $r-$rectangles in each $(r+1)$-rectangle where the 
faults where not $(r,k)-$sparse.
By the fact that the $(r+1)-$working period $r-$simulates 
a sequence of gates with spread $l$,
these errors can effect only $kl\le d$   $r-$blocks in each $(r+1)-$block,
 at the end of the $(r+1)-$stage,
so we have that the final density matrix at the end of the $(r+1)-$stage is
$(r+1,d)-$deviated from the correct one, $\phi^{r+1}(\rho_0)$.

\ref{totalcor}.
We consider one stage of $(r+1)-$corrections on
the $n$ $(r+1)-$blocks in an arbitrary density matrix.
Again, let us assume that the faults in all the  $r-$rectangles
are $(r,k)-$sparse.
By the induction stage on claim \ref{totalcor},
after one $r-$working period,
the density matrix is 
$(r,d)-$deviated from some $\phi^{r}(\rho_1)$.
Let us consider the trajectory of $\phi^{r}(\rho_1)$
in the correction part of the working period. 
It is an $r-$simulation of an error correction procedure, which performs the computation which 
 takes 
the density matrix $\rho_1$  to some word $\phi(\rho_0)$. 
As before, we can prove by induction on the $w$  $r-$stages that 
at the end of the correction 
 part of the $(r+1)$-working period, which is an $r-$simulation 
of the above error correction, we end up with a matrix which 
is $(r,d)-$deviated from $\phi^{r}(\phi(\rho_0))$. 
Taking into account the $r-$rectangles with faults which are not 
$(r,k)-$sparse, we end up with a density matrix which is 
$(r+1,d)-$deviated from $\phi^{r+1}(\rho_0)$.
$\Box$

Sparse errors are indeed ``good'', because if the error is sparse
at the end of the computation, the majority of the bits will give the 
correct answer.

\begin{lemm}\label{maj}
Let $2d+1\le m$. 
If the final  density matrix of $M_r$ is $(r,d)-$deviated from 
the correct one, then the distribution 
on the strings which are obtained when 
taking   recursive majority on each $r-$block of its output is correct. 
\end{lemm}

{\bf Proof:}
Let $\rho_0$ be the correct final density matrix of $M_0$, 
Let  $\rho_r$ be the correct final density matrix of $M_r$,
and let $\rho'_r$ be the final density matrix which 
is $(r,d)-$deviated from $\rho_r$. 
First, we note that if we discard all the $r-$blocks of qubits which are 
not measured for output, the resulting two density matrices 
are still $(r,d)-$deviated. 
This remains true if we apply a measurement of all the qubits. 
Moreover, the output distribution remains the same after the measurements. 
We can thus assume that $\rho_r$ and $\rho'_r$ are density matrices of the 
output qubits, which are mixtures of basic states, and $\rho'_r$
is $(r,d)-$deviated from $\rho_r$. 
$\rho_r$ can thus be written as $\{p_i,|\alpha_i\ra\}$, where 
$|\alpha_i\ra=|i_1^{m^r}i_2^{m^r}...i_n^{m^r}\ra$.
All pits in an $r-$block are equal, since the matrix is correct.
The probability that the recursive majority string is $i$
is exactly $p_i$.    
$\rho'_r$ can be written as $\{q_j,|\beta_j\ra\}$, 
where $|\beta_j\ra$ are basis vectors. 
Let $A$ be  subset of qubits such that $\rho_r|_A=\rho'_r|_A$
and $A$ contains all the qubits on which $\rho_r$ operates on, 
except an $(r,d)-$sparse set.
Note, that $\rho'_r$ is supported only on $|\beta_j\ra\}$
in which all the pits in $A$ agree, and that
\beq  
\sum_{j,\beta_j|_A=\alpha_i|_A} q_j=p_i.
\enq
because the reduced density matrix on $A$ is the same as that of $\rho_r$.   
The probability that the recursive majority string 
is $l$ is exactly the sum of probabilities 
of $\beta_j$ which give the string $l$, but the only $\beta_j$
that appear in the mixture agree on $A$, and we already know that
the sum of the probabilities of those is correct. $\Box$

\subsection{The Bad Part: Non-Sparse Fault Paths are Rare Below the
 Threshold}\label{badbad}
Let us first find the  effective error rate in $M_1$. 
It is the probability for one rectangle to have more faults than the code 
can handle. 
If this effective error rate is smaller than the actual error rate $\eta$, 
this achieves an improvement in reliability from $M_0$ 
to $M_1$, and in fact from 
$M_s$ to $M_{s+1}$. 
Let the code $C$ correct $d$ errors, and have spread $l$.  
Let $A$ be the maximal  number of locations in a rectangle. 
We require that the probability for more than 
$\lfloor d/l\rfloor=k$ faults in $A$ locations is smaller than $\eta$. 
This is indeed the case if the following condition is satisfied. 
\begin{deff}
The threshold condition: 
\end{deff} 
\beq
\left(\begin{array}{c}A\\k+1\end{array}\right)\eta^{k+1}<\eta
\enq

We can thus define the threshold of a code:
\begin{deff}
Let $C$ be a quantum computation code, which corrects 
$d$ errors, has spread
 $l$, and where 
 $A$ is the maximal number of locations in a rectangle
for simulations using this code. Let 
  $k=\lfloor d/l \rfloor$. 
The threshold for probabilistic noise for the code $C$ 
is
\end{deff}
 \beq\label{thre}
\eta_c(C)=\left(\begin{array}{c}A\\k+1\end{array}\right)^{-k}.
\enq
It is easy to see that any $\eta<\eta_c$ satisfies the  threshold  condition.
The threshold can be computed given the parameters of the code. 
However, it is quite complicated to calculate the  parameter $A$ exactly. 
Moreover, in our constructions of the procedures we did not
 attempt to optimize 
their size. 
In our constructions we estimate the threshold 
 to be $\approx 10^{-6}$ in both cases of CSS codes and 
 polynomial codes, using codes of length $m=7$, which can correct 
one error. This estimation is done assuming that 
measurements can be applied during the computation, 
classical operations in the procedures are error free, 
and this allows us to save some of the operations in the fault 
tolerant procedures, which we do not describe here. 
Optimizations of the suggested procedures 
are certainly required and presumably can reduce
this threshold by several orders of magnitudes.

We now show that below the threshold bad fault paths are rare.

\begin{lemm}\label{bad}
If $\eta<\eta_c$, $\exists\delta>0$ such that 
the probability  $P(r)$,
 for the faults in an $r-$rectangle to be $(r,k)$ sparse
is larger than  \(1-\eta^{(1+\delta)^r}\). 
\end{lemm}

{\bf Proof:} 
Let $\delta$ be as follows. 
\beq
\left(\begin{array}{c}A\\k+1\end{array}\right)\eta^{k+1}<\eta^{1+\delta}.
\enq
Such $\delta$ exists for $\eta$ below the threshold. 
The proof will follow by  induction on $r$.
The probability for an $0-$rectangle, i.e. one location, 
 to have faults which are $(0,k)$ sparse, i.e. that in this location 
a fault did not occur, is $1-\eta$.
Assume for $r$, and let us prove for $r+1$.
For the faults in an $(r+1)$-rectangle not to be $(r+1,k)$ sparse,
there must be at least $k+1$ $r-$rectangles in which the fault is not 
$(r,k)$ sparse.
So \(P(r+1)\ge 1-\left(\begin{array}{c}A\\k+1\end{array}\right)
(1-P(r))^{k+1}\ge 1-\eta^{(1+\delta)^{r+1}}\),
using the induction assumption, and the fact that 
$\eta^{(1+\delta)^{r}}<\eta_c$, so it satisfies the threshold 
condition. 
$\Box$

We can now prove the threshold theorem. 

\begin{theo}\label{threshold} {\bf The Threshold Theorem for Probabilistic Noise:}

Let $\epsilon>0$.   Let $C$ be a computation code with gates $\calG$. 
There exists a threshold $\eta_c>0$, and constants $c_1,c_2,c_3$
 such that the following holds. 
Let $Q$ be a 
 quantum circuit, with $n$ input qubits (qupits), which operates $t$ time steps,  
uses $s$ gates from 
${\cal G}$, and has $v$ locations.
There exists a quantum circuit  $Q'$  which 
operates on $n\rm{log^{c_1}}(\frac{v}{\epsilon})$ qubits (qupits), 
for time $ t \rm{log^{c_2}}(\frac{v}{\epsilon})$, and 
uses $ s \rm{log^{c_3}}(\frac{v}{\epsilon})$ gates from ${\cal G}$
such that  in the presence of 
probabilistic noise with error rate  $\eta< \eta_c$, 
$Q'$ computes a function  which is $\epsilon$-close to that 
computed by $Q$. 
\end{theo}

{\bf Proof:}
$Q$ will be taken as $M_0$, and we 
generate $M_r$ according to the above scheme, where $r$ is chosen so that 
$v\eta^{(1-\delta)^r}<\epsilon$. 
By lemma \ref{bad}, we have that the probability for a fault path to be 
bad is smaller than $\epsilon.$ By lemmas \ref{Theproof} and \ref{maj}
the sparse fault paths give correct outputs. $\Box$

%{\bf Remark:} Theorem \ref{Thetheo} requires
%that the code can correct $d>1$ errors.
%A similar result holds for $d=1$, with the threshold 
%  \(\left(\begin{array}{c}b\\2\end{array}\right)\eta< 1\)
%where $b$ is the maximal  size of slightly different  rectangles,
%defined to  contain   a computation and a correction procedure together.
%The proof is almost the same. 
%In some cases this threshold is better. \bbox

\section{The Threshold Result for General Noise}\label{noi}\label{sec8}
We generalize the result to general local noise. 
The threshold condition is slightly different. 
Again, we fix a computation code and a set of gates $\calG$. 
The proof consists of showing that the reliable 
circuit constructed for the case of probabilistic noise is 
robust against general noise, but the  
 error rates which can be tolerated  are slightly worth. 
Again, the proof consists of dealing with bad fault paths 
and good fault paths separately. 
The proof that good fault paths are indeed good uses a reduction to 
the case of the probabilistic noise. 
The proof that the bad part is negligible is more involved than in the 
probabilistic case.

\subsection{Fault Paths in the Case of General Noise}
The notion of fault paths is less clear in the case of general noise. 
To define fault paths, write the final density matrix of the noisy circuit as follows:
\beq\label{fin}
\rho(t)={\cal E}(t)\cdot{\cal L}(t)\cdot {\cal E}(t-1)\cdot {\cal L}(t-1)\cdots{\cal E}(0)\cdot {\cal L}(0)\rho(0).
\enq
In the above equation, ${\cal E}(t)$ is the noise operator operating at time $t$, and ${\cal L}(t)$ is the computation 
operator at time $t$. According to our noise model, equation \ref{localcond1}, 
${\cal E}(t)$ can be written as a 
tensor product of operators, operating on the possible locations
of faults at time $t$, $A_{i,t}$. Each such operator can be written as a sum of two operators, using equation 
\ref{localcond2}:
\beq\label{app}
{\cal E}_{A_{i,t}}(t)=(1-\eta) I+{\cal E}'_{A_{i,t}}(t)~~,~~~ 
\|{\cal E}'_{A_{i,t}}(t)\|\le 2\eta.
\enq

We can replace all the error operators in equation \ref{fin} by
the products of operators of the form \ref{app}. 
We get:
  \begin{eqnarray}\label{fpaths}
\rho(t)&=& \left(\otimes_{A_{i,t}}((1-\eta) I+{\cal E}'_{A_{i,t}})\right)\cdot {\cal L}(t)\cdot
 \left(\otimes_{A_{i,t-1}}((1-\eta) I+{\cal E}'_{A_{i,t-1}})\right)\cdot{\cal L}(t-1)\cdots\\\nonumber
&&\cdots\left(\otimes_{A_{i,0}}((1-\eta) I+{\cal E}'_{A_{i,t-1}})\right) {\cal L}(0)\rho(0).
\end{eqnarray}

We can open up the brackets in the above expression.  We get a sum of 
terms, where in each term, for each set of qubits $A_{i,t}$ at time $t$ we either 
 operate $(1-\eta)I$ or ${\cal E}'_{A_{i,t}}(t)$.
Thus, each term in the sum corresponds to a certain fault path. 
More precisely, a fault path is a subset of the locations 
 $\{A_{i,t}\}_{i,t}$, and the term in the sum which corresponds to this fault path 
is exactly the term in which we apply ${\cal E}'_{A_{i,t}}(t)$ on all locations in the fault path, 
and apply $(1-\eta)I$ on the rest.
As was done in the probabilistic case, we can now  divide the above sum \ref{fpaths} to two parts:
the sum over the {\it good} fault paths, and the sum over the {\it bad} fault paths. 
We define the good fault paths
 to be those which are $(r,k)$-sparse, and the bad ones are all the rest. 
We write:
\beq
\rho(t)=  {\cal L}_g\cdot \rho(0)+{\cal L}_b \cdot \rho(0) 
\enq

We will treat each part separately.
 
\subsection{The Bad Part: Non-Sparse
 Fault Paths are Negligible Below the Threshold}
We show
 the trace norm of the {\it bad} part is negligible, when $\eta$ 
is below the threshold for general noise.

\begin{deff} The threshold condition for general noise,  $\eta'_c>0$ is defined such that
\end{deff}
\beq
e\left(\begin{array}{c}A\\k+1\end{array}\right)(2\eta)^{k+1}<
2\eta
\enq
Where as before, $d$ is the number of errors which the code corrects, 
 $l$ is the spread,  $k=\lfloor d/l \rfloor$ and 
 $A$ is the maximal number of locations in a rectangle. 
We can thus define the threshold of a code, for general noise:
\begin{deff}
Let $C$ be a quantum computation code, which corrects 
$d$ errors, has spread
 $l$, and where 
 $A$ is the maximal number of locations in a rectangle
for simulations using this code. Let 
  $k=\lfloor d/l \rfloor$. 
The threshold for general noise for the code $C$ 
is
\end{deff}
 \beq
\eta'_c(C)=\frac{1}{2} e^{-k}\left(\begin{array}{c}A\\k+1\end{array}\right)^{-k}.
\enq
It is easy to see that any $\eta<\eta_c'$ satisfies the  threshold  condition.
Note that there is a slight difference from 
 the threshold in the case of probabilistic noise. 
A factor of $2$ is added to $\eta$, 
and the factor of $e$ is added to the whole definition. 
These differences are due to slight technical difficulties which 
rise from the fact that the norm of the good operators in not smaller than 
$1$, but can also be slightly larger than $1$. 
$\eta'_c$ will be smaller than the threshold for probabilistic noise,  $\eta_c$.

\begin{lemm}\label{small}
Let $\eta< \eta'_c, \epsilon>0$. Let $M_0$  use $v$ locations. 
Let $r=\rm{polyloglog}(\frac{v}{\epsilon})$. Then in $M_r$ 
\[\|{\cal L}_b\cdot \rho(0)\|\le \epsilon~~,~~\|{\cal L}_g\cdot \rho(0)\|\le 1+\epsilon\]
\end{lemm}

\noindent {\bf Proof:}
We shall rewrite the sum over all fault paths, by collecting together all the operations 
according to which  $r-$procedures they were done in. 
We denote by  ${\cal L}_b(i)$, the sum over all operators on the $i$'th procedure, 
when applying errors on bad fault paths. 
${\cal L}_g(i)$ will thus be  the sum over all operators on the $i$'th procedure, 
when applying errors on good fault paths.
If there are $v$ procedures, we can write:
 \beq\label{procedu}
\rho(t)=({\cal L}_g(v)+{\cal L}_b(v))\cdot ({\cal L}_g(v-1)+{\cal L}_b(v-1))
\cdots ({\cal L}_g(1)+{\cal L}_b(1))\rho(0)
\enq
We first prove that 
\beq\label{tinyy}
\forall 1\le i\le v~~,~~ 
\|{\cal L}_b(i)\|\le  (2\eta)^{(1+\delta)^r}~~,~~~ \|{\cal L}_g(i)\|\le 1+(2\eta)^{(1+\delta)^r}
\enq
Where $\delta $ is defined by 
\beq
e\left(\begin{array}{c}A\\k+1\end{array}\right)(2\eta)^{k+1}<
(2\eta)^{1+\delta}.
\enq
Such $\delta$ exists because $\eta$ is below the threshold. 
The proof of inequality \ref{tinyy} for $\eta<\eta'_c$, 
  follows the lines of lemma \ref{bad}. 
We use induction on $r$, and denote by ${\cal L}^r_b(i)$
the sum over all bad fault paths in an $r-$procedure.  For $r=0$, a
  $0$-procedure is simply one location, and 
 ${\cal L}^0_b(i)$ is the sum over all {\it bad} fault paths 
in this procedure. For one location, 
there is only one term in this sum: the identity, or the gate
applied in this location, 
followed by one noise operator. 
By equation \ref{app}, and the properties of the norm 
on super-operators in section \ref{normproperties}  $\|{\cal L}^0_b(i)\|\le 2\eta$, and 
$\|{\cal L}^0_g(i)\|\le 1+2\eta$
Now assume for $r$ and  prove for $r+1$.
For the faults in an $(r+1)$-rectangle {\it not}
 to be $(r+1,k)$ sparse, there must be at least $k+1$ 
$r-$rectangles in which the fault is not 
$(r,k)$ sparse.
So by the induction assumption on both $\calL_b(i)$ and  $\calL_g(i)$
\beq
\|{\cal L}^{r+1}_b(i)\|\le \left(\begin{array}{c}A\\k+1\end{array}\right)
 ((2\eta)^{(1+\delta)^{r}})^{k+1}(1+ (2\eta)^{(1+\delta)^{r}})^{A-k-1}
\le e \left(\begin{array}{c}A\\k+1\end{array}\right)((2\eta)^{(1+\delta)^{r}})^{k+1}.
\enq
where we have used the fact $(1+ (2\eta)^{(1+\delta)^{r}})^{A-k-1}<e$
since $(2\eta)^{(1+\delta)^{r}}<2\eta$, and $2\eta A\le 1.$ 
The right hand side is $\le
 (2\eta)^{(1+\delta)^{r+1}}$ using the fact that $(2\eta)^{(1+\delta)^{r}}<2\eta$, and the threshold condition  
is satisfied for  $\eta\le \eta'_c$. This proves the induction step 
for ${\cal L}_b(i)$. Using 
$\|{\cal L}_g(i)+ {\cal L}_b(i)\|=1 $
 proves the induction step also for   $\calL_g(i)$.
To prove the statement, we consider bad fault paths, 
i.e. at least  one $r-$rectangle is bad.  
If there are $v$ rectangles, we have:
\beq
\|{\cal L}_b\|\le v\cdot ( 1+ (2\eta)^{(1+\delta)^r})^{v-1}(2\eta)^{(1+\delta)^r}
\enq
 taking  $r=\rm{polyloglog}(\frac{v}{\epsilon})$ we get the desired result.
$\Box$

\subsection{The Good Part: Sparse Fault Paths Give Almost Correct Outputs}

\begin{lemm}\label{gooood}
Let $\eta\le \eta'_c, \epsilon>0$. Let $M_0$  use $v$ locations. 
Let  $r=\rm{polyloglog}(\frac{v}{\epsilon})$, as in lemma \ref{small}. Let 
 $M_0$ output $f(i)$ for a given $i$  with probability of error $\epsilon'$.   
Then for $M_r$, 
\[\sum_{bad~j}[{\cal L}_g\cdot \rho(0)]_{j,j}\le (1+\epsilon)\epsilon'\] 
where bad $j'$s are those basis states with 
the majority of the result qubits not equal to $f(i)$.  
\end{lemm}

\noindent{\bf Proof:}
For any density matrix $\rho$  which is $(r,d)-$deviated from the
correct final density matrix of $M_r$
 we have \(\sum_{bad~i}\rho_{i,i}= \epsilon\). 
We will write ${\cal L}_g$ as a linear sum (not necessarily positive) of sparse 
physical fault paths, for which lemma \ref{Theproof}
can be applied. We will get the desired result from linearity. 
 We write 
\beq\label{bimi}
{\cal L}_g\cdot \rho(0)=\sum_E {\cal L}_E\cdot \rho(0) 
\enq
where $E$ runs over all  general sparse  fault paths, 
and  ${\cal L}_E$ is the operator corresponding to the computation 
done in the presence of the general  fault path $E$. 
Now each general fault in $E$ is a linear sum of physical operators: 
\beq
{\cal E}_{A_{i,t}}'= {\cal E}_{A_{i,t}}-(1-\eta)I
\enq
Inserting this to equation \ref{bimi}, and we get a linear sum over terms 
which correspond to physical fault paths, 
with sparse faults. 
\beq
{\cal L}_g\cdot \rho(0)=\sum_{f} \lambda_f  {\cal L}_f\cdot \rho(0) 
\enq
Lemma \ref{Theproof} applies to each term in the sum. 
We get that each density matrix ${\cal L}_E\cdot \rho(0)$
is $(r,d)-$deviated from correct. 
This means that 
\beq 
\sum_{bad~i}[{\cal L}_g\cdot \rho(0)]_{i,i}=
\sum_f \lambda_f \sum_{bad~i} [{\cal L}_f\cdot \rho(0)]_{i,i}=(\sum_f \lambda_f) \epsilon'=
\Tr( {\cal L}_g\cdot \rho(0))\epsilon' \le \|{\cal L}_g\|\epsilon \le (1+\epsilon)\epsilon',
\enq
using lemma \ref{small} in the last inequality. $\Box$

\subsection{The Threshold Theorem for General Noise}

We can now prove the threshold result for general noise:

\begin{theo} {\bf The Threshold Theorem for General Noise:}

Let $\epsilon>0$.   Let $C$ be a computation code with gates $\calG$. 
There exists a threshold $\eta'_c>0$, and constants $c_1,c_2,c_3$
 such that the following holds. 
Let $Q$ be a 
 quantum circuit, with $n$ input qubits, which operates $t$ time steps,  
uses $s$ gates from 
${\cal G}$, and has $v$ locations.
There exists a quantum circuit  $Q'$  which 
operates on $n\rm{log^{c_1}}(\frac{v}{\epsilon})$ qubits, 
for time $t \rm{log^{c_2}}(\frac{v}{\epsilon})$, and 
uses $s \rm{log^{c_3}}(\frac{v}{\epsilon})$ gates from ${\cal G}$
such that  in the presence of 
general noise with error rate  $\eta< \eta'_c$, 
$Q'$ computes a function  which is $\epsilon$-close to that 
computed by $Q$. $\Box$. 
\end{theo}

\noindent {\bf Proof:}
Let $Q$ compute $f$ with accuracy $\epsilon'$, meaning 
that for any input $i$, the output is $f(i)$ with probability at least 
$1-\epsilon'$. 
We construct  $Q'$ 
which computes $f$ with accuracy $\epsilon+\epsilon'$.
$Q'$ will be the $r-$simulation of $Q$, where  
 $r$,  is chosen such that 
the requirements of lemma \ref{small} are satisfied with $\epsilon''=\epsilon/(1+\epsilon')$. 
We write the probability to measure 
a bad basis state as the sum of the diagonal elements of the density matrix $\rho(t)$ corresponding to bad states. 
\beq
\sum_{bad ~i} \rho(t)_{i,i}=\sum_{bad ~i}[{\cal L}_b\cdot \rho(0)]_{i,i}+
\sum_{bad~i}[{\cal L}_g\cdot \rho(0)]_{i,i}
\enq
And hence
\beq
~~~~~~~~~|\sum_{bad ~i} \rho(t)_{i,i}|\le 
\sum_{bad ~i}|[{\cal L}_b\cdot \rho(0)]_{i,i}|+
|\sum_{bad~i}[{\cal L}_g\cdot \rho(0)]_{i,i}|\le \|{\cal L}_b\cdot \rho(0)\|+|\sum_{bad~i}[{\cal L}_g\cdot \rho(0)]_{i,i}|\le \epsilon''+(1+\epsilon'')\epsilon'
\enq
The first inequality follows from the fact that $\sum_i |\rho_{i,i}|\le \|\rho\|$ for any 
Hermitian matrix (see \cite{aharonov3}.  
 Lemmas \ref{small} and \ref{gooood} are used to derive the second inequality.
Due to the choice of $\epsilon$''
This probability is indeed smaller than $\le \epsilon+\epsilon'$.
$\Box$

\section{Fault Tolerance with  Any Universal Set of Gates}\label{sec9}
So Far, the reliable circuits which we have constructed can use only 
universal set of gates associated with a quantum computation code, such 
as the sets $\calG_1$ and$ \calG_2$. This is an undesirable situation, both theoretically and 
practically. Theoretically, we would like to be able to show that the fault tolerance result 
is robust, meaning that quantum computation can be performed  fault tolerantly
 regardless of the universal set of gates which we use.
Practically, it is likely that the sets of gates ${\cal G}_1$ or  ${\cal G}_2$
are difficult to implement in the laboratory, since they contain gates 
involving three qubits,  and we would like to be able to use 
other, perhaps simpler, universal sets of gates. 
Indeed,  in this chapter we provide the desired generalization, and 
show that the threshold result holds for any 
universal set of gates $\calG$. 
We require, however, that $\calG$ contains 
a gate which discards a qubit, and a gate 
which adds a blank qubit to the circuit.

Let $\calG$ be a set of gates of a computation code,  $\calG'$
be a set of gates to be used for computation. 
After constructing the reliable circuit using $\calG$, 
 we will approximate 
each gate by gates from   $\calG'$, such that 
the sum of the errors of each gate is smaller than the 
allowed error. The threshold is of course  worse, and  will depend on the set of 
gates $\calG'$. 
\begin{deff}
The Threshold condition for General unitary gates.

Let  $S(\delta)$ be the number of gates required  to approximate
a gate from $\calG$ by gates from $\calG'$ to accuracy $\delta$. 
For any $\delta<\eta'_c$, 
we define  $\eta''_c$ to be 
\end{deff}
\beq
 \eta''_c=\max_{\delta} \frac{\eta'_c-\delta}{S(\delta)}
\enq
$\eta<  \eta''_c$ guarantees that  $S(\delta_0)\eta+\delta_0 < \eta'_c$, 
thence the total error of all gates approximating one gate 
of the computation code will not exceed $\eta'_c$,
so the fault tolerance scheme for general noise will apply.

\begin{theo}\label{full} {\bf The Threshold Result in Full Generality:}

Let $\epsilon>0$.  Let ${\cal G}',{\cal G}''$ 
be two universal sets of quantum gates.
There exists a threshold $\eta''_c>0$, and constants $c_1,c_2,c_3,c_4$
 such that the following holds. 
Let $Q'$ be a 
 quantum circuit, with $n$ input qubits, which operates $t$ time steps,  
uses $s$ gates from 
${\cal G'}$, and has $v$ locations.
There exists a quantum circuit  $Q''$  which 
operates on $c_1 n\rm{log^{c_2}}(\frac{v}{\epsilon})$ qubits, 
for time $c_3 t \rm{log^{c_2}}(\frac{v}{\epsilon})$, and 
uses $c_4 s \rm{log^{c_2}}(\frac{v}{\epsilon})$ gates from ${\cal G''}$
such that  in the presence of 
general noise with error rate  $\eta< \eta''_c$, 
$Q''$ computes a function  which is $\epsilon$-close to that 
computed by $Q'$. $\Box$. 
\end{theo}

{\bf Proof:}
We first approximate $Q'$ by
 a circuit $M_0$  which uses only gates from 
the a set of gates ${\cal G}$ of a computation code.  
Due to the Kitaev-Solovay theorem, this approximation can be done, with polylogarithmic cost
of space and time, to an arbitrary accuracy.
We will require that each gate is approximated up to $1/T$, so that 
$M_0$ computes a function which is $\epsilon/2$ close to that 
computed by $Q'$. We then construct $M_r$, 
the $r-$simulation of $M_0$, which again uses 
 gates from $\calG$. 
$r$ is chosen such that the general threshold scheme will give an $\epsilon/2$ 
error when the error rate in is taken to be  
 $\eta_e=S(\delta_0)\eta+\delta_0$. 
To construct $Q''$,  we replace each gate in $M_r$ by $S(\delta_0)$ gates 
from $\calG''$, up to $\delta_0$. $\Box$

\section{Robustness Against Exponentially Decaying Correlations}\label{robust}

We would now like to show how the above results hold also 
in the case of exponentially decaying correlations between the 
noise processes, in both space and time, as is described in 
subsection \ref{subaddcor}.  
We observe that all the lemmas which we use to prove the 
threshold theorems hold in this case, except 
for one step which fails.
It is the step that  shows that bad fault paths are rare,
in the case of probabilistic noise (lemma \ref{bad}).
% or that the contribution of the bad fault paths is
% negligible, in the case of general noise (lemma \ref{small}.)
The proof of this lemma rely on the independence of faults. 
%or the fact the noise operators on different locations 
%are in a tensor product.  
We observe that the proof of lemma \ref{bad}
% and \ref{small} 
is actually a union bounds. 
In this section we show that the same threshold as is used for 
probabilistic noise, 
(definition \ref{thre}) guarantees that
 the bad fault paths are negligible also is the presence of  
 exponentially decaying correlations, for 
probabilistic noise.

We use a union bound argument, 
as follows. 
Consider fault paths in  $v$ $r-$rectangles. 
If a fault path is bad, there must be at least 
one $r-$rectangle in which it is bad. 
In this rectangle, let us first count 
the number of minimal bad fault paths.
If $F_r$ is this number, we have:
 \beq
F_{1}\le
\left(\begin{array}{c}A\\k+1\end{array}\right)
\enq
and 

\beq
F_{r+1}\le
\left(\begin{array}{c}A\\k+1\end{array}\right) F_r^{k+1}
\enq
We can solve the recursion, to get 

\beq
F_{r}\le
\left(\begin{array}{c}A\\k+1\end{array}\right)^{\frac{(k+1)^r-1}{k}}
\enq

A minimal bad fault path contains exactly $(k+1)^r$ locations. 
Now, $v$ $r$-rectangles contain $v A^r$ locations. 
We can bound the number of bad fault paths in $v$ $r$-rectangles
 consisting of 
$(k+1)^r+i$ locations by choosing one $r-$rectangle (This gives 
a factor of $v$).
In this rectangle  
we pick one of the possible minimal bad fault paths (This gives a factor 
or $F_r$). We can then choose the rest of the locations arbitrarily. 
This gives that the number of   bad fault paths in $v$ $r$-rectangles
 consisting of 
$(k+1)^r+i$ locations is at most:
\beq
v\left(\begin{array}{c}A\\k+1\end{array}\right)^{\frac{(k+1)^r-1}{k}}
 \left(\begin{array}{c}v A^r-(k+1)^r\\i\end{array}\right)
\enq
 
Using the assumption \ref{expdecnoi0} 
%and \ref{expdecnoi}
 on the exponentially decaying 
probability 
%(or norm)
 of a fault path 
consisting of $k$ locations, we have that 
the overall probability 
%(or norm)
 of the bad fault paths is at most:

\begin{eqnarray}
\sum_{i=0}^{v A^r-(k+1)^r}
v\left(\begin{array}{c}A\\k+1\end{array}\right)^{\frac{(k+1)^r-1}{k}}
 \left(\begin{array}{c}v A^r-(k+1)^r\\i\end{array}\right) 
c\eta^{(k+1)^r+i}(1-\eta)^{v A^r-(k+1)^r-i}\le
\\\nonumber
cv(\left(\begin{array}{c}A\\k+1\end{array}\right)^{\frac{1}{k}}\eta)^{(k+1)^r}
\sum_{i=1}^{v A^r-(k+1)^r} \left(\begin{array}{c}v A^r-(k+1)^r\\i\end{array}\right) 
\eta^i(1-\eta)^{v A^r-(k+1)^r-i}=\\\nonumber
cv(\left(\begin{array}{c}A\\k+1\end{array}\right)^{\frac{1}{k}}\eta)^{(k+1)^r}
\end{eqnarray}

This completes the proof for this model of noise, 
since the expression above decays exponentially fast to zero with $r$, 
if $\eta$ is below the threshold for probabilistic noise from definition 
\ref{thre}:
 
\beq
\left(\begin{array}{c}A\\k+1\end{array}\right)\eta^k<1.
\enq

\section{Fault Tolerance in a d-Dimensional Quantum Computer}\label{sec10}
So far, we allowed a gate to operate on any set of qubits, regardless on the 
actual location of these qubits in space. 
We call this case the no-geometry case. 
We consider also another case, in which 
the qubits are located a one array, 
and the gates can be applied only on nearest neighbor qubits. 
 For more dimensions, the gates are restricted to operate
on qubits which are nearest neighbors and lie in the same line. 
We show that the threshold result holds in full 
generality for $d-$dimensional quantum computers, 
for any $d\ge 1$.  
Note that for the geometry of the circuit to be preserved, we require 
that the number of qubits is preserved throughout the computation.
Hence discarding and adding qubits are not allowed.  
However, we still need a gate which allows to put entropy out of the system.
This will be the RESTART gate, which is constructed by discarding a qubit and then 
adding a blank qubit instead of the discarded qubit. 
We also need the SWAP gate, which switches between two qubit, i.e. it applies 
the unitary transformation
$|i\ra|j\ra\leftrightarrow |j\ra|i\ra$.  
We thus require that the universal set of gates which is used in the
final circuit, $\calG''$,  
 contains the SWAP and the RESTART gates. 

\begin{theo}\label{dd} {\bf Threshold theorem for $d$-dimensional circuits::}

Let $\epsilon>0$. Let $d\ge 1$. Let ${\cal G}',{\cal G}''$ 
be two universal sets of quantum gates.
There exists a threshold $\eta''_c>0$, and constants $c_1,c_2,c_3,c_4$
 such that the following holds. 
Let $Q'$ be a d-dimensional 
 quantum circuit, with $n$ input qubits, which operates $t$ time steps,  
uses $s$ gates from 
${\cal G'}$, and has $v$ locations.
There exists a d-dimensional quantum circuit  $Q''$  which 
operates on $c_1 n\rm{log^{c_2}}(\frac{v}{\epsilon})$ qubits, 
for time $c_3 t \rm{log^{c_2}}(\frac{v}{\epsilon})$, and 
uses $c_4 s \rm{log^{c_2}}(\frac{v}{\epsilon})$ gates from ${\cal G''}\cup \{SWAP\}$
such that  in the presence of 
general noise with error rate  $\eta< \eta''_c$, 
$Q''$ computes a function  which is $\epsilon$-close to that 
computed by $Q'$. 
\end{theo}

\noindent{\bf Proof:} 
The only difference from the proof of the threshold theorem for 
circuits without geometry is that the fault tolerant procedures need to 
apply gates on nearest neighbors only, and that 
all the qubits remain in the circuit throughout the computation.
Here is how one level of the simulation is done.  
We first pick a preferred direction, and each qubit will be extended to 
an array of qubits lying in that direction. 
The simulation will ``stretch'' the simulated circuit only in the preferred direction, 
by a constant factor.  
Let $a$ be the maximal number of ancilla qubits used in all the procedures 
of the computation code ${\cal G}$. Let $m$ be the size of the block. 
A qubit in $Q_1$ will be replaced by $m+a$ qubits, 
placed in a line along the preferred direction.
The ancilla qubits will serve as working space, but we will 
also SWAP qubits with ancilla qubits, in order to bring computation qubits closer 
and operate gates on them. 
The fault tolerant procedures are modified as follows. 
First, instead of  adding ancilla qubits during the procedure, 
we only use the ancilla qubits that are already there, and  
apply a RESTART 
gate on an ancilla qubit one step before we use it
in the procedure.  
Also, any gate $g$ 
in the procedure, which operates on 
qubits which are far apart, is replaced by  
 a sequence of SWAP gates 
 which bring the qubits $g$ operates on to nearest neighbor sites, 
 followed by $g$, which is then followed by another sequence 
of  SWAP gates to bring  the qubits back to their original sites.
Since the simulated circuit applies gates on nearest neighbors,
say on $k$ qubits in a raw,  
the number of SWAP gates required is at most $2(k+1)m$, 
i.e. a constant.  

The claim is that the procedure is still fault tolerant. 
This might seem strange since the SWAP gates operate on many qubits, 
 and seem to help in propagation of errors. 
 However, note that a SWAP gate which operates on a faulty qubit 
and an unaffected qubit does not propagate the error 
to the two qubits, but keeps it confined to the original qubit, 
which is now in a new site. 
Hence, a SWAP gate which is not faulty
 does not 
cause propagation of error. 
If an error occurred in a SWAP gate, then the two qubits participating 
in it are contaminated. This will cause contamination of all the qubits 
on which $g$ operates on, (in the worst case), so an error in a SWAP gate 
is equivalent to an error in all the original 
sites of the qubits participating in the gate, 
and the final site of the other qubit participating in the SWAP gate. 
This adds a factor of $2$ to the original 
spread of the procedure. 
All other aspects of the theorem remain the same. $\Box$.

\section{Conclusions and Open Problems}\label{sec11}
The result implies that quantum computation might be practical if the 
noise in the system can be made very small.
We hope these results 
motivate physicists to achieve lower error rates, and
theoreticians to develop codes with 
better parameters, in order to push the threshold as high as possible.
The point at which the physical data meets the theoretical threshold
 is where quantum computation becomes
practical.

We did not attempt to optimize the threshold 
in the fault tolerant scheme. 
The reason is that this threshold depends on a lot of parameters, such as 
whether the system is $d-$dimensional or whether it has no geometry; 
Whether measurements are allowed or not, and which set of gates 
can be implemented in the laboratory. 
We did give a rough estimate of the threshold in one case, 
in which the length of the code 
is  $m=7$, in both cases of CSS and polynomial codes, in 
 a circuit with no geometry, and 
with noiseless classical operations and measurements allowed. 
The threshold then is estimated to be $\approx 10^{-6}$. 
This threshold is far from being practical, according to the
 state of the art, and optimization is certainly required. 
 One can save space by distinguishing 
two types of qubits: regular qubits, and ``classical'' 
 qubits which we do not need to protect from phase flips, but only from bit-flips. 
Such are the qubits involved in the classical computation of the error 
given the syndrome. These qubits can be encoded using 
classical error corrections.
In practical cases, it might be possible to allow measurements 
on such qubits, during the computation,
 and apply noiseless classical computation instead of 
quantum computation on these qubits, followed by quantum gates 
conditioned on the classical results of these computations. 
Such ideas might reduce the threshold by several orders of magnitude.

%We are grateful to I. Cirac and P. Zoller for the following idea:
%In most quantum systems, it is reasonable to assume that different 
%types of faults occur with different frequencies. 
%In such systems, one can improve the bound on the noise rate significantly,
%by using in most levels of the simulation a quantum code which can correct
%only for the most frequent errors,
%while for less frequent errors it is enough to correct only once in
%a few levels.
%(For example, a code which 
%has block size $m=3$, and
%corrects against bit errors is suggested in  \cite{},  
%while the smallest code which corrects against any
% single qubit error must have $m\ge 5$.) 
%Examples will be given in the final version.
%It is therefore important  to have  good understanding of the
% different noise rates for
% different types of faults,
%for specific potential physical realizations of quantum computers.

%Quantum block codes can correct noise when it is independent 
%for different qubits, or even when the noise is correlated
% in small
% sets of qubits.
%It was argued \cite{} that in quantum
% systems there are other types of noise, 
%with long correlations, that quantum block  codes can not cope with.
%We\cite{} intend to generalize the results, 
% using a quantum version of multi-linear codes\cite{},
%an extension which we hope will be noise resistant even when  
%long correlations exist.

Our scheme requires a polylogarithmic blow-up in the 
depth of the circuit.
It might be possible to use a 
quantum analogue of  multi linear codes\cite{spielman}, 
to reduce the  multiplicative factor of $O(log(n))$ to  a factor of 
$O(log(log(n))$. 
An open question is whether it is possible to reduce 
the time cost  to a constant, as in the classical case. 
We conjecture that the answer is negative.

An interesting direction to pursue  is  to consider 
different assumptions on the noise, such as very strong correlations 
in time and space between the noise process. 
Results in this direction were found by 
Lidar Chuang and Whaley\cite{lidar}. 

The threshold result might have an impact on a long standing question in 
quantum physics, regarding the transition from quantum to classical
physics\cite{zurek1}.
Traditionally, this transition is treated by 
taking the limit of Planck's constant to $0$, 
and it is viewed as a gradual transition (but see \cite{tsirelson1}). 
In \cite{aharonov2} it
 was shown that for  a very high noise rate, the quantum circuit
behaves in a classical way.
It is interesting to consider a different point of view, in which 
the definition of quantum versus classical behavior is computational. 
In this paper we show that for very small noise rate, quantum systems 
can 
maintain their quantum nature.
In a previous paper, we have shown that quantum systems
can be simulated efficiently by a classical Turing machine 
 if the noise is large\cite{aharonov2}.
Suppose that  indeed quantum behavior cannot be simulated efficiently by classical 
systems, an idea suggested by  Feynmann\cite{feynman1} which originated 
quantum computation. In other words, suppose  $BPP\ne BQP$. 
Then, increasing the noise, a transition from the quantum computational
 behavior
 to classical computational behavior occurs.
Does this transition happen at a critical error rate?
Indications for a positive answer are already shown in a previous paper 
of ours \cite{aharonov2}, which might suggest that the transition 
from quantum to classical physics occurs via a phase transition. 
We view this connection between quantum complexity and quantum physics
as extremely interesting.

\section{Acknowledgments}
We wish to thank  Peter Shor for discussions about his result, 
 Thomas Beth for helping to construct the fault tolerant
Toffoli gate without measurements for the CSS codes,
Richard Cleve for asking the question of nearest neighbors 
 and suggesting the solution, 
Noam Nisan for a fruitful discussion regarding the correct model to use,  
Prasad Gospel, Denis Gaitsgori and Erez Lapid for 
helpful discussions regarding lemmas \ref{one} and \ref{corkitaev}. 
We are grateful to Bob Solovay for detecting and correcting 
crucial errors in lemmas  \ref{one}, \ref{corkitaev} and \ref{bridge}
in an early draft of this paper. 
Finally, we thank Alesha Kitaev 
for a very helpful remark regarding the 
generalization to the general noise model. 
 This research was supported by The Israel Science Foundation, grant number
 $69/96$, 
and the Minerva Leibniz Center at the Hebrew University. 

\small
\bibliographystyle{plain}

\normalsize

\end{document}